\documentclass[aps,pra,twocolumn,floats,superscriptaddress,tighten,letterpaper,floatfix,eqsecnum]{revtex4}
\usepackage{bbm,mathtools}
\usepackage[utf8]{inputenc}
\usepackage{amssymb}
\usepackage{amsbsy}
\usepackage{amsmath}
\usepackage{graphicx}
\usepackage{graphics}
\usepackage{setspace}
\usepackage{array}
\usepackage{color}
\usepackage{xcolor}
\usepackage{fontenc}
\usepackage{textcomp}
\usepackage{rotating}
\usepackage{bm}
\usepackage{braket}
\usepackage[colorlinks=true,linkcolor=black,bookmarksopen=false,urlcolor=blue,citecolor=blue]{hyperref}
\hypersetup{pdfpagemode=UseNone}
\usepackage{chngcntr}
\usepackage{subfigure}

\DeclareMathOperator{\e}{e}
\DeclareMathOperator{\sech}{sech}

\newcommand{\vex}[1]{\bm{\mathrm{#1}}}

\newcommand{\bsub}{\begin{subequations}}
\newcommand{\esub}{\end{subequations}}
\newcommand{\hsig}{\hat{\sigma}}
\newcommand{\hSig}{\hat{\Sigma}}
\newcommand{\htau}{\hat{\tau}}
\newcommand{\tilE}{\tilde{\varepsilon}}
\newcommand{\varE}{\tilde{\varepsilon}}
\newcommand{\Gel}{\Upsilon_{\mathsf{el}}}
\newcommand{\sfT}{\mathsf{T}}
\newcommand{\sfP}{\mathsf{P}}

\newcommand{\sfs}{\mathsf{s}}
\newcommand{\sfb}{\mathsf{b}}
\newcommand{\sfbb}{\mathsf{bb}}
\newcommand{\sfss}{\mathsf{ss}}
\newcommand{\sfsb}{\mathsf{sb}}
\newcommand{\sfsp}{\mathsf{sp}}
\newcommand{\sfd}{\mathsf{d}} 
\newcommand{\qmin}{q_{\mathsf{min}}}
\newcommand{\cl}{\mathsf{cl}}
\newcommand{\LO}{\mathsf{LO}}
\newcommand{\q}{\mathsf{q}}
\newcommand{\swave}{\mathsf{s-wave}}
\newcommand{\bdg}{\mathsf{BdG}}

\newcommand{\ext}{\mathsf{ext}}

\newcommand{\Kb}{K_{\sfb}}

\newcommand{\lcoh}{l_{\mathsf{coh}}}

\newcommand{\LL}{\lambda_L}

\newcommand{\Qsp}{\hat{Q}_{\mathsf{sp}}}
\newcommand{\QspSwave}{\hat{Q}_{\mathsf{sp}}^{\mathsf{s-wave}}}

\newcommand{\tel}{\tau_{\mathsf{el}}}
\newcommand{\Gdress}{\hat{{\cal G}}_{\Upsilon}}

\newcommand{\hG}{\hat{{\cal G}}}
\newcommand{\BigA}{\bm{\hat{{\cal A}}}}
\newcommand{\frakc}{\mathfrak{c}} 
 
\newcommand{\Tr}{\mathsf{Tr}}
\newcommand{\sign}{\mathsf{sgn}}
\newcommand{\re}{\text{Re} \,}
\newcommand{\im}{\text{Im} \,}

\let\normalint\int 
\def\nint{\displaystyle\normalint} 

\let\normalsum\sum 
\def\nsum{\displaystyle\normalsum} 

\begin{document}

\title{
Topological Anomalous Skin Effect in Weyl Superconductors
}

\author{Tsz Chun Wu}
\affiliation{Department of Physics and Astronomy, Rice University, Houston, Texas 77005, USA}

\author{Hridis K. Pal}
\affiliation{Department of Physics, IIT Bombay, Powai, Mumbai 400076, India}

\author{Matthew S. Foster}
\affiliation{Department of Physics and Astronomy, Rice University, Houston, Texas 77005, USA}
\affiliation{Rice Center for Quantum Materials, Rice University, Houston, Texas 77005, USA}

\date{\today}

\begin{abstract}
We show that a Weyl superconductor can absorb light via a novel surface-to-bulk mechanism,
which we dub the topological anomalous skin effect. This occurs even in the absence of disorder for a single-band 
superconductor, and is facilitated by the topological splitting of the Hilbert space into bulk and chiral surface Majorana states. 
In the clean limit, the effect manifests as a characteristic absorption peak due to surface-bulk transitions. 
We also consider the effects of bulk disorder, using the Keldysh response theory. 
For weak disorder, the bulk response is reminiscent of the Mattis-Bardeen result for $s$-wave superconductors, 
with strongly suppressed spectral weight below twice the pairing energy, despite the presence of gapless Weyl points.
For stronger disorder, the bulk response becomes more Drude-like and the $p$-wave features disappear.  
We show that the surface-bulk signal survives when combined with the bulk in the presence of weak disorder. 
The topological anomalous skin effect can therefore serve as a fingerprint for Weyl superconductivity. 
We also compute the Meissner response in the slab geometry, incorporating the effect of the surface states.  
\end{abstract}

\maketitle

\tableofcontents

\section{Introduction}
\label{intro}

Despite displaying perfect dissipationless conduction at zero frequency, 
superconductors absorb electromagnetic radiation at finite frequencies. 
The classical skin depth of a metal is given by 
$\delta(\omega) = c / \sqrt{2 \pi \sigma_{\mathsf{dc}} \omega}$,
where $\omega$ is the radiation frequency and $\sigma_{\mathsf{dc}}$ is
the static, zero-frequency conductivity due to impurity scattering \cite{LLv10}. 
The classical skin depth vanishes in the clean limit. The absorption in a superconductor
is associated to a nonzero field penetration depth at finite frequencies due to 
the pairing, giving rise to an anomalous skin effect and associated optical conductivity \cite{Tinkham,Abrikosov}.

\begin{figure}[b]
	\centering
	{\includegraphics[width=0.48\textwidth]{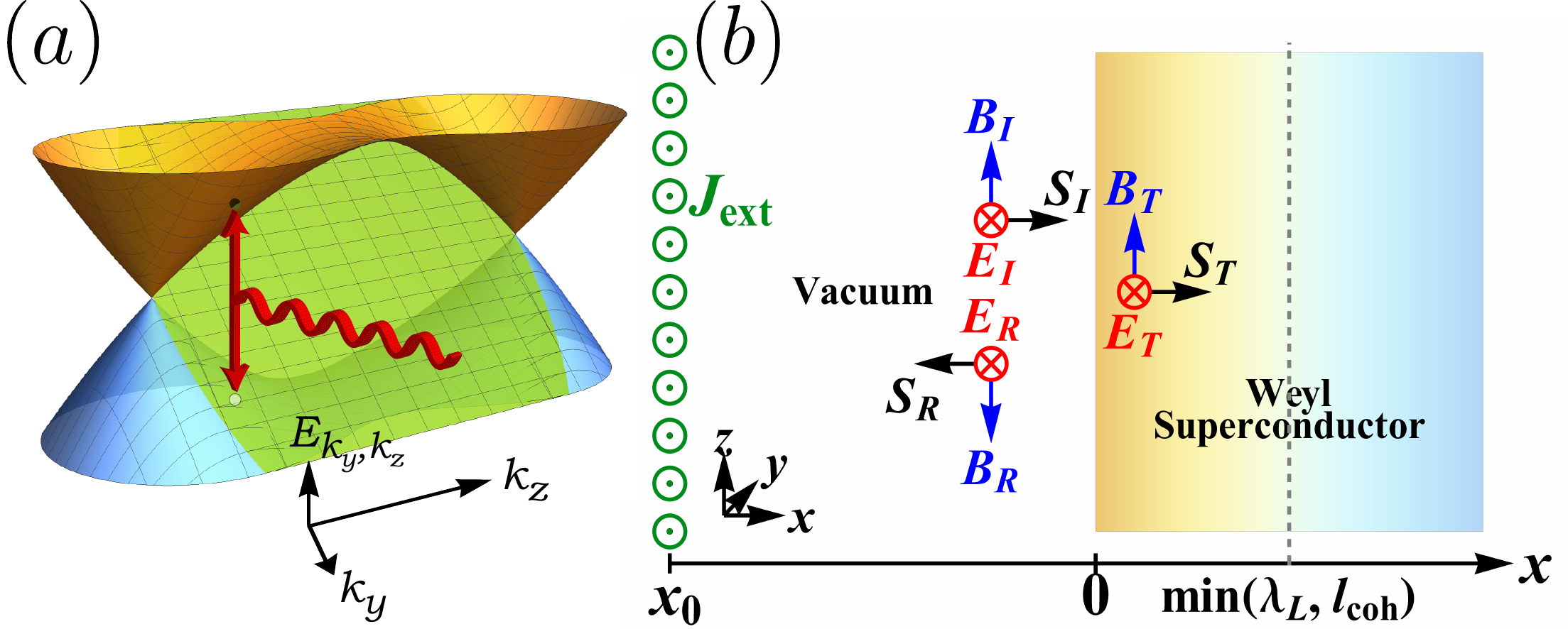} }
	\caption{Geometry for the topological anomalous skin effect.
		(a) We consider a Weyl superconductor described by Eq.~(\ref{eq:HBdG}). 
		Due to bulk $k_x + i k_y$ pairing, a single pair of Weyl points arises at $k_z = \pm k_F$. 
		In the figure, the green plane depicts the dispersion of the chiral surface state $E^{\sfs}_{k_y} = -\Delta k_y$,
		for a superconductor occupying the half space $x \geq 0$. 
		Since we consider the surface only at $x = 0$, there is only one branch for the surface states.  
		The topological anomalous skin effect arises via absorption due to 
		optical transitions between surface and bulk states, as indicated by the red vertical arrow
		in the sketch. 
		(b) 
		We consider a plane electromagnetic wave impacting the Weyl superconductor occupying the 
		$x \geq 0$ half-space at normal incidence. The label $I$, $R$ and $T$ denote respectively the incident, 
		reflected and transmitted component of the radiation. 
		$(\textbf{E},\textbf{B},\textbf{S})$ stands for (E-field, B-field, Poynting vector). 
		A polarization along $\hat{y}$ (i.e. perpendicular to the plane of incidence) 
		is assumed for the electric field $\textbf{E}$ in this sketch. 
		The topological anomalous skin effect mainly takes place in the yellow region, extending up to the scale of $\min(\LL,\lcoh)$.
		Here $\LL$ denotes the London depth, and $\lcoh$ is the coherence length (which is the minimal confinement depth for the
		chiral Majorana surface fluid). 
	}  	
	\label{fig:Geometry}
\end{figure}

In a topologically trivial one-band superconductor, the dissipative part of the bulk optical conductivity 
vanishes in the clean limit \cite{s_wave_opt_resp_Mahan}. This result obtains due to the orthogonality
of positive- and negative-energy bulk states and the lack of matrix structure for the current operator,
despite the particle-hole hybridization induced by pairing. However, most low-temperature
superconductors are measured in the dirty limit, where $\Delta_0 \ll 1/\tau_{\mathsf{el}} \ll \varepsilon_F$.
Here $\Delta_0$ is the pairing energy, $\tau_{\mathsf{el}}$ is the lifetime due to elastic impurity scattering,
and $\varepsilon_F$ is the Fermi energy. For an $s$-wave superconductor at zero temperature, absorption turns on 
at $\omega = 2 \Delta_0$ according to the famous Mattis-Bardeen result for the dirty limit
\cite{s_wave_MB_original,Tinkham,s_wave_opt_resp_Mahan}. 

What happens when the Hilbert space of a one-band superconductor is split into bulk and surface states by 
non-trivial topology? In this paper, we will show that a novel surface-bulk absorption can occur
in a topological superconductor \cite{classificiation_gapless_Ryu,classificiation_Ryu_RMP}, 
even for a one-band system in the \emph{clean limit}. 
We dub this the topological anomalous skin effect. 
We consider optical absorption by Weyl superconductors (WSCs) 
\cite{WSC_Balents,WSC_Sau,WSC_Sato_review,WSM_Ashvin_review,
Silaev_Fermi_arc_He3,
Moore_WSC_from_WSM,
Pallab_axionic_FT,
Fan_WSC_3D,
Liu_WSC_Fermi_gas,
Brydon_review_nodal_SC,
Volovik_Weyl_fermions,
Yuan_WSC_NbBiSe,
Yanase_MobiusTSC_UPt3,
Kallin_UPt3,
Pacholski_WSC_LL,
Okugawa_WSC_phase_diagram,
Yanase_classification2,
Yanase_classification,
Ishihara_He3A,
Nakai_WSC_engineering}.
WSCs can arise due to bulk $p + i p$ pairing, as exemplified by ${}^3$He$A$ \cite{He3_book_Vollhardt,He3_book_Volovik,He3B_review_Sato_Machida}. 
In static mean-field theory, the quasiparticle spectrum of a WSC exhibits pairs of gapless Weyl nodes in the bulk.
Each momentum-separated nodal pair gives rise to a chiral, two-dimensional (2D) Majorana surface fluid, displaying a Fermi arc connecting the nodes
\cite{WSM_Ashvin_review,He3B_review_Sato_Machida,WSC_Sato_review,classificiation_Ryu_RMP,classificiation_gapless_Ryu}. 
WSCs could serve as a platform for realizing Majorana zero modes and topological quantum computation \cite{review_quantum_computation}. 

\begin{figure*}[t]
{\includegraphics[width=16cm]{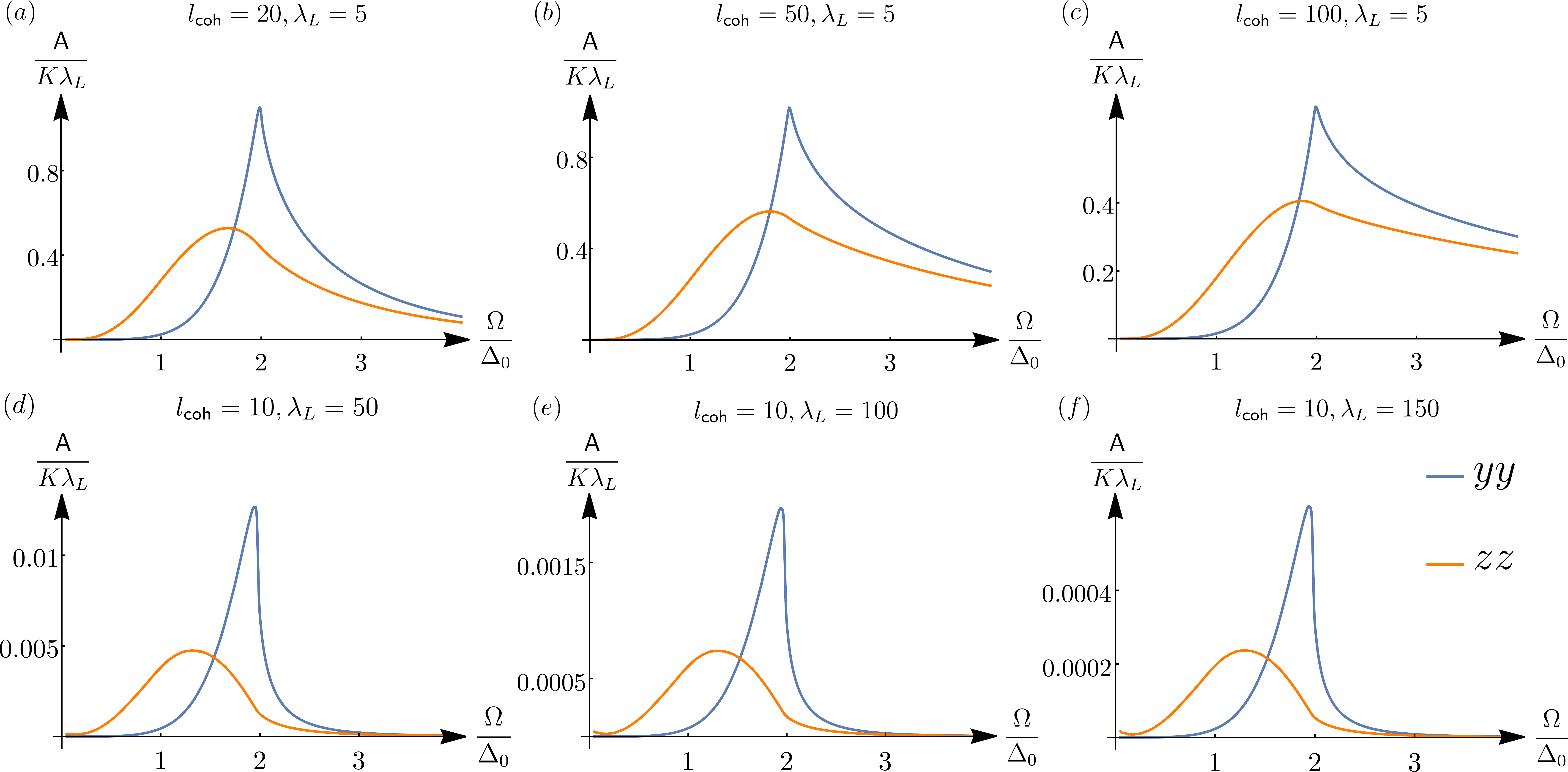} }
	\caption{The topological anomalous skin effect in the clean limit. 
		The panels in this figure plot the frequency ($\Omega$)-dependence of the absorbance 
		$\mathsf{A}^{yy}({\Omega})$ 
		(blue) 
		and 
		$\mathsf{A}^{zz}({\Omega})$ 
		(orange) 
		due to the surface-bulk transition,
		Eq.~(\ref{eq:absorbance_sb2--Intro}).
		Panels show results at zero temperature 
		for different combinations of the coherence length $\lcoh$ and the diamagnetic (London) penetration depth $\LL$. 
		The Weyl nodes lie along the $k_z$-axis in our model, so that 
		$\mathsf{A}^{yy}({\Omega})$ 	
		$\left[   \mathsf{A}^{zz}({\Omega})   \right]$ 
		encodes absorption for electric-field polarization perpendicular (parallel) to the surface Majorana Fermi arc.
		Parameters used in panels (a)--(c) correspond to the type-I superconductor regime, 
		while 
		those used in panels (d)--(f) correspond to the type-II regime. 
		Here, the frequency is normalized by the pairing gap energy $\Delta_0 = \Delta k_F$;
		$\Delta$ is the $p+ip$-wave pairing amplitude in the model [Eq.~(\ref{eq:HBdG})].  
		We set $k_F = 1$ for all the plots.}  	
	\label{fig:hatF}
\end{figure*}

We show that WSCs absorb radiation through optically driven surface-bulk transitions, see Fig.~\ref{fig:Geometry}.
We compute the surface-bulk absorption coefficient for a clean, spinless WSC with $p+ip$ pairing in the slab geometry. 
For a plane electromagnetic wave with normal incidence upon a crystal face with chiral surface states, we find
a relatively narrow (broad) peak around $2 \Delta_0$ (below $2 \Delta_0$) for electric polarization perpendicular to (along) the 
Fermi arc, see Fig.~\ref{fig:hatF}. 

\begin{figure}[b]
	\centering
	{\includegraphics[width=0.475\textwidth]{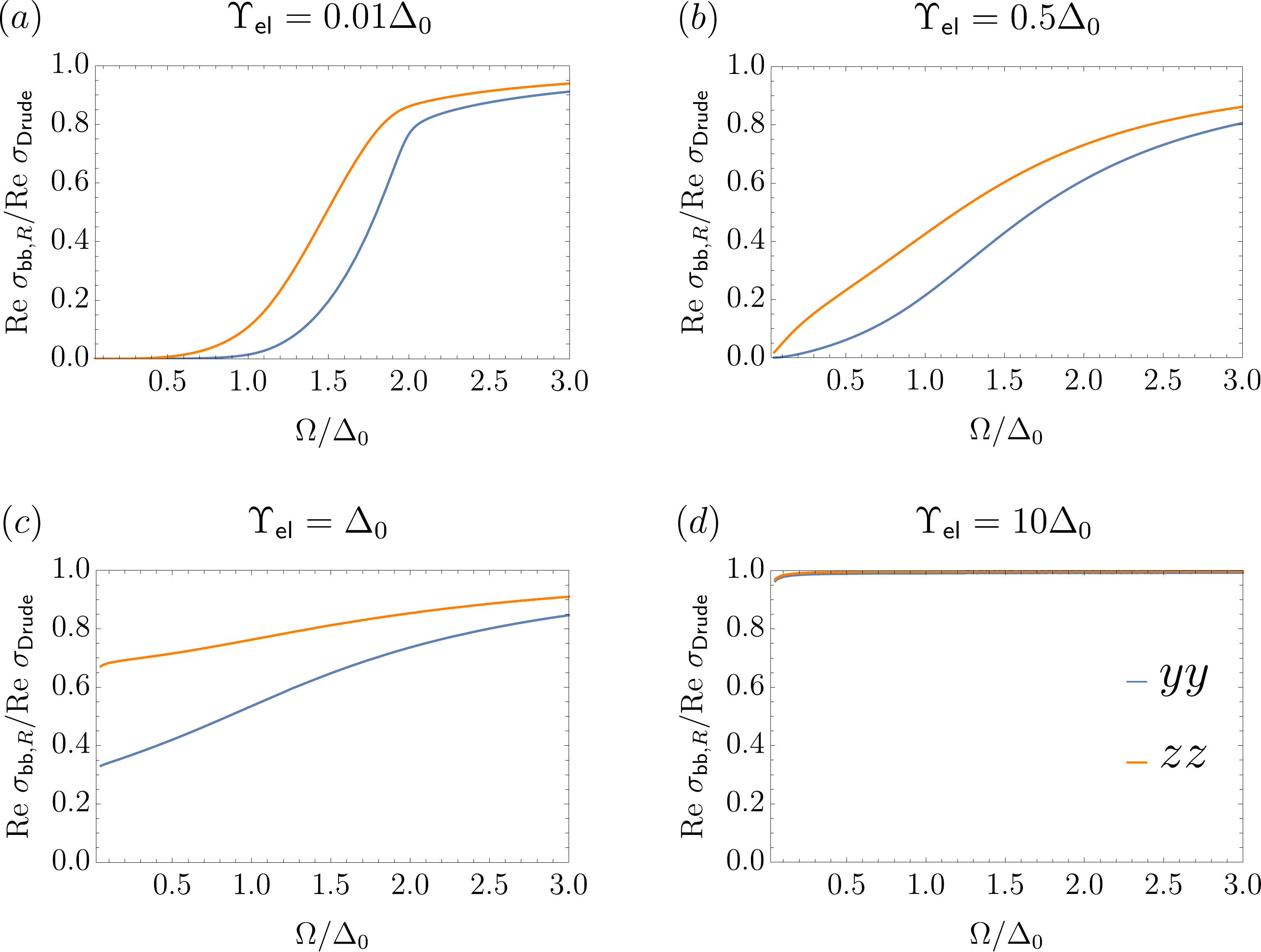} }
	\caption{The bulk optical conductivity of the dirty Weyl superconductor. 
		The panels in this figure plot the normalized optical conductivity 
		$\re \,\sigma_{\sfbb,R}^{\mu \nu}/\re \,\sigma_{\mathsf{Drude}}^{\mu \nu}$ 
		as a function of the reduced frequency $\Omega/\Delta_0$ at 
		zero temperature, 
		with disorder strength 
		(a) $\Gel = 0.01 \Delta_0$ 
		(b) $\Gel = 0.5 \Delta_0$
		(c) $\Gel =  \Delta_0$
		(d) $\Gel = 10 \Delta_0$.
		Here $\Gel = 1/(2 \tau_{\mathsf{el}})$, where $\tau_{\mathsf{el}}$ is the elastic lifetime due to impurity scattering in the normal state, 
		and $\sigma_{\mathsf{Drude}}^{\mu \nu} = \delta^{\mu \nu} \sigma_{\mathsf{dc}}/\left[1 + (\omega \tau_{\mathsf{el}})^2\right]$. 
		In the weak disorder regime, the frequency dependence of the optical conductivity is reminiscent of the Mattis-Bardeen result for a 
		dirty $s$-wave superconductor. The gapless excitations around the Weyl points cause a finite zero-temperature response for $\Omega < 2\Delta_0$. 
		As the disorder strength increases, the optical conductivity gradually approaches the Drude result. 
		This is consistent with the formation of a thermal metallic phase in the dirty limit 
		\cite{rare_WSC_Jed6,WSC_thermal_Hall_Bitan_Sayed,Sau2017}.}  	
	\label{fig:p_wave_sigma_bulk}
\end{figure}

We also consider the effects of disorder on the bulk $p$-wave state.
Using Keldysh response theory \cite{s_wave_NLSM1_Larkin,s_wave_NLSM4_Kamenev,s_wave_NLSM5_Yunxiang}, we 
derive the semiclassical optical conductivity for a Weyl superconductor.
Disorder is treated at the saddle-point level. The surface-bulk transitions giving rise to the topological
anomalous skin effect largely involve states away from the Weyl nodes, and therefore we do not expect rare
region effects
\cite{rare_WSM_Rahul_Huse_instanton,rare_WSM_eps_exp_Syzranov,rare_WSM_Jed3,rare_WSM_Jed4,rare_WSM_Jed5,rare_WSC_Jed6,rare_WSM_Jed1,rare_WSM_Jed2,rare_WSM_Jed7}
to play an important role.
For weak disorder, the bulk response is reminiscent of the Mattis-Bardeen result for $s$-wave superconductors, 
with strongly suppressed spectral weight below twice the pairing energy, despite the presence of gapless Weyl points.
This is consistent with canonical results for dirty anisotropic superconductors \cite{p_wave_opt_resp_Hirschfeld}.
For stronger disorder, the bulk response becomes more Drude-like and the $p$-wave features disappear.
Results are displayed in Fig.~\ref{fig:p_wave_sigma_bulk}. 

\begin{figure*}[t]
	\centering
	{\includegraphics[width=18cm]{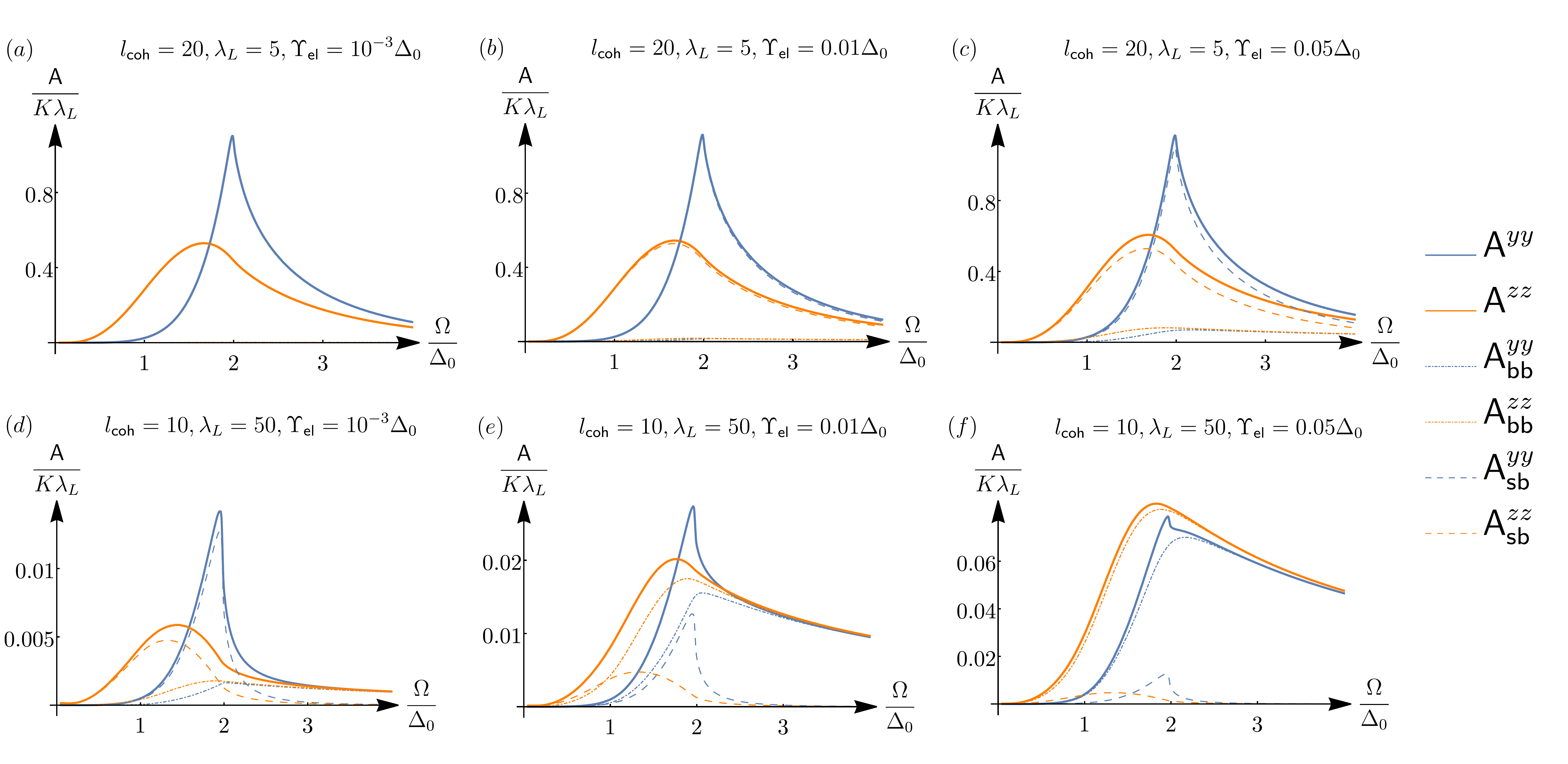} }
	\caption{Topological anomalous skin effect in the presence of weak disorder. 
		The panels in this figure plot the total absorbance ($\mathsf{A}$) 
		as a function of the reduced frequency 
		$\hat{\Omega} = \Omega / \Delta_0$, in the weak-disorder limit with different 
		coherence lengths $\lcoh$, 
		London penetration depths $\LL$,
		and disorder strengths $\Gel$  
		in the
		(a)--(c) type I and 
		(d)--(f) type II superconductor regimes.
		In all the panels, orange and blue curves represent the $yy$ and $zz$ components of the absorbance respectively,
		compare to Fig.~\ref{fig:hatF}.
		Long dashed lines correspond to the surface-bulk contribution, while short dashed lines correspond to the disordered 
		bulk contribution. 
		The total absorbance is depicted by the solid lines. 
		We use $k_F = 1$ for all the plots.
		In the type I case, the features from the surface-bulk absorption survive up to a relatively large amount of disorder. 
		In the type II case, those features are gradually suppressed by the bulk absorption as the disorder strength 
		increases.}  	
	\label{fig:combine_A}
\end{figure*}

We show that the surface-bulk signal survives when combined with the bulk in the presence of weak disorder,
see Fig.~\ref{fig:combine_A}. The topological anomalous skin effect can therefore serve as a fingerprint for Weyl superconductivity. 
We also compute the Meissner response in the slab geometry, incorporating the effect of the surface states. 
In the case of a strong topological superconductor analogous to ${}^3$He$B$ \cite{He3_book_Vollhardt,He3_book_Volovik,He3B_review_Sato_Machida},
an anomalous power-law-in-temperature dependence was found for the penetration depth, due to the paramagnetic surface state response 
\cite{He3B_Meissner}. The latter result is surprising because the bulk is fully gapped in that case. 
For the WSCs studied here, we do not find qualitatively new behavior in the Meissner effect arising from the presence of surface states.
This is because power-law temperature-dependence is already expected due to the bulk Weyl nodes.   

The Majorana surface fluid in WSCs could also be detected by 
scanning tunneling microscopy (STM), 
angle-resolved photoemission spectroscopy (ARPES) \cite{WSC_Sau}, 
and the anomalous thermal Hall effect \cite{WSC_Balents,WSC_thermal_Hall_Bitan_Sayed,WSC_thermal_Hall_Goswami_Andriy,WSC_SrPtAs_Sigrist}. 
Effects due to the axial anomaly have also been proposed as signatures for WSCs, 
including negative thermal magnetoresistance \cite{WSC_anomaly_Sato} and a $T^2$ temperature dependence 
of the axial current \cite{WSC_anomaly_Nissinen,WSC_anomaly_Nissinen_Volovik}.

A key difference between Weyl semimetals and Weyl superconductors is that the bulk optical conductivity of the former is
nonzero even in the clean limit, due to interband transitions \cite{Hosur2012,Bera2016,Roy2016,Roy2018}. 
Another key distinction concerns Fermi arc transport \cite{Potter2014,Baum2015}. 
For both fully gapped and gapless topological superconductors lacking spin SU(2) symmetry 
(classes DIII and D \cite{classification_Ludwig,classificiation_Ryu_RMP,WSC_Sato_review}), 
the coupling of the surface Majorana fluid to electromagnetism is effectively \emph{gravitational},
i.e.\ the electric density and current operators are formed from components of the stress tensor for the surface Majorana fluid \cite{He3B_Meissner,Sayed_QGD,GravityFootnote}.
For the chiral fluid at the surface of a WSC, this has the consequence that no thermal current flows along the 
Fermi arc in the presence of perpendicular magnetic flux, different from the surface states of Weyl semimetals \cite{Potter2014,Baum2015}. 

There are only a few bulk candidate materials that might exhibit Weyl superconductivity in nature.  
Older candidates of WSCs include the Uranium-based ferromagnetic superconductors 
UGe$_2$ \cite{WSC_UGe2}, 
URhGe \cite{WSC_URhGe}, 
and 
UCoGe \cite{WSC_UCoGe}, in which spin-triplet $p$-wave pairing is expected. 
More recent works suggest that 
the $B$ phase of UPt$_3$ \cite{WSC_thermal_Hall_Goswami_Andriy,WSC_Sato_review,WSC_UPt3_Yanase,Yanase_MobiusTSC_UPt3,Kallin_UPt3}, 
SrPtAs \cite{WSC_SrPtAs_Sigrist},
Praseodymium-based compounds (e.g.\ PrOs$_4$Sb$_{12}$ and PrPt$_4$Ge$_{12}$ \cite{WSC_Pr_family}) 
and
YiPtBi \cite{WSC_YiPtBi_JP,WSC_thermal_Hall_Bitan_Sayed,Brydon2016,Savary2017}
may host gapless Majorana surface states. 

Superconductivity was very recently observed in UTe$_2$ \cite{WSC_UTe2_JP2_Science}. 
There is already extensive experimental evidence pointing towards Weyl superconductivity in this compound
\cite{WSC_UTe2_Aoki_NMR,
	WSC_UTe2_JF_high_field,
	WSC_UTe2_JF_specific_heat,
	WSC_UTe2_JP1_surface_resistivity,
	WSC_UTe2_JP2_Science,
	WSC_UTe2_JP3_thermal_transport,
	WSC_UTe2_JP4_Kerr_rotation,
	WSC_UTe2_angular_specific_heat_Machida,
	WSC_UTe2_STM_Jiao_Nature,
	WSC_UTe2_JP5_muSR,
	WSC_UTe2_JP6_high_field}.
UTe$_2$ shows a relatively high transition temperature of $1.6$ K \cite{WSC_UTe2_JP2_Science}
and its superconductivity is suspected to be mediated by ferromagnetic spin-fluctuations 
in proximity to a quantum critical point \cite{WSC_UTe2_JP5_muSR,WSC_UTe2_JP6_high_field}. 

The nearly temperature-independent Knight shift across the transition temperature \cite{WSC_UTe2_JP2_Science,WSC_UTe2_Aoki_NMR} 
and an extremely high upper critical field exceeding the Pauli limit \cite{WSC_UTe2_JF_high_field} strongly support the spin-triplet 
pairing scenario. 
Penetration depth, thermal transport, spin-relaxation, and specific heat measurements together demonstrate strong evidence for 
point nodes lying along the crystallographic $a$ axis 
\cite{WSC_UTe2_JF_specific_heat,
	WSC_UTe2_JP3_thermal_transport,
	WSC_UTe2_JP2_Science,
	WSC_UTe2_angular_specific_heat_Machida}. 
Direct evidence for chiral Majorana surface states was observed in STM experiments with a step-edge setup
\cite{WSC_UTe2_STM_Jiao_Nature}; indirect evidence was also suggested by measurements of the surface impedance
\cite{WSC_UTe2_JP1_surface_resistivity} at microwave frequencies. 
Kerr rotation experiments further suggest time-reversal symmetry breaking in UTe$_2$
\cite{WSC_UTe2_JP4_Kerr_rotation}. 

Despite substantial effort in experimental measurements and first-principle calculations based on density functional theory (DFT)
\cite{WSC_UTe2_DFT1_Andriy,WSC_UTe2_DFT2_Ishizuka,WSC_UTe2_DFT3_Xu,WSC_UTe2_Wray_ARPES_DFT,WSC_UTe2_Agterberg,WSC_UTe2_Yanase}, 
consensus on the exact pairing symmetry in UTe$_2$ is yet still to be reached. 
On one hand, angular field dependence measurements of specific heat \cite{WSC_UTe2_angular_specific_heat_Machida} and 
a Ginzberg-Landau (GL) free energy analysis \cite{WSC_UTe2_GL_analysis_pairing_Machida} suggest a $d$-vector order parameter of 
the form $\bm{\sfd}(\textbf{k}) = (\textbf{b} + i\textbf{c})(k_b + ik_c)$. This would imply pairing of a single spin
species (while the other remains unpaired), as in ${}^3$He$A_1$ \cite{He3_book_Vollhardt,He3_book_Volovik,He3B_review_Sato_Machida}.
On the other hand, there are also analyses based on DFT, GL free energy and point-group symmetries suggesting more complicated 
scenarios \cite{WSC_UTe2_DFT1_Andriy,WSC_UTe2_DFT2_Ishizuka,WSC_UTe2_DFT3_Xu,WSC_UTe2_JP4_Kerr_rotation,WSC_UTe2_Agterberg,WSC_UTe2_Yanase}. 
Further investigations are necessary to clarify the underlying pairing symmetry in UTe${}_2$.

\subsection{Summary of main results \label{sec:ResultsSum}}

In superconductors, key electromagnetic (EM) responses are the Meissner effect and optical absorption.
In the following, we elucidate the interplay of the surface and bulk responses by considering a spinless $p + ip$ 
model, which is a solid-state analog of $^3$He\textit{A} \cite{He3_book_Vollhardt,He3_book_Volovik,He3B_review_Sato_Machida}.
This model, with a Hamiltonian described by Eq.~(\ref{eq:HBdG}), has a pair of Weyl nodes lying along 
$k_z$ and no time-reversal symmetry. As a result, it only possesses one branch of chiral surface states, as sketched in 
Fig.~\ref{fig:Geometry}(a). It serves as a minimal model for WSCs.

We now summarize the main results of this paper.


\subsubsection{Meissner effect}
\label{sec:intro_Mesiser}

In a previous work, we predicted a power-law temperature ($T$)-dependence in the magnetic penetration depth 
$\Delta \LL(T) \equiv \LL(T) - \LL(0) \sim T^3$ due to surface states in a strong topological superconductor (TSC), 
based on a solid-state model analog of ${}^3$He$B$ \cite{He3B_Meissner}. 
In that case, since the bulk is fully gapped, 
in the absence of magnetic impurities \cite{Prozorov2006,Cooper1996}
a power-law $T$-dependence in $\Delta \LL$ can only arise from the surface states. 
The combination of power-law temperature dependence in the magnetic penetration depth 
along with exponential suppression of the specific heat (due to the fully gapped bulk) are hallmarks for strong TSCs \cite{He3B_Meissner}. 

On the other hand, a superconductor with bulk nodes should exhibit power-law temperature dependence in \emph{both}
the magnetic penetration depth and the specific heat \cite{Tinkham}. 
Power-law $T$-dependence in $\Delta \LL$ has also been experimentally observed in WSC candidates \cite{WSC_YiPtBi_JP,WSC_UTe2_JP3_thermal_transport}. 
However, for WSCs, because of the gapless excitations in the bulk, this is not necessarily indicative for the presence of surface states. 
Nevertheless, for WSCs with Weyl nodes lying along a particular $k$ axis (say $k_z$), we can show by power-counting that the power-law 
$T$-dependence from the surface states may still override that from the bulk, depending on the direction of the external magnetic field $\textbf{B}$ (see Appendix \ref{sec:app_Meissner}).
In fact, by considering the model WSC in Eq.~(\ref{eq:HBdG}) (with one pair of Weyl nodes at $k_z = \pm k_F$),
occupying the $x > 0$ half-space and using the same framework outlined in Ref.~\cite{He3B_Meissner}, 
our calculations reveal that the bulk contribution is
\begin{equation}\label{Meissner:BB}
\Delta \LL 
\sim
\begin{cases}
 T^2, 
\quad \textbf{B}\parallel \hat{y},
\\
T^4, 
\quad \textbf{B}\parallel \hat{z},
\end{cases}
\end{equation}
whereas surface state correction is
\begin{equation}\label{Meissner:SS}
\delta \LL 
\sim
\begin{cases}
T^2, 
\quad \textbf{B}\parallel \hat{y},
\\
T^2, 
\quad \textbf{B}\parallel \hat{z}.
\end{cases}
\end{equation}
This suggests that at low enough temperature, a $T^2$ dependence in penetration depth can be an indicator for the presence of surface states,
for a magnetic field orientation parallel to the line joining the Weyl nodes.  
In fact, this could be the case for UTe$_2$, since recent experimental evidence suggests that Weyl nodes lie along the $a$-axis
\cite{WSC_UTe2_JF_specific_heat,
	WSC_UTe2_JP3_thermal_transport,
	WSC_UTe2_JP2_Science,
	WSC_UTe2_angular_specific_heat_Machida}.

Nevertheless, more generally, for WSCs with nodes lying along arbitrary directions and/or an arbitrary orientation
of the magnetic field, we expect identical $T^2$ contributions from both the bulk and surface. 
This calls for an alternative EM response that can possibly distinguish features of the surface and bulk 
states in WSCs.


\subsubsection{Optical absorption}

We consider a plane EM wave normally incident upon the surface of a WSC at $x = 0$, 
as shown in Fig.~\ref{fig:Geometry}(b). The Weyl nodes in our model [Eq.~(\ref{eq:HBdG})] 
lie along the $k_z$-axis, giving rise to a chiral surface Majorana band 
with dispersion $E^{\sfs}_{k_y} = -\Delta k_y$ [Fig.~\ref{fig:Geometry}(a)]. 
The reflectance and absorbance can be obtained by solving the Maxwell's equation (\ref{eq:Maxwell0}). 
The material response is encoded in the current-current correlation function in this equation,
with 
diamagnetic 
and
three different paramagnetic contributions, 
originating
(i) purely from the bulk,
(ii) purely from the surface, and
(iii) from the surface-bulk cross terms. 
The topological anomalous skin effect arises from the combination of the diamagnetic response and (iii). 
   
For the clean one-band WSC model in Eq.~(\ref{eq:HBdG}), the dissipative part of the optical conductivity in the bulk 
vanishes, consistent with the standard result for clean $s$-wave superconductors \cite{s_wave_opt_resp_Mahan}. 
Meanwhile, because we only have one branch of surface states [Fig.~\ref{fig:Geometry}(a)], 
surface intraband transitions that conserve energy and (nearly) preserve the transverse momentum are impossible. 
This leaves surface-bulk transitions to dominate in the clean limit. 
We evaluate the surface-bulk optical conductivity by exploiting the exact eigenstates of the system in the slab geometry.  

The zero temperature surface-bulk absorbance $\mathsf{A}^{\mu \mu}$
can be written compactly as 
\begin{equation}\label{eq:absorbance_sb2--Intro}
\begin{aligned}
	\mathsf{A}^{\mu \mu}
	=&\,
	\frac{\Omega}{k_F^2 \lcoh c}
	\,
	\hat{{\cal F}}^{\mu \mu}
	(\hat{\Omega}),
	\quad
	\hat{\Omega}
	\equiv
	\frac{\Omega}{\Delta_0},
\end{aligned}
\end{equation}
where 
$\Omega$ is the radiation frequency, 
$k_F$ is half the separation between the Weyl points,
and
$\lcoh$ is the coherence length of the superconductor
(which determines the \emph{minimum} for the chiral Majorana surface-state confinement to the WSC-vacuum interface).
In Eq.~(\ref{eq:absorbance_sb2--Intro}),
$\hat{{\cal F}}^{\mu \mu}$ is a dimensionless function of the reduced frequency $\hat{\Omega}$,
which is the frequency relative to the pairing energy $\Delta_0$. 
The behavior of $\mathsf{A}$ is shown in Fig.~\ref{fig:hatF}, for different combinations of $\lcoh$
and the diamagnetic penetration depth $\LL$. 
The components $\mathsf{A}^{yy}({\Omega})$ and $\mathsf{A}^{zz}({\Omega})$ correspond to the case with $y$- and $z$-polarized incident electric fields, 
respectively; the Weyl nodes lie along the $k_z$-axis in our model [Fig.~\ref{fig:Geometry}(a)]. 
In the type-I regime ($\LL \ll \lcoh$) 
[Fig.~\ref{fig:hatF}(a)--(c)], 
both the $yy$ and $zz$ components of $\mathsf{A}$ demonstrate a peak centered around $\Omega = 2\Delta_0$. 
The broadness of the peaks increases with the coherence length $\lcoh$. 
Larger $\lcoh$ means more deconfined surface states.
In the type-II regime ($\LL \gg \lcoh$) 
[Fig.~\ref{fig:hatF}(d)--(f)], 
the peaks for different polarizations are more distinct. 
In particular, the $yy$ component still manifests a peak at around $2\Delta_0$, 
whereas the $zz$ component is peaked slightly above $\Delta_0$.

We note that the results in Eq.~(\ref{eq:absorbance_sb2--Intro}) and Fig.~\ref{fig:hatF}
are appropriate for the strong type-I and type-II limits, where the absorbance 
$0 < \mathsf{A}^{\mu \mu} \ll 1$, but not the intermediate regime with $\lcoh \sim \LL$. 
The absorption is small in the strong type-I limit, because the field penetration is 
limited to $\LL$, which is much smaller than the spatial extent of the surface
states (bounded from below by $\lcoh$). 
The absorption is also small in the strong type-II limit, due to the orthogonality of
bulk and surface states. The case with $\lcoh \sim \LL$ requires the inversion
of an integral equation to determine the electric field profile and the absorbance, 
as discussed in Sec.~\ref{sec:Classical_EM}. 
Combining the type-I and type-II cases, the surface-bulk absorption only extends up to the scale 
of $\min(\LL,\lcoh)$, see Fig.~\ref{fig:Geometry}(b).

Quenched disorder due to impurities and other defects is inevitable in real materials, 
and gives rise to a dissipative bulk optical conductivity. 
Although low-temperature superconductors are typically good metals with $\varepsilon_F \tau_{\mathsf{el}} \gg 1$,
where $\varepsilon_F$ is the Fermi energy and $\tau_{\mathsf{el}}$ is the lifetime due to elastic impurity scattering,
the small $T_c$ means that the superconducting phase typically occurs in the dirty limit with 
$\Delta_0 \ll 1/\tau_{\mathsf{el}}$. The optical absorption is governed by the classic Mattis-Bardeen 
result \cite{s_wave_MB_original,Tinkham}.

We use the Keldysh formalism \cite{s_wave_NLSM4_Kamenev,s_wave_NLSM5_Yunxiang}
to rederive the $s$-wave Mattis-Bardeen result via the saddle-point of the matrix field integral 
\cite{s_wave_NLSM1_Larkin,s_wave_NLSM2_Lerner,s_wave_NLSM3_Mirlin}.
We then derive the matrix field theory appropriate for the finite-frequency response of a disordered $p+ip$ WSC. 
We compute the bulk Kubo optical conductivity for a WSC with disorder in the saddle-point approximation, equivalent to the 
self-consistent Born approximation. The zero-temperature bulk conductivity is shown in Fig.~\ref{fig:p_wave_sigma_bulk},
for different values of the normal-state scattering rate $\Gel \equiv 1 / (2 \tau_{\mathsf{el}})$.

For weak disorder, the optical conductivity shows a Mattis-Bardeen-like frequency dependence [Fig.~\ref{fig:p_wave_sigma_bulk}(a)], 
despite being nonzero even for $\Omega < 2\Delta_0$ due to the low-energy excitations around the Weyl nodes.
Here $\Delta_0 = \Delta k_F$ is the pairing energy; $\Delta$ is the $p+ip$-wave pairing amplitude in the model [Eq.~(\ref{eq:HBdG})].
Our results are consistent with previous calculations for anisotropic superconductors \cite{p_wave_opt_resp_Hirschfeld}.  
As the disorder strength $\Gel$ increases relative to the pairing energy $\Delta_0$, 
the ``pseudogap'' in the optical conductivity gradually fills in, approaching
the normal-state Drude optical conductivity. Unlike the $s$-wave case, the dirty limit for the WSC 
($\Delta_0 \ll \Gel$) reduces to that of the 
normal state, consistent with formation of a thermal diffusive
metal for sufficiently strong disorder \cite{rare_WSC_Jed6,WSC_thermal_Hall_Bitan_Sayed,Sau2017}. 

Finally, we consider the combined effect of the bulk and the surface-bulk response on the absorbance in the weak disorder limit.
Since the topological anomalous skin effect is nonvanishing in the clean limit, 
we expect weak disorder to modify the surface-bulk response only by slightly broadening its features. 
To leading order, it is sufficient to compare the contribution from the clean surface-bulk response versus that from the weakly disordered bulk. 
The result is
\begin{equation}
\label{eq:absorbance_combine}
	{\mathsf{A}}^{\mu\mu}
	=
	\frac{\Omega \LL}{c}
	\left\lbrace 
		\frac{\hat{{\cal F}}^{\mu \mu}
		(\hat{\Omega})}{k_F^2 \LL \lcoh}
		+
		\left[
		\frac{
			2 \Omega\tau_{\mathsf{el}}
		}{
			1 + (\Omega \tau_{\mathsf{el}})^2
		}
		\right]
	\frac{
		\re \,
		\sigma^{\mu \mu}_{\sfb \sfb,R}(\Omega) 
	}{
		\re \, \sigma_{\mathsf{Drude}}(\Omega)
	}
	\right\rbrace\!,
\end{equation}
where
\[
	\re \, \sigma_{\mathsf{Drude}}(\Omega)
	=
	{\sigma_{\mathsf{dc}}}/{\left[1 + (\Omega \tau_{\mathsf{el}})^2\right]}
\] 
is the normal-state optical conductivity. 
The first term in Eq.~(\ref{eq:absorbance_combine}) is the surface-bulk absorption [Eq.~(\ref{eq:absorbance_sb2--Intro})],
while $\re \, \sigma^{\mu \mu}_{\sfb \sfb,R}(\Omega)$ is the bulk 
optical conductivity of the WSC with disorder.  

In Fig.~\ref{fig:combine_A}, we plot the combined absorbance $\mathsf{A}$ for both type I and type II superconductors. 
The component 
$\mathsf{A}^{yy}$ 
($\mathsf{A}^{zz}$) 
corresponds to the absorbance with an incident electric field polarized along the $y$ ($z$) direction, respectively. 
The overall shape of the absorbance versus frequency depends on the relative magnitude of 
the London penetration depth $\LL$, 
coherence length $\lcoh$,
and the disorder strength $\Gel$. 
In the type-I case, since the external field can only penetrate up to a shallow region at the proximity of the surface, 
the surface-bulk absorption dominates over that of the bulk for weak enough disorder. 
The frequency dependence of the absorbance thus mainly follows that of $\mathsf{A}({\Omega})$ in the clean case, shown in Fig.~\ref{fig:hatF}.
In the type-II case, the external field can penetrate deep into the sample and therefore the bulk contribution starts to overtake 
surface-bulk one as the amount of disorder increases.
One would therefore need a relatively clean sample in order to observe the surface-bulk effect in type II WSCs.
Due to the sharpness of the $yy$ surface-bulk absorption peak at $2\Delta_0$, its feature can still be seen in the overall absorbance.


\subsection{Outline}

The rest of this paper is organized as follows. 
In Sec.~\ref{sec:Classical_EM}, we illustrate the geometry of our problem and derive a 
formal expression for the absorbance via classical electrodynamics. 
In Sec.~\ref{sec:resp_clean}, we evaluate the optical conductivity in the clean limit using linear response theory.  
In Sec.~\ref{sec:resp_dirty}, we derive the optical conductivity for the disordered bulk using the Keldysh theory.
The main results of this paper appear already in Sec.~\ref{sec:ResultsSum} and Figs.~\ref{fig:hatF}--\ref{fig:combine_A}.
We further discuss these and conclude in Sec.~\ref{sec:discussion_conclusion}.
Technical details are relegated to the Appendices.


\section{Setup of the problem and optical absorbance}\label{sec:Classical_EM}

For simplicity, we assume that the magnetic permeability $\mu_{\mathsf{m}} $ of the material is the same as that for vacuum.
Consider an incident EM plane wave that is linearly polarized and propagates along $x$, as schematically sketched in Fig.~\ref{fig:Geometry}(b). 
In the temporal gauge, the electric field is related to the vector potential via $\textbf{E} = -c^{-1} \partial_t \textbf{A}$. The electric field of the system is governed by the Maxwell's equation
\begin{equation}\label{eq:Maxwell0}
\begin{aligned}
\left(
-\partial_x^2 
- K^2 
\right)
E^{\mu}(x)
&=
\frac{4\pi}{c} iK \, J_{\ext}^{\mu}(x)
\\
&-
\frac{4\pi}{c^2}
\int_0^{\infty}dx' \,
\Pi^{\mu \nu}_R(\Omega;x,x') E^{\nu}(x'),
\end{aligned}
\end{equation}
where 
$\mu \in \left\lbrace y,z\right\rbrace$,  
$K = \Omega/c$ is the wavevector of the incident EM wave, 
$J^{\mu}_{\ext}(x) = -(cE_0^{\mu}/2\pi) \delta(x-x_0)$ is a source current that generates the EM radiation at $ x_0 < 0$, 
and 
$\Pi^{\mu \nu}_R(\Omega;x,x')$ is the total retarded current-current correlation function of the material, including both the diamagnetic and paramagnetic terms.
The second term on the right-hand-side captures the response from the system, consisting of contributions due to bulk-bulk, surface-bulk and surface-surface transitions. 
For the bulk response, we write 
\begin{equation}
\begin{aligned}
&
\frac{4\pi}{c^2}
\Pi^{\mu \nu}_{\sfbb,R}(\Omega;x,x')
\\
&=
\delta(x-x')
\theta(x)
\theta(x')
\left[\frac{1}{\LL^2} \delta^{\mu\nu}
-
\frac{4\pi i \Omega}{c^2} 
\, 
\re
\sigma^{\mu \nu}_{\sfbb,R}(\Omega)
\right],
\end{aligned}
\end{equation}
where 
$\theta(x)$ is the Heaviside step function, 
$\LL$ is the London penetration depth, 
and 
$\sigma^{\mu \nu}_{\sfbb,R}$ is 
the paramagnetic
optical conductivity from the bulk. 
Since our focus is optical absorption, $\im\sigma^{\mu \nu}_{\sfbb,R}$ will be neglected in the following calculations. 
For the bulk-bulk paramagnetic response,
we neglect the difference between the slab and homogeneous geometries
and take $\sigma^{\mu \nu}_{\sfbb,R}(\Omega)$ to be independent of $x$. 
This is justified for terahertz (THz) radiation in 
low-transition-temperature superconductors with modest disorder, 
since the light wavelength $\lambda \gg \{\lcoh,\LL,v_F \tau_{\mathsf{el}}\}$,
where $\lcoh$ is the coherence length
and 
$v_F \tau_{\mathsf{el}}$ is the bulk mean free path due to impurity scattering
\cite{s_wave_opt_resp_Mahan}.

As we will see in the next section, 
for the $p + ip$ WSC that we are going to study, 
the paramagnetic response function is purely diagonal and the 
contribution from surface-surface transitions vanishes. 
In this case, Eq.~(\ref{eq:Maxwell0}) 
reduces to
\begin{equation}\label{eq:Maxwell1}
\begin{aligned}
&
\left[
-\partial_x^2 
-K^2
+
\Kb^2
\,
\theta(x)
\right]
E^{\mu}(x) 
\\
&=
\frac{4\pi}{c}
iK
\,
J_{\mathsf{ext}}^\mu
-
\frac{4\pi }{c^2}
\int_0^{\infty}
dx'
\,
\Pi^{\mu \nu}_{1,\sfsb,R}(\Omega; x,x')
\,
E^{\nu}(x'),
\end{aligned}
\end{equation}
where 
$
	\Kb^2(\Omega)
	=
	\frac{1}{(\LL)^2} 
	-
	\frac{4\pi i \Omega}{c^2} 
	\, \sigma^{\mu \mu}_{\sfb \sfb,R}(\Omega)
$
captures the response from the 
bulk, 
and
$\Pi^{\mu \nu}_{1,\sfsb,R}(\Omega; x,x')
\equiv
\Pi^{\mu \nu}_{1,\sfsb,R}(\Omega,\vex{q} \rightarrow 0; x,x')
$ 
is the paramagnetic current-current correlation function due to the transitions between surface and bulk states.
Here $\vex{q} \rightarrow 0$ is the photon momentum parallel to the interface, which vanishes
for normal incidence. 
To solve for $E^{\mu}(x)$, we seek a Green's function $G(x,x')$ satisfying 
\begin{equation}\label{eq:Maxwell_GF_PDE}
\left[
-\partial_{x}^2 - K^2 + \Kb^2 \theta(x)
\right]
G(x,x')
=
\delta(x-x'),
\end{equation}
where $G(x,x')$ is subjected to the boundary conditions
\begin{eqnarray}
G(0^+,x') &=& G(0^-,x'),
\\
\partial_x G(x,x')
\vert_{x = 0^+}
&=&
\partial_x G(x,x')
\vert_{x = 0^-}.
\end{eqnarray}
The Green's function $G(x,x')$ can be solved by the method of images. 
The result is
\begin{equation}
\!\!\!
\begin{aligned}[b]
&
	G(x,x')
\\
&
	=
	\frac{i \theta(-x) \theta(-x')}{2K}
	\left[
	e^{iK|x - x'|}
	+
	\frac{
		K - i{\cal K}
	}{
		K + i{\cal K}
	}
	e^{-iK(x+x')}
	\right]
\\
&
	+
	\frac{\theta(x) \theta(x')}{2{\cal K}}
	\left[
	e^{-{\cal K}|x-x'|}
	-
	\frac{K - i{\cal K}}{K + i{\cal K}}
	e^{-{\cal K}(x+x')}
	\right]
\\
&
	+
	\frac{\theta(x)\theta(-x')}{
		-iK + {\cal K}
	}
	e^{-{\cal K}x - iK x'}
	+
	\frac{\theta(-x)\theta(x')}{
		-iK + {\cal K}
	}
	e^{-{\cal K}x' - iK x},
\end{aligned}
\end{equation}
where ${\cal K} \equiv \sqrt{\Kb^2 - K^2}$.
The electric field can now be expressed as
\begin{equation}\label{eq:E_field_formal0}
\begin{aligned}
E^{\mu}(x)
&=
E^{(0)\mu}(x)
+
\delta E^{\mu}(x),
\end{aligned}
\end{equation}
where the bare electric field is
\begin{equation}
E^{(0)\mu}(x)
=
-2iK
\,
E_0^{\mu} G(x,x_0)
\end{equation}
and the correction due to the surface-bulk response is
\begin{equation}\label{eq:delta_E}
\delta E^{\mu}(x)
=
-
\frac{4\pi}{c^2}
\int_{x_{1,2}>0}
G(x,x_1)
\,
\Pi^{\mu\nu}_{1,\sfsb,R}(\Omega; x_1,x_2)
\,
E^{\nu}(x_2).
\end{equation}

In general, the integral equation in (\ref{eq:E_field_formal0})--(\ref{eq:delta_E})
cannot be solve exactly. However, in the strong type-I or type-II limits for superconductors, 
the contribution from the surface-bulk response is perturbatively small and explicit results obtainable.
For a strong type-I WSC, 
the coherence length $\lcoh$ is much greater than $\LL$. 
Since the spatial extent of the surface states is roughly $\sim \lcoh$ except when they merge into the bulk, 
the limited penetration of the EM wave means that the response involving surface states cannot contribute much. 
On the other hand, for a strong type-II WSC, the coherence length $\lcoh$ is much smaller than $\LL$. 
As a result, the field decays very slowly within the spatial extent of the surface-bulk correlation function. 
Owing to the orthogonality between bulk and surface states, the contribution from the surface-bulk response is 
again small, 
and would vanish in the limit ${\cal K} \rightarrow 0$.

We now focus on the strong type-I and -II limits such that corrections from the surface-bulk term 
can be treated perturbatively. Formally, Eq.~(\ref{eq:E_field_formal0}) can be written as
\begin{equation}
\left[
\hat{1}
+
\frac{4\pi}{c^2}
\hat{G} 
\,
\hat{\Pi}^{\mu\nu}_{1,\sfsb,R}
\right]
\ket{E^{\mu}}
=
\ket{E^{(0){\mu}}}
\end{equation}
such that
\begin{equation}
\ket{E^{\mu}}
=
\left\lbrace 
\left[
\hat{1}
+
\frac{4\pi}{c^2}
\hat{G} 
\,
\hat{\Pi}_{1,\sfsb,R}
\right]^{-1}
\right\rbrace^{\mu \nu}
\ket{E^{(0){\nu}}}.
\end{equation}
To leading order, the electric field is just
\begin{equation}
\ket{E^{(1)\mu}}
=
\ket{E^{(0){\mu}}}
-
\frac{4\pi}{c^2}
\hat{G} 
\,
\hat{\Pi}^{\mu\nu}_{1,\sfsb,R}
\ket{E^{(0){\nu}}},
\end{equation}
that is,
\begin{equation}\label{eq:E_field}
\begin{aligned}
&E^{(1)\mu}(x)
\\
&\simeq
E_0^{\mu}
\begin{bmatrix}
\theta(-x)
\left(
e^{iKx} + r \, e^{-iK x}
\right)
\\
+
t
\,
\theta(x)
e^{-{\cal K}x}
\end{bmatrix}
e^{-iK x_0}
+
\delta E^{(1)\mu}(x),
\end{aligned}
\end{equation}
where
\begin{eqnarray}
r
&=&
\frac{
K - i {\cal K}
}{
K + i {\cal K}
},
\quad
t
=
\frac{
	2K
}{
	K + i {\cal K}
},
\end{eqnarray}
and
\begin{equation}
\begin{aligned}
&
\delta E^{(1)\mu}(x)
\\
&=
-
\frac{4\pi}{c^2}
\int_{x_{1,2}>0}
\hat{G}(x,x_1)
\,
\hat{\Pi}^{\mu\nu}_{1,\sfsb,R}(\Omega;x_1,x_2)
\,
E^{(0){\nu}}(x_2).
\end{aligned}
\end{equation}
In Eq.~(\ref{eq:E_field}),  $\delta E(x)$ is the correction term arising from the surface-bulk response.
Here, $r$ and $t$ carry the physical meanings of bare reflection and transmission coefficients respectively.
$\delta E^{(1)\mu}(x)$ is the surface-bulk correction in first order.
As we will demonstrate in the next section, the correlation function $\hat{\Pi}^{\mu\nu}_{1,\sfsb,R}(\Omega;x_1,x_2)$ is non-vanishing only for $ \mu = \nu$. 
Thus, at the proximity of the surface ($x \rightarrow 0$), the correction takes a simple form
\begin{align}
	\delta E^{\mu}(x \rightarrow 0)
	=&\,
	E_0^{\mu}
	\left[
		\delta r
		\,
		\theta(-x)
		+
		\delta t
		\,
		\theta(x)
	\right]
	e^{-i K x_0},
\end{align} 
where
\begin{align}
	\delta r
	=
	\delta t
	=
	-\frac{4\pi i }{c^2}
	\frac{
		t
	}{
		K + i{\cal K}
	}
	{\cal Q}^{\mu \mu}(\Omega),
\end{align}
\begin{align}
\!\!\!\!
	{\cal Q}^{\mu \mu}(\Omega)
	\equiv
	\int_{x_{1,2}> 0}
		e^{-{\cal K} x_1}
		\,
		\Pi^{\mu \mu}_{1,\sfsb,R} (\Omega; x_1,x_2)
		\,
		e^{-{\cal K} x_2}.
\!\!\!\!
\end{align}
To leading order, the absorbance in the weak disorder limit is simply
\begin{equation}\label{eq:absorbance}
\begin{aligned}[t]
\mathsf{A}^{\mu \mu}
&=
1 - |r + \delta r|^2
\\
&\simeq
K (\LL)^3
\frac{8\pi \Omega}{c^2}
\re \,
\sigma^{\mu \mu}_{\sfb \sfb,R} 
-
\frac{16\pi }{c^2}
\frac{
	K
}{
	\LL^{-2}
}
\im \, {\cal Q}^{\mu \mu},
\end{aligned}
\end{equation}
where we have assumed $\LL^{-1} \gg K$. 
This is a valid approximation in the THz regime, which is appropriate for probing features of pairing energies $\Delta_0$
in WSC candidate materials. 
In the above, we also used the fact that 
$(\Omega/c^2) \, \re \, \sigma^{\mu \mu}_{\sfb \sfb,R} \ll \LL^{-2}$ for weak disorder. 	
We will corroborate this point by explicit calculations in Secs.~\ref{sec:resp_clean} and \ref{sec:resp_dirty}.
Eq.~(\ref{eq:absorbance}) is the key result of this section.
Combined with the results of Secs.~\ref{sec:resp_clean} and \ref{sec:resp_dirty}, 
it gives the absorbance expression in Eq.~(\ref{eq:absorbance_combine}) of the Introduction.


\section{Optical response in the clean limit}\label{sec:resp_clean}

\subsection{Model and some comments on the effective surface theory}\label{subsec:model}

We consider a minimal model for a Weyl superconductor (WSC), 
consisting of spinless electrons with $p+ip$ pairing. 
This is represented by the following static,
mean-field Bogoliubov-de Gennes Hamiltonian
\bsub\label{eq:HBdG}
\begin{eqnarray}
\label{eq:hamiltonian}
H	 
&=&
\frac{1}{2}
\int_{\textbf{k}} 
\Psi_{\textbf{k}}^{\dagger}
\,
\hat{h}_{\textbf{k}} 
\,
\Psi_{\textbf{k}},
\\
\label{eq:model_ham}
\hat{h}_{\textbf{k}}
&=&
\tilE_{\textbf{k}} \hsig^3
+
\Delta k_x \hsig^1
+
\Delta k_y \hsig^2,
\\
\tilE_{\textbf{k}}
&=&
\frac{\textbf{k}^2 - k_F^2}{2m},
\nonumber
\end{eqnarray}
\esub
where $\int_{\textbf{k}} = \int \frac{d^3 k}{(2\pi)^3}$, 
$k_F$ is the Fermi momentum, $\Delta$ is the superconducting order parameter amplitude, 
$\hsig^i$ are the Pauli matrices in particle-hole space,
and 
$\Psi_{\textbf{k}}^{\sfT} = 
\left[
c_\textbf{k}^{\sfT}, 
\,\, 
c_{-\textbf{k}}^{\dagger} 
\right]$.
This Hamiltonian possesses the particle-hole symmetry
\begin{eqnarray}
	-
	\hat{M}_{\sfP} 
	\,
	\hat{h}^{\mathsf{T}}(-\textbf{k})
	\,
	\hat{M}_{\sfP}
	&=&
\hat{h}(\textbf{k}),
\quad
\hat{M}_{\sfP} = \hat{\sigma}^1,
\end{eqnarray}
but has no time-reversal symmetry due to the explicit appearance of ``$i$'' in the pairing. 
It belongs to class D according to the ten-fold classification scheme \cite{WSC_Sato_review,classification_Ludwig,classificiation_Ryu_RMP,classificiation_gapless_Ryu}. 
The Weyl nodes of this system are located at 
$\textbf{k}_{\mathsf{nodes}} = (0,0,\pm k_F)$.
At the proximity of the Weyl nodes, we can linearize $\hat{h}_{\textbf{k}}$ such that
\begin{equation}
\label{eq:hamiltonian_Weylpoint}
	\hat{h}_{\pm}(\delta \textbf{k}) 
	=
	v_{\pm}^x \delta k_x  \hat{\sigma}^1
	+
	v_{\pm}^y \delta k_y  \hat{\sigma}^2
	+
	v_{\pm}^z \delta k_z  \hat{\sigma}^3,
\end{equation}
where $\pm$ correspond to the two Weyl nodes
with opposite chirality \cite{WSM_Ashvin_review}, 
$\textbf{v}_{\pm} 
= 
\left(
\Delta,\Delta,\pm k_F/m
\right)$ 
and 
$\delta \textbf{k} 
=
\textbf{k} - \textbf{k}_{\mathsf{nodes}}$.
However, we do not employ the linearized theory in the rest of this Section. 

In order to analyze the response related to the surface, we solve for the surface states 
with dispersion 
$E^{\sfs}_{k_y} = -\Delta k_y$ by replacing 
$k_x \rightarrow -i\partial_x$ in Eq.~(\ref{eq:model_ham}), 
which is appropriate for the geometry that we are considering in Fig.~\ref{fig:Geometry}(b).
By imposing a hard-wall boundary condition at $x = 0$, we obtain the chiral surface state
\begin{equation}\label{eq:eigenstate_surf}
\psi^{\sfs}(\textbf{k};x)
=
\frac{
\theta(k_F^2 - \textbf{k}^2)
}{
\sqrt{N_{\textbf{k}}^{\sfs}}
}
\e^{-x/\lcoh}
\sinh(\kappa_{\textbf{k}} x)
\ket{\sigma^2 = -1}
,
\end{equation}
where 
$\textbf{k} = (k_y,k_z)$ is the in-plane momentum parallel to the surface, 
$\lcoh \equiv 1/ m\Delta$ is the coherence length,
\begin{align}\label{eq:kappa_k_Def}
	\kappa_{\textbf{k}} \equiv \sqrt{\lcoh^{-2} + \textbf{k}^2 - k_F^2} 
\end{align}
is the wavenumber that controls the confinement of the surface state to the vacuum-WSC interface, 
$N_{\textbf{k}}^{\sfs}
\equiv
\frac{
\lcoh^{-2} + \textbf{k}^2  - k_F^2
}{
	4(k_F^2 - \textbf{k}^2) \lcoh^{-1}
}
$ is the normalization factor, 
and
$\ket{\sigma^2 = -1}$ denotes the eigenstate of $\hsig^2$ with eigenvalue $-1$. 
Notice that there is only one branch of surface states due to time-reversal symmetry breaking. 
The Heaviside step function $\theta(k_F^2 - \textbf{k}^2)$ ensures that $\psi_{\sfs}(\textbf{k};x\rightarrow \infty) = 0$
for physical surface state solutions. 
For $|\vex{k}| \rightarrow k_F$ (including the Weyl nodes at $k_z = \pm k_F$),  
the confinement length for the surface states diverges, signifying the merger with the bulk. 
The surface states exhibit a ``Majorana Fermi arc'' for $k_y = 0$, where $E^{\sfs}_{k_y} = 0$ for $0 \leq |k_z| \leq k_F$.  

\begin{figure}
	\centering
	{\includegraphics[width=9cm]{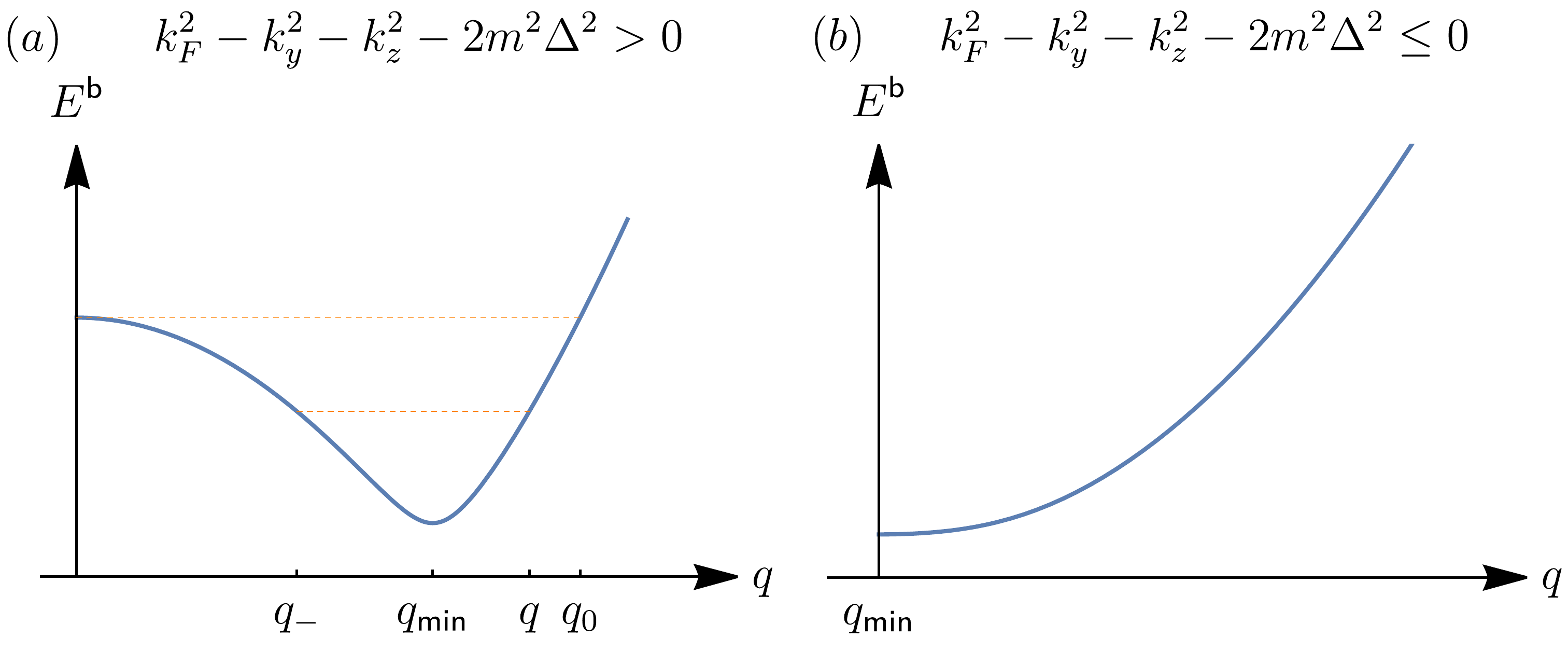} }
	\caption{Schematic illustration of the degeneracy in the bulk band,
	for bulk eigenenergies $E^{\sfb}_{q,\textbf{k}}$ [Eq.~(\ref{eq:BulkEE})]
	plotted versus the standing wave momentum $q \geq 0$. 
	Eigenstates are semi-infinite standing waves in the $x$-direction [due to the slab geometry, Fig.~\ref{fig:Geometry}(b)], plane waves transverse to this with momenta $\vex{k}=(k_y,k_z)$.  
	(a) For $k_F^2 - k_y^2 - k_z^2 - 2m^2 \Delta^2 > 0$, there exists a $q_- < \qmin$ such that $E^{\sfb}_{q,\textbf{k}} = E^{\sfb}_{q_-,\textbf{k}}$ for each $q \in (\qmin,q_0]$. 
	The minimum of $E^{\sfb}_{q,\textbf{k}}$ is located at $\qmin = \sqrt{k_F^2 - k_y^2 - k_z^2 - 2m^2\Delta^2}$.
	In this region, the two degenerate bulk states are $\Psi^{\sfb (1)}_{\lambda }(q,\textbf{k};x) $ and $\Psi^{\sfb (2)}_{\lambda}(q,\textbf{k};x)$. 
	Beyond $q_0$, the bulk states are given by $\psi^{\sfb>}_{\lambda }(q,\textbf{k};x)$. 
	(b) For $k_F^2 - k_y^2 - k_z^2 - 2m^2 \Delta^2 \le 0$, $\qmin = 0$ and there is no degeneracy for all $q$. The bulk states are just $\psi^{\sfb>}_{\lambda }(q,\textbf{k};x)$.}  	
	\label{fig:bulk_band_degeneracy}
\end{figure}

The bulk states are denoted by $\psi^{\sfb}_{\lambda = \pm 1}(q,\textbf{k};x)$, 
which are labeled by the transverse momentum $\vex{k}$ and standing wave $x$-momentum $q \geq 0$. 
Such a state has eigenenergy
\begin{equation}\label{eq:BulkEE}
	\lambda E^{\sfb}_{q,\textbf{k}}
	=
	\lambda 
	\sqrt{
	\varE_{q,\textbf{k}}^2
	+
	\Delta^2 (q^2 + k_y^2)
	}
	,
	\quad
	\varE_{q,\textbf{k}}
	=
	\frac{q^2 + \textbf{k}^2 - k_F^2}{2m}.
\end{equation}
The index $\lambda = \pm $ labels the positive and negative energy bulk states that are related by particle-hole symmetry. 
Since we consider the weak-pairing BCS limit appropriate to low-temperature solid-state superconductors \cite{Foster2013}, 
we have to consider two cases.
\vspace{1mm}
\\
(i) For $k_F^2 - k_y^2 - k_z^2 - 2m^2 \Delta^2 > 0$, the bulk scattering states are two-fold degenerate for 
$q \le q_0 = \sqrt{2k_F^2 - 2k_y^2 - 2k_z^2 - 4m^2 \Delta^2}$, 
as illustrated in Fig.~\ref{fig:bulk_band_degeneracy}(a). 
For each $\qmin < q \le q_0$, where $\qmin = q_0 /\sqrt{2}$ minimizes $ E^{\sfb}_{q,\textbf{k}}$, we have $ E^{\sfb}_{q,\textbf{k}} =  E^{\sfb}_{q_-,\textbf{k}}$.
In this region, we identify two orthonormal states $\Psi^{\sfb (1)}_{\lambda }(q,\textbf{k};x) $ and $\Psi^{\sfb (2)}_{\lambda}(q,\textbf{k};x)$.
For $q > q_0$, degeneracy no longer exists and there is only one bulk state $\psi^{\sfb>}_{\lambda }(q,\textbf{k};x)$ for each $q$.
\vspace{1mm}
\\
(ii) For $k_F^2 - k_y^2 - k_z^2 - 2m^2 \Delta^2 \le 0$, there is no degeneracy for all $q$ [Fig.~\ref{fig:bulk_band_degeneracy}(b)].
In this case, the bulk states are just given by $\psi^{\sfb>}_{\lambda }(q,\textbf{k};x)$.
\vspace{1mm}
\\
The explicit expressions of the bulk scattering states are complicated and thus relegated to 
Appendix~\ref{sec:app_eigenstates_bulk}.

The effect of external fields on the surface can be intuited by
incorporating an vector potential 
$\textbf{A}$ in Eq.~(\ref{eq:hamiltonian}), 
and then projecting it to the low-energy surface states, as in Refs.~\cite{He3B_Meissner,Sayed_QGD}. 
The result is
\begin{equation}\label{eq:SurfHam}
	H_{\sfs}
	=
	\int_{\textbf{r}}
	\left[
		\frac{1}{2}
		\eta(\textbf{r})
		\,
		\Delta (i{\partial}_y)
		\,
		\eta(\textbf{r})
		-
		\frac{1}{c}
		\textbf{A} 
		\cdot 
		\textbf{J}
	\right],
\end{equation}
where 
$\textbf{r} = (y,z)$, 
$\eta = \eta^\dagger$ is the one-component (chiral) Majorana fermion 
operator on the surface, 
and 
\begin{equation}
	\textbf{J}(\textbf{r})
	=
	\frac{e}{4m}
	\int_{\textbf{r}}
	\eta(\textbf{r})
	i
	\overleftrightarrow{\bm{\nabla}}
	\eta(\textbf{r}),
\end{equation}
where $\overleftrightarrow{\bm{\nabla}} 
\equiv
\overrightarrow{\nabla}
-
\overleftarrow{\nabla}
$ is the left-right derivative. 
The pairing amplitude $\Delta$ sets the ``speed of light'' 
for the surface chiral modes. 

The Hamiltonian in Eq.~(\ref{eq:SurfHam}) with $\vex{A} = 0$ describes a collection of 
1+1-D chiral Majorana fermions, labeled by the continuous index $z$ (since there is no dispersion 
in this direction). Alternatively, we can view this as a many-channel Majorana wire, with
channels labeled by the transverse momentum $k_z$. 
Formally, the coupling to $J^y$ in the surface theory given by Eq.~(\ref{eq:SurfHam}) is \emph{gravitational}, 
i.e.\ the vector potential couples to 
\[
	J^y 
	\propto
	\left(
	T^{-+}
	-
	T^{--}
	\right),
\]
where $T^{-\pm} \equiv - \pi \eta \, i \left(\partial_t \mp \Delta \, \partial_y\right) \eta$ are
stress tensor components for the 1+1-D chiral Majorana fermions, expressed in lightcone coordinates \cite{BYB}. 
The operator $J^z$ on the other hand takes the form of a non-abelian current in the space of $k_z$ channels, 
\[
	J^z(y,z)
	=
	\frac{e}{2m}
	\int_{k_z,k_z'}
	\eta(y,k_z)
	\,
	\left(- k_z' \right)
	\,
	\eta(y,k_z')
	\,
	e^{i (k_z + k_z')z}.
\]

Although the interpretation in terms of 1+1-D relativistic quantum field theory is interesting, 
we emphasize that the surface states in Eq.~(\ref{eq:SurfHam}) cannot be treated in isolation,
owing to the gapless nature of the bulk \cite{Gorbar2016}. In particular, we use the full eigenstate
spectrum of the 3D Hamiltonian in Eq.~(\ref{eq:model_ham}) in the slab geometry to compute
the electromagnetic response.


\subsection{Bulk-bulk transitions}

For the clean bulk, the paramagnetic current-current correlation function is given by the bubble 
\begin{equation}
\begin{aligned}
\Pi^{\mu \nu}_{1,\sfb \sfb}(i\Omega_m,\textbf{q})
&=
-
\frac{1}{2}
\left(\frac{e}{m}\right)^2
T 
\sum_{\omega_n} 
\int_{\textbf{k}}\,
\left(
\textbf{k} + \frac{\textbf{q}}{2}
\right)^{\mu}
\left(
\textbf{k} + \frac{\textbf{q}}{2}
\right)^{\nu}
\\
&
\times
\Tr \left[
\hG_{\sfb} (i\omega_n + i\Omega_m, \textbf{k} + \textbf{q}) 
\hG_{\sfb} (i\omega_n , \textbf{k} ) 
\right],
\end{aligned}
\end{equation}
where 
$T$ is temperature, 
$\omega_n = \pi T(2n + 1)$ is the Matsubara frequency,
$i\Omega_m$ is the external bosonic frequency,
$\textbf{q} = (q_x,q_y,q_z)$ is the external momentum,
and $\textbf{k} = (k_x,k_y,k_z)$. 
The bulk Green's function is defined as 
\begin{equation}
\begin{aligned}
\hG_{\sfb}(i\omega_n,\textbf{k})
&=
\frac{1}{
	-i\omega_n + \hat{h}_{\textbf{k}}
}
=
\frac{1}{
	-i\omega_n + \textbf{b}_{\textbf{k}}\cdot \bm{\hsig}
},
\end{aligned}
\end{equation}
where $\textbf{b}_{\textbf{k}} = (\Delta k_x, \Delta k_y, \varE_{\textbf{k}})$. 
The trace can be evaluated readily as
\begin{equation}
\begin{aligned}
&\Tr \left[
\hG_{\sfb} (i\omega_n + i\Omega_m, \textbf{k} + \textbf{q}) 
\hG_{\sfb} (i\omega_n , \textbf{k} ) 
\right]
\\
&=
\frac{
	2\left[
	( i\omega_n + i\Omega_m) (i\omega_n) 
	+
	\textbf{b}_{\textbf{k}+\textbf{q}} \cdot \textbf{b}_{\textbf{k}}
	\right]
}{
	\left[
	( i\omega_n + i\Omega_m)^2 - b_{\textbf{k}+\textbf{q}}^2
	\right]
	\left[
	(i\omega_n)^2 - b_{\textbf{k}}^2
	\right]
}.
\end{aligned}
\end{equation}
The Matsubara summation can be performed using standard contour integral technique, resulting in
\begin{equation}
\begin{aligned}
&\Pi^{\mu \nu}_{1,\sfb \sfb}(i\Omega_m,\textbf{q})
=
\left(\frac{e}{m}\right)^2
\int_{\textbf{k}}\,
\left(
\textbf{k} + \frac{\textbf{q}}{2}
\right)^{\mu}
\left(
\textbf{k} + \frac{\textbf{q}}{2}
\right)^{\nu}
\\
&
\times
\sum_{\lambda = \pm, \lambda' = \pm}
\frac{1}{4}
\left(
1
+
\frac{
	\textbf{b}_{\textbf{k}+\textbf{q}} 
	\cdot 
	\textbf{b}_{\textbf{k}}
}{
	b_{\textbf{k},\lambda} b_{\textbf{k}+\textbf{q},\lambda'}
}
\right)
\frac{
	f(b_{\textbf{k}+\textbf{q},\lambda'})
	-
	f(b_{\textbf{k},\lambda})
}{
	i\Omega_m
	-
	\left(
	b_{\textbf{k}+\textbf{q},\lambda'}
	-
	b_{\textbf{k},\lambda}
	\right)
},
\end{aligned}
\end{equation}
where $f(E) = 1/(1 + e^{E/T})$ is the Fermi distribution function,
and 
$b_{\textbf{k},\pm} \equiv \pm |\vex{b}_{\textbf{k}}|$. 
In the $\textbf{q} \rightarrow \textbf{0}$ limit, appropriate for the optical conductivity in the THz regime, 
the above expression vanishes. Since the optical conductivity is related to the current correlation function via
\begin{equation}
	\sigma^{\mu \nu}_{\sfbb}(i\Omega_m,\textbf{q})
	=
	\frac{-1}{i\Omega}
	\Pi^{\mu \nu}_{1,\sfb \sfb}(i\Omega_m,\textbf{q}),
\end{equation}
this implies that the
paramagnetic contribution to the
optical conductivity of the WSC is zero in the clean limit. 
This conclusion holds regardless of the detailed form of $\textbf{b}_{\textbf{k}}$, and is thus valid for all 
one-band superconductors with Hamiltonian of the form $\hat{h}_{\textbf{k}} = \textbf{b}_{\textbf{k}}\cdot \bm{\hsig}$.  
This is consistent with the standard result for clean $s$-wave superconductors \cite{s_wave_opt_resp_Mahan}.


\subsection{Surface-surface transitions}

For the model that we are considering, there is only one branch of chiral surface states, with 
eigenenergy $E^{\sfs}_{k_y} = -\Delta k_y$. 
As a result, intraband transitions amongst surface states is not possible 
and optical absorption purely from the surface is vanishing.


\subsection{Surface-bulk transitions \label{sec:TASE}}

In order to capture the interplay between the surface and bulk states, 
we have to implement the open boundary condition along $x$ explicitly 
and 
make use of the exact bulk and surface eigenstates given in Appendix~\ref{sec:app_eigenstates_bulk} 
and 
by Eq.~(\ref{eq:eigenstate_surf}).

The position-dependent paramagnetic current-current correlation function 
due to transitions between bulk and surface states can be expressed as
\begin{equation}\label{eq:correlation_fn_sb}
\!\!\!\!
\begin{aligned}
&\Pi^{\mu \nu}_{1,\sfsb}(i\Omega_m, \textbf{0};x,x')
=
-
\frac{1}{2}
\left(\frac{e}{m}\right)^2
T \sum_{\omega_n} \int_{\textbf{k}}
k^{\mu} k^{\nu}
\\
&
\times
\Tr
\begin{bmatrix}
\hG_{\sfs}(i\omega_n,\textbf{k};x,x')
\,
\hG_{\sfb}(i\omega_n + i\Omega_m,\textbf{k};x',x)
\\
+
\hG_{\sfb}(i\omega_n,\textbf{k};x,x')
\,
\hG_{\sfs}(i\omega_n + i\Omega_m ,\textbf{k};x',x)
\end{bmatrix},
\end{aligned}
\!\!\!\!
\end{equation}
where 
$\mu ,\nu \in \left\lbrace y,z\right\rbrace$,
$\hG_{\sfb}(i\omega_n,\textbf{k};x,x')$ 
and 
$\hG_{\sfs}(i\omega_n,\textbf{k};x,x')$ 
respectively denote the bulk and surface Green's functions, 
and 
$\textbf{k} = (k_y,k_z)$. 
The transverse photon momentum has been sent to zero, appropriate for normal incidence. 
The Green's functions can be expressed via spectral functions,
\begin{align}\label{eq:GF_bulk_pos_dept}
\!\!\!\!
	\hG_{\sfb,\sfs}(i\omega,\textbf{k};x,x')
	=
	\int
	\frac{d \omega'}{2 \pi(i\omega - \omega')}
	\hat{{\cal A}}_{\sfb,\sfs}(\omega',\textbf{k};x,x'),
\!\!\!\!
\end{align} 
where the bulk spectral function
\begin{equation}
\label{eq:spectral_fn_bulk}
\begin{aligned}
&\hat{{\cal A}}_{\sfb}(\omega,\textbf{k};x,x')
\\
&=
\int_{q}
\sum_{\lambda = \pm}
\frac{
	-2 \kappa_{\sfb}(\omega)
}{
	(\omega - \lambda E^{\sfb}_{q,\textbf{k}})^2
	+ 
	\kappa_{\sfb}^2(\omega)
}
\psi^{\sfb}_{\lambda} (q,\textbf{k};x)
\psi^{\sfb \dagger}_{\lambda}(q,\textbf{k};x')
\end{aligned}
\end{equation}
and the surface spectral function
\begin{equation}
\label{eq:spectral_fn_surf}
\begin{aligned}
\hat{{\cal A}}_{\sfs}(\omega,\textbf{k};x,x')
&=
\frac{
	-2 \kappa_{\sfs}(\omega)
}{
	(\omega - E^{\sfs}_{k_y})^2
	+ 
	\kappa_{\sfs}(\omega)^2
}
\psi^{\sfs} (\textbf{k};x)
\psi^{\sfs \dagger}(\textbf{k};x')
.
\end{aligned}
\end{equation}
In Eq.~(\ref{eq:spectral_fn_bulk}), the summation $\sum_{\lambda = \pm}$ sums over the eigenstates with eigenenergy $\pm E^{\sfb}_{q,\textbf{k}}$ and
$\int_q = \int \frac{dq}{2\pi}$ integrates over all scattering wave momenta. 
$\kappa_{\sfb}$ and $\kappa_{\sfs}$ are respectively the bulk and surface impurity scattering rates, which can be taken to be $0^+$ in the clean limit. 
Notice that the surface-bulk current-current correlation in Eq.~(\ref{eq:correlation_fn_sb}) depends separately on $x$ and $x'$ due to translational symmetry breaking.

We then substitute 
Eqs.~(\ref{eq:GF_bulk_pos_dept})--(\ref{eq:spectral_fn_surf}) 
into 
Eq.~(\ref{eq:correlation_fn_sb}) such that
\begin{equation}
\begin{aligned}
&\Pi^{\mu \nu}_{1,\sfsb}(i\Omega_m, \textbf{0};x,x')
=
\frac{1}{2}
\left(\frac{e}{m}\right)^2
\int_{q,\textbf{k}}
\int d\omega' \, 
\int d\omega'' \,
k^{\mu} k^{\nu}
\\
&\times
\left[
\frac{f(\omega'') - f(\omega')}{i\Omega_m - (\omega'' - \omega')}
\right]
{\cal I}_{q,\textbf{k}}(\omega',\omega'';x,x'),
\end{aligned}
\end{equation} 
where
\begin{equation}\label{eq:kernel_I}
\begin{aligned}
&{\cal I}_{q,\textbf{k}}(\omega',\omega'';x,x')
\\
&\;\;\equiv
\sum_{\lambda = \pm}
\begin{bmatrix}
{\cal D}^{\sfs}(\omega', E^s_{k_y})
\,\,
{\cal D}^{\sfb}(\omega'',\lambda E^{\sfb}_{q,\textbf{k}})
\,
\Sigma^{\sfsb}_{\lambda}(q,\textbf{k};x',x)
\\
+
{\cal D}^{\sfs}(\omega'', E^{\sfs}_{k_y})
\,\,
{\cal D}^{\sfb}(\omega',\lambda E^{\sfb}_{q,\textbf{k}})
\,
\Sigma^{\sfsb}_{\lambda}(q,\textbf{k};x,x')
\end{bmatrix},
\end{aligned}
\end{equation}
with the broadened Dirac $\delta$-functions
\begin{align}
{\cal D}^{\sfs}(\omega', E^{\sfs}_{k_y})
=&\,
\frac{
	\kappa_{\sfs}(\omega') /\pi
}{
	(\omega' - E^{\sfs}_{k_y})^2
	+ 
	\kappa_{\sfs}(\omega')^2
},
\\
\label{eq:broad_delta_fn_bulk}
{\cal D}^{\sfb}(\omega'',\lambda E^{\sfb}_{q,\textbf{k}})
=&\,
\frac{
	\kappa_{\sfb}(\omega'')/\pi
}{
	(\omega'' - \lambda E^{\sfb}_{q,\textbf{k}})^2
	+ 
	\kappa_{\sfb}(\omega'')^2,
}
\end{align}
and 
the double-overlap between surface and bulk states
\begin{equation}
\label{eq:double_overlap_surface_bulk}
\Sigma^{\sfsb}_{\lambda}(q,\textbf{k};x,x')
=
\psi^{\sfb \dagger}_{\lambda}(q,\textbf{k};x)
\psi^{\sfs} (\textbf{k};x)
\psi^{\sfs \dagger}(\textbf{k};x')
\psi^{\sfb}_{\lambda} (q,\textbf{k};x').
\end{equation}

Specifically, from the bulk states we obtained in Appendix \ref{sec:app_eigenstates_bulk}, we have to consider two cases,
due to the double-degeneracy of some bulk states, see Fig.~\ref{fig:bulk_band_degeneracy}. 

For $k_F^2 - k_y^2 - k_z^2 - 2m^2 \Delta^2 > 0$,
\begin{widetext}
\begin{equation}
\begin{aligned}
\Sigma^{\sfsb}_{\lambda}(q,\textbf{k};x,x')
=
\begin{cases}
0, 
&\quad 0 < q < \qmin,
\\\\
\begin{bmatrix}
\Psi^{\sfb(1) \dagger}_{\lambda} (q,\textbf{k};x)
\psi^{\sfs} (\textbf{k};x)
\psi^{\sfs \dagger}(\textbf{k};x')
\Psi^{\sfb (1)}_{\lambda } (q,\textbf{k};x')
\\\\
+
\Psi^{\sfb(2) \dagger}_{\lambda} (q,\textbf{k};x)
\psi^{\sfs} (\textbf{k};x)
\psi^{\sfs \dagger}(\textbf{k};x')
\Psi^{\sfb(2)}_{\lambda} (q,\textbf{k};x')
\end{bmatrix}
, 
&\quad  \qmin < q \le q_0,
\\\\
\psi^{\sfb> \dagger}_{\lambda}(q,\textbf{k};x)
\psi^{\sfs} (\textbf{k};x)
\psi^{\sfs \dagger}(\textbf{k};x')
\psi^{\sfb>}_{\lambda } (q,\textbf{k};x')
, 
&\quad  q > q_0,
\end{cases}
\end{aligned}
\end{equation}
\end{widetext}
and for $k_F^2 - k_y^2 - k_z^2 - 2m^2 \Delta^2 \le 0$,
\begin{equation}
\begin{aligned}
&\Sigma^{\sfsb}_{\lambda}(q,\textbf{k};x,x')
\\
&=
\psi^{\sfb> \dagger}_{\lambda}(q,\textbf{k};x)
\psi^{\sfs} (\textbf{k};x)
\psi^{\sfs \dagger}(\textbf{k};x')
\psi^{\sfb>}_{\lambda } (q,\textbf{k};x').
\end{aligned}
\end{equation}
By performing analytical continuation 
\begin{equation}
\Pi^{\mu \nu}_{1,\sfsb,R}(\Omega,\textbf{0};x,x') 
= 
- \Pi^{\mu \nu}_{1,\sfsb}(i\Omega_m \rightarrow \Omega + i\eta,\textbf{0};x,x')
\end{equation}
where $\eta \rightarrow 0^+$, 
we obtain the real part of the retarded optical conductivity due to surface-bulk transitions
\begin{equation}
\begin{aligned}
&\re
\,
\sigma^{\mu \nu}_{\sfsb,R}(\Omega;x,x')
=
\frac{-1}{\Omega}
\im\,
\Pi^{\mu \nu}_{1,\sfsb,R}(\Omega,\textbf{0};x,x')
\\
&=
\frac{\pi}{2} \left(\frac{e}{m}\right)^2
\int_{q,\textbf{k}}
\int d\omega
\,
k^{\mu} k^{\nu}
\left[
\frac{f(\omega) - f(\omega + \Omega)}{\Omega}
\right]
\\
&
\times
\re
\,
{\cal I}_{q,\textbf{k}}(\omega,\omega+\Omega;x,x').
\end{aligned}
\end{equation}
The real part of $\Pi^{\mu \nu}_{1,\sfsb,R}(\Omega,\textbf{0})$ is irrelevant for optical absorption and therefore will be neglected for simplicity. 

In the clean limit, we can convert 
${\cal D}^{\sfs}(\omega_1,\omega_2) \rightarrow \delta(\omega_1 - \omega_2)$ 
and
${\cal D}^{\sfb}(\omega_1,\omega_2) \rightarrow \delta(\omega_1 - \omega_2)$.
Using the fact that
\begin{equation}
\Sigma^{\sfsb}_{\lambda= -1}(q,\textbf{k};x,x')
=
\Sigma^{\sfsb}_{\lambda= +1}(q,-\textbf{k};x',x)
\end{equation}
and performing the $\omega$ integral, 
$\re
\,
\sigma^{ij}_{\sfsb,R}(\Omega;x,x')$
can be simplified as
\begin{equation}
\!\!\!\!
\begin{aligned}[b]
	&\re
	\,
	\sigma^{\mu \nu}_{\sfsb,R}(\Omega;x,x')
	=
	\frac{\pi}{2} \left(\frac{e}{m}\right)^2
	\frac{1}{\Omega}
	\int_{q,\textbf{k}}
	\,
	k^{\mu} k^{\nu}
\\
	&
	\times
	\left\{
	\begin{aligned}
		&\,\left[
			f(E^{\sfs}_{k_y}) - f(E^{\sfs}_{k_y} + \Omega)
		\right]	
	\\
	&\,
	\times
		\delta(E^{\sfs}_{k_y} + \Omega - E^{\sfb}_{q,\textbf{k}})
		\,
		\Sigma^{\sfsb}_{+1}(q,\textbf{k};x',x)
	\\
	+
	&\,
		\left[
			f(-E^{\sfb}_{q,\textbf{k}}) - f(-E^{\sfb}_{q,\textbf{k}} + \Omega)
		\right]
	\\
	&\,
	\times
		\delta(-E^{\sfb}_{q,\textbf{k}} +\Omega -  E^{\sfs}_{k_y})
		\,
		\Sigma^{\sfsb}_{+1}(q,-\textbf{k};x',x)
	\end{aligned}
	\right\}\!\!.
\end{aligned}
\!\!
\end{equation}
We can proceed further by considering the $T = 0$ limit, in which the Fermi function becomes
$f(E) = \theta(-E)$. 
In this case, the surface-bulk optical conductivity becomes
\begin{equation}
\begin{aligned}
&
\re
\,
\sigma^{\mu \nu}_{\sfsb,R}(\Omega;x,x')
=
2\pi \left(\frac{e}{m}\right)^2
\frac{1}{\Omega}
\int_0^{\infty} \frac{dq}{2\pi}
\int_0^{k_F} \frac{k \,dk}{2 \pi}
\\
&
\times
\delta^{\mu \nu}
\int_{0}^{\pi/2} 
\frac{d\theta}{2\pi}
\,
k^{\mu} k^{\nu}
\,
\delta(E^{\sfs}_{k_y} +\Omega  -E^{\sfb}_{q,\textbf{k}}  )
\,
{\Sigma}^{\sfsb}_{+1}(q,\textbf{k};x',x),
\end{aligned}
\end{equation}
where we made a change of coordinates
$(k_y, k_z ) = k(\sin\theta, \cos\theta)$ and repeated indices are not summed over.
The constraint 
$k_y^2 + k_z^2 \le k_F^2$ for the surface states is now automatically satisfied. 
By expressing the $\delta$ function as
\begin{equation}
\delta(E^{\sfs}_{k_y} +\Omega  -E^{\sfb}_{q,\textbf{k}}  )
=
\sum_{\substack{\alpha = \pm 1,
		\\
		{q_{\textbf{k},\Omega}^{(\alpha)} \in \mathbb{R}^+}}
}
\Bigg|
\frac{	\partial E^{\sfb}_{q,\textbf{k}}}{\partial q}
\Bigg|^{-1}_{q_{\textbf{k},\Omega}^{(\alpha)}}
\delta(q - q^{(\alpha)}_{\textbf{k},\Omega} ),
\end{equation}
where
\begin{equation}\label{eq:q_root}
\begin{aligned}
	q^{(\pm 1)}_{\textbf{k},\Omega}
	=&\,
	\sqrt{
		-m^2\Delta^2 - \kappa_{\textbf{k}}^2 \pm 2 m \sqrt{ \Delta^2 \kappa_{\textbf{k}}^2 - 2  \Delta k_y \Omega + \Omega^2}
	}
\end{aligned}
\end{equation}
solve
$E^{\sfs}_{k_y} +\Omega  -E^{\sfb}_{q,\textbf{k}} = 0$,
we can eliminate the $q$ integral, yielding
\begin{equation}\label{eq:sec4_sigma_clean}
\begin{aligned}
	&\re
	\,
	{\sigma}^{\mu \nu}_{\sfsb,R}(\Omega;x,x')
	\\
	&=
	\frac{	\delta^{\mu \nu}}{\Omega}
	\left(\frac{e^2}{m}\right)
	\int_0^{k_F} \frac{k \,dk}{2 \pi}
	\int_{0}^{\pi/2} 
	\frac{d\theta}{2\pi}
	\,
	\frac{k^{\mu} k^{\nu}
	}{
		\sqrt{\Delta^2 \kappa_{\textbf{k}}^2 - 2\Delta k_y \Omega + \Omega^2}
	}
	\\
	&\times
	\sum_{\substack{\alpha = \pm 1,
			\\
			{q_{\textbf{k},\Omega}^{(\alpha)} \in \mathbb{R}^+}}
	}
	\frac{
		E^{\sfb}_{q_{\textbf{k},\Omega}^{(\alpha)},\textbf{k}}
	}{
		q_{\textbf{k},\Omega}^{(\alpha)}
	}
	\,
	\re
	\,
	{\Sigma}^{\sfsb}_{+1}(q_{\textbf{k},\Omega}^{(\alpha)},\textbf{k};x',x),
\end{aligned}
\end{equation}
where the sum over $\alpha =  \pm 1$ takes the two roots in Eq.~(\ref{eq:q_root}) into account.
The constraint ${q_{\textbf{k},\Omega}^{(\alpha)} \in \mathbb{R}^+}$ on allowed
standing wave momenta is crucial, and restricts the domains of the remaining
$k$ and $\theta$ integrations [via Eq.~(\ref{eq:q_root})].
In Eqs.~(\ref{eq:q_root}) and (\ref{eq:sec4_sigma_clean}),
the wavenumber $\kappa_{\textbf{k}}$ controls the confinement depth of the
surface states, see Eqs.~(\ref{eq:eigenstate_surf}) and (\ref{eq:kappa_k_Def}). 

From Eq.~(\ref{eq:absorbance}), the absorbance due to surface bulk transition is
\begin{equation}\label{eq:absorbance_sb}
\begin{aligned}
\mathsf{A}^{\mu \mu}_{\sfsb}
&=
-
\frac{16\pi }{c^2}
\frac{
	K
}{
	\LL^{-2}
}
\im \, {\cal Q}^{\mu \mu},
\end{aligned}
\end{equation}
where
\begin{equation}
\im {\cal Q}^{\mu \mu}(\Omega)
=
-\Omega
\int_{x,x'>0}
e^{-{\cal K} x}
\,
\re
\,
{\sigma}^{\mu \mu}_{\sfsb,R}(\Omega;x,x')
\,
e^{-{\cal K} x'}
\end{equation}
and ${\cal K} = \sqrt{\LL^{-2} - K^2}$ in the clean limit.
Introducing the dimensionless quantities 
$\hat{k}^{\mu} = k^{\mu}/k_F$, 
$\hat{q} = q/k_F$, 
$\hat{\kappa}_{\hat{\textbf{k}}} = \kappa_{\textbf{k}}/k_F$
[Eq.~(\ref{eq:kappa_k_Def})],
and 
$\hat{\Omega} = \Omega /\Delta k_F = \Omega /\Delta_0$,
where $\Delta_0$ is the pairing energy,
the absorbance can be expressed as
\begin{equation}\label{eq:absorbance_sb2}
\begin{aligned}
	\mathsf{A}^{\mu \mu}
	=&\,
	\frac{
		K
	}{
		k_F^2 \lcoh
	}
	\hat{{\cal F}}^{\mu \mu}(\hat{\Omega}),
\end{aligned}
\end{equation}
where the dimensionless integral $\hat{{\cal F}}^{\mu \mu}$ is given by
\begin{equation}\label{eq:kernelF_dimensionless}
\begin{aligned}
	\hat{{\cal F}}^{\mu \mu}
	(\hat{\Omega})
	=&\,
	24 \pi^2 
	\int_0^{1} \frac{\hat{k} \,d\hat{k}}{2 \pi}
	\int_{0}^{\pi/2} 
	\frac{d\theta}{2\pi}
	\,
	\frac{\hat{k}^{\mu} \hat{k}^{\mu}
}{
	\sqrt{
		\hat{\kappa}_{\hat{\textbf{k}}}^2 
		- 
		2 \hat{k}_y\hat{\Omega} 
		+ 
		\hat{\Omega}^2
	}
}
\\
&\times
	\sum_{\substack{
		\alpha = \pm 1,
		\\
		{\hat{q}_{\textbf{k},\Omega}^{(\alpha)} \in \mathbb{R}^+}
		}
	}
	\left[
	\frac{
		-\hat{k}_y + \hat{\Omega}
	}{
		\hat{q}_{\hat{\textbf{k}},\hat{\Omega}}^{\alpha}
	}
	\right]
	\re
	\,
	\hat{J}^{\sfsb}_{\hat{q}_{\textbf{k},\Omega}^{(\alpha)},\hat{\textbf{k}}},
\end{aligned}
\end{equation}
in which repeated indices are again not summed over. The dimensionless function $\hat{J}^{\sfsb}_{\hat{q}_{\textbf{k},\Omega}^{(\alpha)},\hat{\textbf{k}}}$ is due to the integral 
\begin{equation}\label{eq:kernel_Jhat}
\begin{aligned}
	\hat{J}^{\sfsb}_{\hat{q},\hat{\textbf{k}}}
	=
	k_F^2 \lcoh \int_{x,x' > 0}
	e^{-{\cal K} x}
	\,
	{\Sigma}^{\sfsb}_{+1}(q,\textbf{k};x',x)
	\,
	e^{-{\cal K} x'},
\end{aligned}
\end{equation}
which is just the integrated double-overlap between the surface and bulk states with a London response weight. 
The detailed form of $\hat{J}^{\sfsb}_{\hat{q},\hat{\textbf{k}}}$ can be found in Appendix \ref{sec:app_double_overlap}.

In Fig.~\ref{fig:hatF}, we plot the absorbance $\mathsf{A}(\Omega)$ with type I and type II parameters based on Eq.~(\ref{eq:absorbance_sb2}).
In the type-I case [Figs.~\ref{fig:hatF}(a)--(c)], 
$\mathsf{A}$ displays relatively broad features. 
Its $yy$ and $zz$ components are both peaked at around $\Omega = 2\Delta_0$.
Since the coherence length $\lcoh$ increases with the spatial extent of the surface state, 
a larger $\lcoh$ results in an enhanced overlap between the bulk and surface states.
However, at the same time, a larger $\lcoh$ 
limits the region of the surface states that can respond to the external field.
The magnitude of $\mathsf{A}$ is controlled by the competition between these two effects.
Nevertheless, the qualitative features of $\mathsf{A}$ are almost unchanged with increasing $\lcoh$ except that the tail region at higher frequency is broadened.

On the other hand, in the type-II case [Figs.~\ref{fig:hatF}(d)--(f)], the feature of $\mathsf{A}$
is relatively sharp, namely there is a peak at around $2\Delta_0$ for its $yy$ component, and slightly above $\Delta_0$ for its $zz$ component.
Interestingly, the positions of these peaks are nearly independent of $\LL$ in the type II limit.
As $\LL$ increases, the integrated double-overlap in Eq.~(\ref{eq:kernel_Jhat}) is gradually suppressed due to the orthogonality between the bulk and surface states, 
resulting in a smaller $\mathsf{A}$ .

In both cases, the peaks in $\mathsf{A}$ are due to the square root van-Hove singularity in its integrand 
[Eq.~(\ref{eq:kernelF_dimensionless})], weighted by various factors, including the current operator and the 
surface-bulk double-overlap. 
Although it is hard to pinpoint the exact position of the peaks analytically, by slicing the integrand at different $k$ 
and partially integrating out ${\theta}$, our numerics reveal that the peaks in $\hat{{\cal F}}$ mainly come from the 
contribution at the vicinity of $k = k_F$, at which the surface states deconfine into the bulk.
Roughly speaking, the peak for the $yy$ component is dominated by the contributions at the proximity of $(k_y,k_z) \simeq (k_F,0)$. 
In this region, the transitions originate from surface band with energy $\sim -\Delta k_F$ to the bulk band with energy $\sim \Delta k_F$, 
resulting in a sharp peak at $\Omega = 2\Delta_0$.  For the $zz$ component, the situation is more complex.

Our results here reveal that the surface-bulk optical conductivity is in general nonvanishing.  
Consequently, in the clean limit, the optical absorption of the system is contributed by the surface-bulk 
transitions.


\section{Disordered bulk optical response}
\label{sec:resp_dirty}

\subsection{Keldysh formalism}\label{subsec:Keldysh_formalism}

In this section, we study the bulk optical conductivity of a WSC with quenched disorder,
based on the finite-temperature Keldysh response theory \cite{s_wave_NLSM4_Kamenev,Keldysh_review}. 
We consider spin-1/2 electrons with $p+ip$ pairing as in  $^3$He\textit{A} \cite{He3_book_Vollhardt,He3_book_Volovik,He3B_review_Sato_Machida}; 
the formalism can be directly applied to other symmetries. 
The Keldysh generating function is \cite{s_wave_NLSM5_Yunxiang}
\begin{widetext}
\begin{equation}\label{eq:ZDef}
Z 
\equiv
\int D\bar{\psi} D\psi
\,
\exp
\left[
\begin{gathered}
	i \nint_{\omega,\textbf{x},\textbf{x}'}
	\bar{\psi}(\omega,\textbf{x}) 
	\,
	\hat{G}^{-1}(\omega; \textbf{x},\textbf{x}')
	\,
	\psi(\omega,\textbf{x}')
\\
	+
	i \dfrac{W}{4}
	\nsum_{
		a 
		\in 
		\left\lbrace 1,2\right\rbrace 
	}
	\htau^3_{a,a}
	\nint_{t,\textbf{x}}
	\left\{
	-i\bar{\psi}_a 
	\left[
	(i\overleftarrow{\nabla}_l)
	\hat{s}^i
	\hat{s}^2 
	\right]
	\bar{\psi}_a^{\sfT}
	\right\}
	\left\{
	i \psi_a^{\sfT} 
	\left[
	\hat{s}^2 
	\hat{s}^i
	(-i\overrightarrow{\nabla}_l)
	\right] 
	\psi_a
	\right\}
\\
	-\dfrac{i}{2} 
	\nint_{\omega,\omega',\textbf{x}}
	\left[
	\textbf{A}_{\cl}(\omega - \omega')
	\,
	\cdot
	\,
	\bar{\psi}(\omega) \htau^3 	
	(-i\overleftrightarrow{\nabla}) 
	\psi(\omega')
	+
	\textbf{A}_{\q}(\omega - \omega')
	\,
	\cdot
	\,
	\bar{\psi}(\omega) 
	(-i\overleftrightarrow{\nabla}) 
	\psi(\omega')
\right]
\end{gathered}
\right],
\end{equation} 
where the $\hat{s}^i$ and $\htau^j$ respectively denote Pauli matrices acting on the spin-1/2 and Keldysh spaces.
In Eq.~(\ref{eq:ZDef}), summations over $i \in \left\lbrace 1,2,3\right\rbrace $ 
and 
$l \in \left\lbrace x,y,z \right\rbrace $  are assumed.
The fermionic field $\psi = \psi_{a,s}(t,\textbf{x})$, where the Keldysh index
$a \in \left\lbrace 1,2\right\rbrace$ corresponds to the $\left\lbrace \text{forward, backward} \right\rbrace$ time contour,
and the spin index
$s\in \left\lbrace \uparrow,\downarrow\right\rbrace$.
The noninteracting Green's function in the space-time basis is given by
\begin{equation}
i\hat{G}(t,t';\textbf{x},\textbf{x}')
=
\begin{bmatrix}
i\hat{G}_T & i\hat{G}_<
\\
i\hat{G}_> & i\hat{G}_{\bar{T}}
\end{bmatrix}
=
\begin{bmatrix}
\left\langle 
T \, \psi(t,\textbf{x}) \, \bar{\psi}(t',\textbf{x}')
\right\rangle_0
&
-\left\langle 
 \bar{\psi}(t',\textbf{x}') \, {\psi}(t,\textbf{x})
\right\rangle_0
\\
\left\langle 
\psi(t,\textbf{x}) \, \bar{\psi}(t',\textbf{x}')
\right\rangle_0
&
\left\langle 
\bar{T} \, \psi(t,\textbf{x}) \, \bar{\psi}(t',\textbf{x}')
\right\rangle_0
\end{bmatrix},
\end{equation}
where $T$ and $\bar{T}$ respectively denote the time-ordering and anti-time-ordering operators.

Spin-triplet pairing is mediated by the attractive interaction $W > 0$ in Eq.~(\ref{eq:ZDef}).
The operator 
\[
	i \psi_a^{\sfT}(t,\vex{x}) 
	\left[
	\hat{s}^2 
	\hat{s}^i
	(-i\overrightarrow{\nabla}_l)
	\right] 
	\psi_a(t,\vex{x})
\]
annihilates a local spin-triplet Cooper pair.

The net vector potential on the forward (backward) part of time contour $\textbf{A}_1$ $(\textbf{A}_2)$ 
in Eq.~(\ref{eq:ZDef})
is expressed in terms of the classical and quantum components of the field, i.e.\
\begin{equation}
	\textbf{A}_1 = \textbf{A}_{\cl} + \textbf{A}_{\q},
	\qquad
	\textbf{A}_2 = \textbf{A}_{\cl} - \textbf{A}_{\q}.
\end{equation}

Decoupling the pairing interaction by Hubbard-Stratonovich transformation, we have
\begin{equation}
	Z
	=
	\int D\bar{\psi} D\psi D \bar{\Delta} D \Delta
	\,
	\exp
	\left\{
	\begin{gathered}
	i \nint_{\omega,\textbf{x},\textbf{x}'}
	\bar{\psi}(\omega,\textbf{x})
	\,
	\hat{G}^{-1}(\omega; \textbf{x},\textbf{x}')
	\,
	\psi(\omega,\textbf{x}')
\\
	-\dfrac{i}{2} 
	\nint_{\omega,\omega',\textbf{x}}
	\left[
		\textbf{A}_{\cl}(\omega - \omega')
		\,
		\bar{\psi}(\omega) \htau^3 
		(-i \overleftrightarrow{\nabla})
		\psi(\omega')
		+
		\textbf{A}_{\q}(\omega - \omega')
		\,
		\bar{\psi}(\omega) 
		(-i\overleftrightarrow{\nabla})
		\psi(\omega')
	\right]
\\
+ 
	i 
	\dfrac{2}{W} 
	\nint_{t,\textbf{x}}
	(
	\Delta_{\q}^{il*} \Delta_{\cl}^{il} + \Delta_{\q}^{il} \Delta_{\cl}^{il*}
	)
\\
	- 	
	\dfrac{i}{2} 
	\nint_{\omega,\omega',\textbf{x}}
	\left\{
	\begin{gathered}
		\Delta_{\cl}^{il}(\omega + \omega')
		\left[
		\bar{\psi}(\omega) 
		(i\overleftarrow{\nabla})_l
		\hat{s}^i 
		\hat{s}^2 \htau^3 
		\bar{\psi}^{\sfT} (\omega')
		\right]
	\\
		+
		\Delta_{\q}^{il} (\omega + \omega')
		\left[
		\bar{\psi}(\omega) 
		(i\overleftarrow{\nabla})_l
		\hat{s}^i 
		\hat{s}^2 
		\bar{\psi}^{\sfT}(\omega')
		\right]
	\\
		- 
		\Delta_{\cl}^{il*}(\omega + \omega')
		\left[
		\psi^{\sfT}(\omega)
		\hat{s}^2 \htau^3 
		\hat{s}^i 
		(-i\overrightarrow{\nabla})_l
		\psi(\omega')
		\right]
\\
		-	
		\Delta_{\q}^{il*}(\omega + \omega')
		\left[
		\psi^{\sfT}(\omega) 
		\hat{s}^2 
		\hat{s}^i 
		(-i\overrightarrow{\nabla})_l
		\psi(\omega')
		\right]
	\end{gathered}
	\right\}
\end{gathered}
\right\}.
\end{equation}

\subsubsection{Keldysh and ``thermal'' rotations}

The inverse Green's function can be written as
\begin{equation}
	\hat{G}^{-1}(\omega; \textbf{x},\textbf{x}')
	=
	\hat{U}_{\LO}
	\,
	\hat{M}_F(\omega)
	\,
	\hat{G}_{\eta}^{-1} (\omega;\textbf{x},\textbf{x}')
	\,
	\hat{M}_F(\omega)
	\,
	\hat{U}_{\LO}^{\dagger} \htau^3,
\end{equation}
where 
\begin{equation}
	\hat{G}_{\eta}(\omega)
	=
	(\omega + i\eta \htau^3 - \hat{h}_0)^{-1}
	,
	\quad
	\hat{U}_{\LO}
	=
	\frac{1}{\sqrt{2}} (1 + i\htau^2)
	,
	\quad
	\hat{M}_F(\omega)
	=
	\begin{bmatrix}
	1 & F(\omega)
	\\
	0 & -1
	\end{bmatrix}_{\tau},
	\text{ and}
	\quad
	F(\omega) = \tanh\left(\frac{\omega}{2T}\right),
\end{equation}
with $\hat{h}_0$ being the static single particle Hamiltonian and $\eta \rightarrow 0^+$. 
The diagonal components of $\hat{G}_{\eta}$ correspond to the retarded and advanced non-interacting Green's functions.
We can remove the distribution function $F(\omega)$ from the non-interacting part of the fermionic action through 
the following non-unitary transformation \cite{s_wave_NLSM5_Yunxiang},
\begin{equation}
	\psi(\omega,\textbf{x})
	\rightarrow
	\htau^3 
	\,
	\hat{U}_{\LO} 
	\,
	\hat{M}_F(\omega) 
	\,
	\psi(\omega,\textbf{x}),
\qquad
	\bar{\psi}
	\rightarrow
	\bar{\psi}(\omega,\textbf{x})
	\,
	\hat{M}_F(\omega)
	\,
	\hat{U}_{\LO}^{\dagger}.
\end{equation}
The Keldysh function then becomes
\begin{equation}
Z
=
\int D\bar{\psi} D\psi D \bar{\Delta} D \Delta
\,
\exp
\begin{Bmatrix}
i \nint_{\omega,\textbf{x},\textbf{x}'}
\bar{\psi}(\omega,\textbf{x}) 
\hat{G}_{\eta}^{-1}(\omega; \textbf{x},\textbf{x}')
\psi(\omega,\textbf{x}')
\\\\
-\dfrac{i}{2} \nint_{\omega,\omega',\textbf{x}}
\begin{bmatrix}
\textbf{A}_{\cl}(\omega - \omega')
\,
\bar{\psi}(\omega) 
\hat{M}_F(\omega) \hat{M}_F(\omega')
(-i\overleftrightarrow{\nabla} )
\psi(\omega')
\\
+
\textbf{A}_{\q}(\omega - \omega')
\,
\bar{\psi}(\omega) 
\hat{M}_F(\omega) \htau^1 \hat{M}_F(\omega')
(-i\overleftrightarrow{\nabla} )
\psi(\omega')
\end{bmatrix}
\\\\
+ 
i \dfrac{2}{W} 
\nint_{t,\textbf{x}}
(
\Delta_{\q}^{il*} \Delta_{\cl}^{il} + \Delta_{\q}^{il} \Delta_{\cl}^{il*}
)
\\\\
- \dfrac{i}{2} 
\nint_{\omega,\omega',\textbf{x}}
\begin{Bmatrix}
\Delta_{\cl}^{il}(\omega + \omega')
\left[
\bar{\psi}(\omega) 
(i \overleftarrow{\nabla})_l
\hat{s}^i
\hat{s}^2
M_F(\omega)
\htau^1
\hat{M}_F^{\sfT}(\omega')
\bar{\psi}^{\sfT} (\omega')
\right]
\\
+
\Delta_{\q}^{il} (\omega + \omega') 
\left[
\bar{\psi}(\omega) 
(i \overleftarrow{\nabla})_l
 \hat{s}^i 
\hat{s}^2 
M_F(\omega)
\hat{M}_F^{\sfT}(\omega')
\bar{\psi}^{\sfT}(\omega')
\right]
\\
-
\Delta_{\cl}^{il*}(\omega + \omega') 
\left[
\psi^{\sfT}(\omega)
\hat{s}^2 
\hat{s}^i 
\hat{M}_F^{\sfT}(\omega)
\htau^1
\hat{M}_F(\omega')
(-i \overrightarrow{\nabla})_l
\htau^3 \psi(\omega')
\right]
\\
-
\Delta_{\q}^{il*}(\omega + \omega')
\left[
\psi^{\sfT}(\omega) 
\hat{s}^2 
\hat{s}^i 
\hat{M}_F^{\sfT}(\omega)
\hat{M}_F(\omega')
(-i \overrightarrow{\nabla})_l
\psi(\omega')
\right]
\end{Bmatrix}
\end{Bmatrix},
\end{equation}
where the thermal matrix $\hat{M}_F(\omega)$ now appears solely in the 
coupling to the superconducting order parameter and the external vector potential. 

At the static mean-field level
for $p$-wave, spin-triplet pairing, 
we take
\begin{eqnarray}
(i\overleftarrow{\nabla})_l \Delta_{\cl}^{il}(\omega + \omega')
&=&
i
 \bm{\sfd}^i
 \,
 \delta_{\omega + \omega',0}
,
\quad
 \Delta_{\cl}^{il*} (-i\overrightarrow{\nabla})_l
=
i(\bm{\sfd}^{i})^*
\,
\delta_{\omega + \omega',0}
,
\quad
\Delta_{\q} (\omega + \omega') 
=
\Delta_{\q}^*(\omega + \omega')  
= 
0
,
\end{eqnarray}
where $\bm{\sfd}$ is just the (unnormalized) $d$-vector order parameter \cite{He3_book_Vollhardt,He3_book_Volovik,He3B_review_Sato_Machida}.
Together with the properties 
\begin{equation}
\htau^1 \hat{M}_F(-\omega) \htau^1 = -\hat{M}_F(\omega),
\quad
\hat{M}_F^{-1}(\omega) = \hat{M}_F(\omega),
\end{equation}
the static mean-field Keldysh function can be recast as
\begin{equation}\label{eq:ZMFT}
Z
=
\int D\bar{\psi} D\psi 
\,
\exp
\begin{Bmatrix}
i \nint_{\omega,\textbf{x},\textbf{x}'}
\bar{\psi}(\omega,\textbf{x}) 
\hat{G}_{\eta}^{-1}(\omega; \textbf{x},\textbf{x}')
\psi(\omega,\textbf{x}')
\\\\
-\dfrac{i}{2} \nint_{\omega,\omega',\textbf{x}}
\left[
\begin{gathered}
\textbf{A}_{\cl}(\omega - \omega')
\,
\bar{\psi}(\omega) 
\hat{M}_F(\omega) \hat{M}_F(\omega')
(-i\overleftrightarrow{\nabla} )
\psi(\omega')
\\
+
\textbf{A}_{\q}(\omega - \omega')
\,
\bar{\psi}(\omega) 
\hat{M}_F(\omega) \htau^1 \hat{M}_F(\omega')
(-i\overleftrightarrow{\nabla} )
\psi(\omega')
\end{gathered}
\right]
\\\\
+\dfrac{i}{2} 
\nint_{\omega,\textbf{x}}
\begin{bmatrix}
i
\bar{\psi}(\omega) 
(\bm{\sfd}\cdot \bm{\hat{s}})
\hat{s}^2 
\htau^1
\bar{\psi}^{\sfT} (-\omega)
\\\\
-
i 
\psi^{\sfT}(-\omega)
\htau^1	
\hat{s}^2 
(\bm{\sfd}^*\cdot \bm{\hat{s}})
 \psi(\omega)
\end{bmatrix}
\end{Bmatrix}.
\end{equation}
\end{widetext}

\subsubsection{Majorana spinor reformulation}

In order to encode both particle-hole and particle-particle channel fluctuations, 
we recast the action in terms of the (real) Majorana spinor \cite{s_wave_NLSM5_Yunxiang}
\begin{eqnarray}
\chi 
&\equiv&
\begin{bmatrix}
\psi
\\
\hat{s}^2 \htau^1 \hSig^1 \bar{\psi}^{\sfT}
\end{bmatrix}
,
\\
\label{eq:chibar}
\bar{\chi}
&=&
[
\bar{\psi}
\,\,
-\psi^{\sfT} \hat{s}^2 \htau^1 \hSig^1
]
=
-\chi^{\sfT} \hat{s}^2  \hsig^1 \htau^1 \hSig^1,
\end{eqnarray}
which carry discrete particle-hole ($\sigma$), Keldysh ($\tau$) and spin $(s)$ indices. 
In addition, $\chi$ possesses a continuous frequency $|\omega|$ index ranging over the positive real axis, 
and a discrete sign index $\Sigma = \sign(\omega) \in \left\lbrace + , -\right\rbrace$. 
The Pauli matrix in Eq.~(\ref{eq:chibar}) is defined as
$
\bra{\omega} \hSig^1 \ket{\omega'} = 2\pi \, \delta(\omega + \omega').
$
Eq.~(\ref{eq:ZMFT}) can be expressed compactly as
\begin{equation}
	Z
	=
	\int D\chi
	\,
	\exp
	\left\lbrace 
	\frac{i}{2}
	\int_{\textbf{x}}
	\bar{\chi}
	\left[
	\hat{G}_{\bdg}^{-1}
	-
	\frac{1}{2}
	\BigA \cdot (-i\overleftrightarrow{\nabla})
	\right]
	\chi
	\right\rbrace\!,
\end{equation}
where
\begin{equation}\label{eq:BdG_G}
	\hat{G}_{\bdg}^{-1}
	\equiv
	\hsig^3 \hat{\omega}
	+
	i\eta \htau^3 \hsig^3
	-
	\hsig^3 \hat{h}_{\bdg},
\end{equation}
\begin{equation}
	\hat{h}_{\bdg}
	\equiv
	\begin{bmatrix}
		\hat{h}_0 &  -i \bm{\sfd}\cdot \bm{\hat{s}} 
	\\
		i\bm{\sfd}^*\cdot \bm{\hat{s}}  & -\hat{s}^2 \hat{h}_0^{\sfT} \hat{s}^2
	\end{bmatrix}_{\sigma},
\end{equation}
and
\begin{align}
	\BigA_{\omega,\omega'}(\textbf{x})
	\equiv&\,
	\textbf{A}_{\cl}(\omega - \omega',\textbf{x})
	\,
	\hat{\gamma}_{\cl}(\omega,\omega')
\nonumber\\
	&\,
	+
	\textbf{A}_{\q}(\omega - \omega',\textbf{x})
	\,
	\hat{\gamma}_{\q}(\omega,\omega'),
\end{align}
with
\begin{eqnarray}\label{eq:def_gamma_matrices}
	\hat{\gamma}_{\cl} (\omega,\omega')
	&=&
	\hat{M}_F(\omega)
	\hsig^3  \hat{M}_F(\omega'),
\\
	\hat{\gamma}_{\q} (\omega,\omega')
	&=&
	\hat{M}_F(\omega) \htau^1 \hsig^3 \hat{M}_F(\omega').
\end{eqnarray}


\subsection{Disorder averaging and saddle-point equation}

For simplicity, we consider only onsite scalar potential disorder and neglect 
spatial fluctuations of the order parameter $\Delta$. 
We assume that the static impurity potential $u(\textbf{x})$ is Gaussian distributed
\begin{equation}
	P[u]
	=
	\exp
	\left[
	-\frac{1}{2g}
	\int_{\textbf{x}} u^2 (\textbf{x})
	\right],
\end{equation}
where $ g = \Gel/\pi \nu_0$ characterizes the width of the distribution. 
Here, 
$\Gel \equiv 1/(2\tau_{\mathsf{el}})$, 
$\tau_{\mathsf{el}}$ is the elastic scattering time, 
and $\nu_0$ is the density of states per spin species at the Fermi surface. 
The disordered part of the action is
\begin{equation}
	S_{\mathsf{dis}}
	=
	\frac{i}{2}
	\int_{\textbf{x}}
	\bar{\chi}(\textbf{x})
	u(\textbf{x})
	\chi(\textbf{x})
\end{equation}
We average over the disorder potential $u$ to get
\begin{equation}
\begin{aligned}
	\left\langle 
	e^{-S_{\mathsf{dis}}}
	\right\rangle
	=&\,
	\int Du
	\,
	P[u]
	\,
	\exp
	\left[
	-\frac{i}{2}
	\int_{\textbf{x}}
	\bar{\chi}(\textbf{x})
	u(\textbf{x})
	\chi(\textbf{x})
	\right]
\\
	=&\,
	\exp
	\left\lbrace 
	-
	\frac{g}{2}
	\int_{\textbf{x}}
	\Tr
	\left[
	\left(
	\frac{\chi
		\bar{\chi}}{2}
	\right)
	\left(
	\frac{\chi
		\bar{\chi}}{2}
	\right)
	\right]
	\right\rbrace 
\\
	=&\,
	\int 
	\!
	D \hat{Q}
	\exp
	\!
	\left\lbrace 
	-
	\frac{1}{4g}
	\int_{\textbf{x}}
	\Tr[\hat{Q}^2]
	-
	\frac{1}{2}
	\int_{\textbf{x}}
	\Tr
	\left[
	\hat{Q}
	\chi
	\bar{\chi}
	\right]
	\right\rbrace\!,
\end{aligned}
\end{equation}
where we have perform a Hubbard-Stratonovich transformation by introducing the 
matrix field $\hat{Q}$. 
By integrating out the fermionic field $\chi$, the action becomes
\begin{equation}\label{eq:action}
	S 
	=
	\frac{1}{4g}
	\int_{\textbf{x}}
	\Tr[\hat{Q}^2]
	-
	\frac{1}{2}
	\Tr
	\log
	\left[
	\hat{G}_{\bdg}^{-1}
	+
	i
	\hat{Q}
	-
	\BigA \cdot \textbf{v}_F
	\right],
\end{equation}
where $\hat{G}_{\bdg}$ is the Bogoliubov-de Gennes (BdG) Green's function defined in Eq.~(\ref{eq:BdG_G}). 

Next, we derive the saddle point equation in the absence of the vector potential. 
By varying the action with respect to $\hat{Q}$ and setting $\delta S/\delta \hat{Q} \vert_{\hat{Q} = \hat{Q}_{\sfsp}} = 0$, we have
\begin{equation}
\frac{1}{2g}
\hat{Q}_{\sfsp}
=
\frac{i}{2}
\bra{\textbf{x}}
\frac{1}{
	\hat{G}_{\bdg}^{-1}
	+
	i
	\hat{Q}_{\sfsp}
}
\ket{\textbf{x}}
,
\end{equation}
or
\begin{equation}
\label{eq:Q_saddle_pt_eqn}
	\frac{-i}{g}
	\hat{Q}_{\mathsf{sp}}
	=
	\int_{\textbf{k}}
	\frac{1}{
		\hsig^3 \hat{\omega}
		+
		i\eta \htau^3 \hsig^3
		-
		\hsig^3 \hat{h}_{\bdg} + i \hat{Q}_{\sfsp}
	}
,
\end{equation}
where $\int_{\textbf{k}} = \int \frac{d^3 k}{(2\pi)^3}$. Eq.~(\ref{eq:Q_saddle_pt_eqn}) is 
equivalent to the self-consistent Born approximation (SCBA). 
In the context of disordered Weyl semimetals, it is known that the SCBA cannot capture the nonperturbative effects 
generated by rare regions \cite{rare_WSM_Jed1,
	rare_WSM_Jed2,
	rare_WSM_Jed3,
	rare_WSM_Jed4,
	rare_WSM_Jed5,
	rare_WSM_Rahul_Huse_instanton,
	rare_WSM_eps_exp_Syzranov}.
For Majorana surface states in WSCs, similar nonperturbative effects induced by rare states are also found numerically \cite{rare_WSC_Jed6}.
However, these rare states are not expected to qualitatively modify the optical conductivity, and thus we restrict our calculation here to the SCBA. 

To solve the saddle point equation [Eq.~(\ref{eq:Q_saddle_pt_eqn})], we employ the \textit{ansatz} 
\begin{equation}
	\hat{Q}_{\sfsp}
	=
	\begin{bmatrix}
	Q_{11} & Q_{12}
	\\
	Q_{21} & -Q_{11}
	\end{bmatrix}_{\sigma}
	\htau^3
	.
\end{equation}
We then compute the matrix elements of Eq.~(\ref{eq:Q_saddle_pt_eqn}) on both sides.
For the WSC that we are studying, the BdG Hamiltonian in Eq.~(\ref{eq:model_ham}) corresponds to a $d$-vector 
$
\bm{\sfd}
=
i\Delta (k_x + ik_y) \hat{x}
$,
as in $^3$He\textit{A} \cite{He3_book_Vollhardt,He3_book_Volovik,He3B_review_Sato_Machida}.
Since the pairing term is odd in $\textbf{k}$, the off-diagonal terms of Eq.~(\ref{eq:Q_saddle_pt_eqn})
must vanish upon angular integration for self-consistency. Thus, we only have to focus on the diagonal components and solve
\begin{align}
	Q_{11}&\,
\nonumber\\
	=&\,
	i g
	\nu_0 \int d\tilE_{\textbf{k}}
	\int \frac{d\Omega_{\hat{\textbf{k}}}}{4\pi}
	\frac{
		-(\omega + iQ_{11}) 
	}{
		-(\omega + iQ_{11})^2 
		+ \Delta_0^2 \sin^2(\theta) + \tilE_{\textbf{k}}^2
	},
\end{align}
where $\Delta_0 = \Delta k_F$ is the pairing energy and we have converted 
\begin{equation}
	\int_{\textbf{k}}
	\simeq
	\nu_0 \int d\varE_{\textbf{k}}
	\int \frac{d\Omega_{\hat{\textbf{k}}}}{4\pi},
\end{equation}
with $d\Omega_{\hat{\textbf{k}}} = \sin\theta \, d\theta \,d\phi$. 
The integrals can be done analytically, the result is
\begin{equation}\label{eq:Q_11_SCBA}
	Q_{11}
	=
	\frac{1}{2\tel}
	\left(
	\frac{\omega + iQ_{11}}{\Delta_0}
	\right)
	\tanh^{-1}
	\left(
	\frac{\Delta_0}{\omega + i Q_{11}}
	\right).
\end{equation}
The real part of $Q_{11}$ carries the physical meaning of impurity scattering rate, 
while the imaginary part of it merely renormalizes the quasiparticle dispersion. 
In the following, we neglect the imaginary part of $\hat{Q}_{\sfsp}$ and just take
\begin{equation}\label{eq:Qsp}
	\hat{Q}_{\mathsf{sp}}(\omega)	
	=
	\frac{1}{2\tel}
	\htau^3 \hsig^3
	\,
	\Gamma(\omega),
\end{equation}
where 
$\Gamma(\omega) = 2\tel \, \re Q_{11}$.
In particular, in the weak disorder limit, $\Gamma(\omega)$ can be approximated as
\begin{equation}
\label{eq:Gamma_weakDisO}
	\Gamma(\omega)
	\simeq
	\re
	\left[
	\frac{\omega + i \eta}{\Delta_0}
	\tanh^{-1}
	\left(
	\frac{\Delta_0}{\omega + i \eta}
	\right)
	\right],
	\quad
	\Gel \ll \Delta_0,
\end{equation}
where $
\eta \rightarrow 0^+
$.
In Fig.~\ref{fig:p_wave_sigma_bulk}, we plot the self-consistent solution of $\Gamma(\omega)$ based on Eq.~(\ref{eq:Q_11_SCBA}).

\begin{figure}
	\centering
	{\includegraphics[width=8cm]{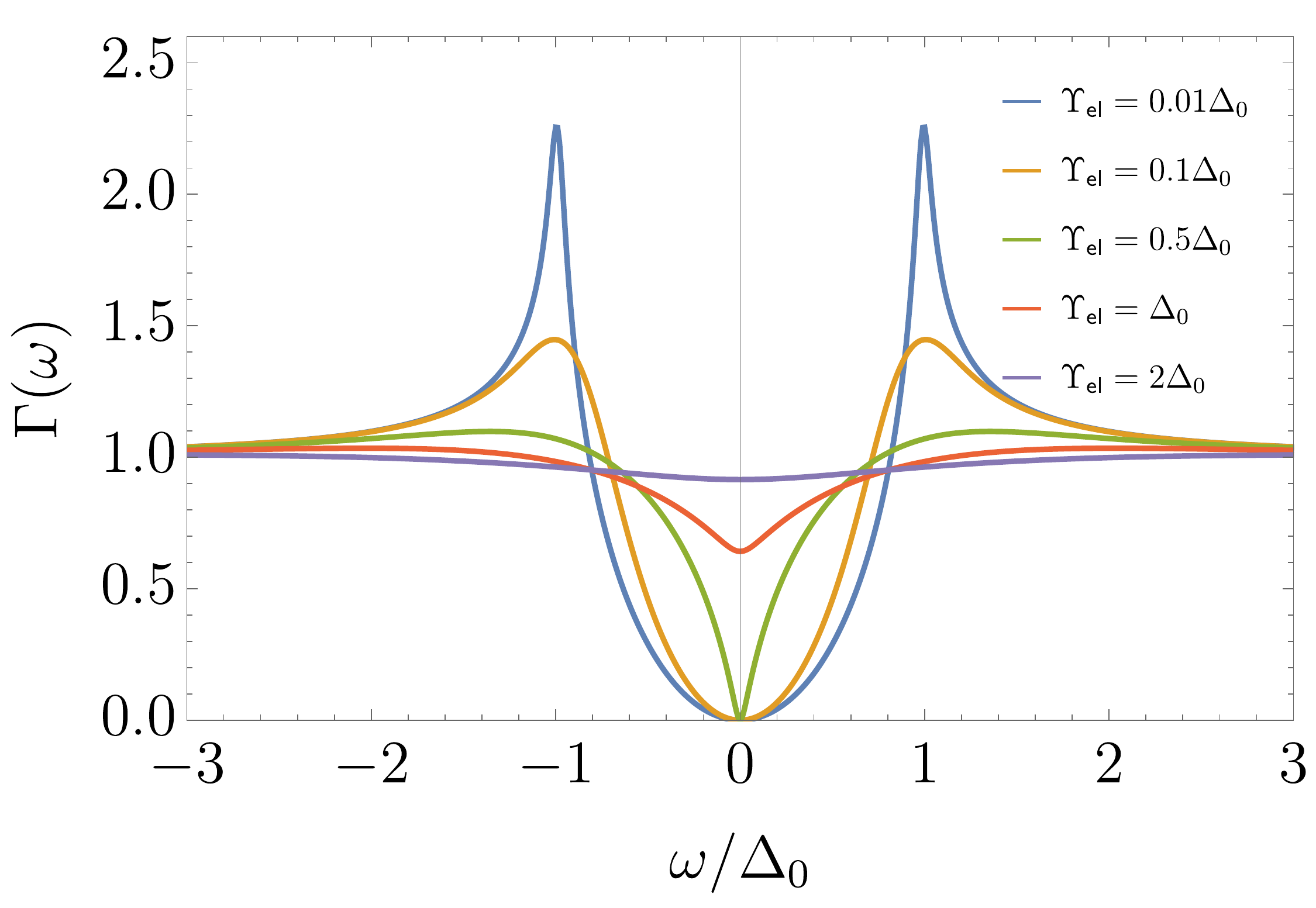} }
	\caption{Plot of the frequency dependence of $\Gamma(\omega)$ based on self-consistently solving Eq.~(\ref{eq:Q_11_SCBA}) at different disorder strengths 
	$\Gel = 1/2\tel$.}  	
	\label{fig:SCBA_p_wave_Gamma}
\end{figure}

For weak disorder, $\Gamma(\omega)$ is well-approximated by Eq.~(\ref{eq:Gamma_weakDisO}) and has two pronounced coherence peaks located at 
$\omega = \pm \Delta_0$.
The gapless nature of $\Gamma(\omega)$ originates from the nodal points of the system. 
As the scattering rate $\Gel$ increases, impurities gradually fill the gap and suppress the coherence peaks. 
In the strong-disorder limit, the $p$-wave features are completely suppressed and we recover the normal metal saddle point
\cite{s_wave_NLSM4_Kamenev,s_wave_NLSM5_Yunxiang}
\begin{equation}
	\hat{Q}_{\mathsf{sp}}(\omega) 
	\rightarrow 
	\frac{1}{2 \tel}
	\htau^3 \hsig^3,
\qquad
	\Gel \gg \Delta_0.
\end{equation}
This is consistent with the formation of a thermal metal phase with a nonzero density of quasiparticle states at zero energy 
\cite{rare_WSC_Jed6,WSC_thermal_Hall_Bitan_Sayed,Sau2017}.
The disorder strength dependence seen here 
is similar to that in the density of states of quasiparticles in $d$-wave superconductors
\cite{d_wave_Dynes_DoS,d_wave_Dynes_Doug}. 
This sharply contrasts with the saddle point solution in $s$-wave superconductors, in which the frequency dependence of 
$\hat{Q}_{\mathsf{sp}}^{\mathsf{s-wave}}$ is independent of $\Gel$ (see Appendix \ref{sec:app_Mattis_bardeen} for a review).


\subsection{Linear response optical conductivity}

The retarded linear response conductivity is defined as
\begin{equation}\label{eq:Kubo_def}
	\sigma_{\sfbb,R}^{\mu \nu}(\Omega)
	=
	-\frac{e^2}{i\Omega}
	\frac{1}{2i}
	\frac{
		\delta^2 Z[\textbf{A}_{\cl},\textbf{A}_{\q}]
	}{
		\delta A^{\mu}_{\cl,\Omega}
		\,
		\delta A^{\nu}_{\q,-\Omega}
	}
	\Bigg\vert_{A=0}.
\end{equation}
In the following, we evaluate $\sigma_{\sfbb,R}^{\mu \nu}(\Omega)$ by expanding
[via Eq.~(\ref{eq:action})]
\begin{equation}
	S_{\chi}[\hat{Q}]	
	=
	-
	\frac{1}{2}
	\Tr
	\log
	\left[
		\hat{G}_{\bdg}^{-1}
		+
		i	
		\hat{Q}
		-
		\BigA \cdot \textbf{v}_F
	\right]
\end{equation}
up to second order in $\BigA$, 
and neglecting fluctuations of $\hat{Q}$ by setting $\hat{Q} = \hat{Q}_{\sfsp}$, as given by Eq.(\ref{eq:Qsp}). 
Then, up to irrelevant constants, $S_{\chi}$ becomes
\begin{equation}
\begin{aligned}
	S_{\chi}[\hat{Q}_{\sfsp}]
	&=
	S_{\chi}^{(1)}[\hat{Q}_{\sfsp}]
	+
	S_{\chi}^{(2)}[\hat{Q}_{\sfsp}]
	+
	{\cal O}(\BigA^3)
\\
	&=
	\frac{1}{2}
	\Tr
	\left[
	\Gdress \BigA \cdot \textbf{v}_F
	\right]
	+
	\frac{1}{4}
	\Tr
	\left[
	\Gdress \BigA \cdot \textbf{v}_F
	\right]^2
	\!\!
	+
	{\cal O}(\BigA^3),
\end{aligned}
\end{equation}
where 
\begin{equation}
	\Gdress^{-1}
\equiv
	\hat{G}_{\bdg}^{-1}
	+
	i
	\hat{Q}_{\mathsf{sp}}.
\end{equation}
By taking partial trace over frequency and momentum, the second order term can be expanded as
\begin{widetext}
\begin{equation}\label{eq:Trlog_expand}
	S_{\chi}^{(2)}[\hat{Q}_{\sfsp}]
	=
	\frac{1}{4}
	\int_{\textbf{k},\textbf{p}}
	\int_{\omega,\omega'}
	\sum_{i,j = \cl,\q}
	\Tr
	\left[	
	\begin{aligned}
	&\,
		\Gdress(\textbf{p},\omega+\omega') 
		\textbf{A}_{i}(\omega,\textbf{k})
		\hat{\gamma}_{i}(\omega+ \omega',\omega')
		\cdot 
		\textbf{v}_F(\textbf{p})
	\\
	\times
	&\,
		\Gdress(\textbf{p},\omega') 
		\textbf{A}_{j}(-\omega,-\textbf{k})
		\hat{\gamma}_{j}(\omega',\omega+\omega')
		\cdot 
		\textbf{v}_F(\textbf{p})
	\end{aligned}
	\right],
\end{equation}
where we have assumed that $\textbf{k}$ is small. Upon taking the derivative with respect to the classical and quantum components of the vector potential, we have
\begin{equation}\label{eq:S_2nd_derivative1}
\frac{
	\delta^2  S_{\chi}^{(2)}[\hat{Q}_{\sfsp}]
}{
	\delta A^{\mu}_{\cl,\Omega}
	\,
	\delta A^{\nu}_{\q,-\Omega}
}
\Bigg\vert_{A=0}
=
2\times 2 \times
\frac{v_F^2}{4}
\int_{\textbf{p},\omega'}
\Tr
\left[
\Gdress(\textbf{p},\Omega+\omega') 
\hat{\gamma}_{\cl}(\Omega+ \omega',\omega')
\Gdress(\textbf{p},\omega') 
\hat{\gamma}_{\q}(\omega',\Omega+\omega')
\right]
\hat{v}_F^{\mu}(\textbf{p})
\hat{v}_F^{\nu}(\textbf{p}),
\end{equation}
where the factor of $2 \times 2$ is due to the two-fold derivative and the trace over the spin degree of freedom.
To simplify the algebra, we introduce the projection operator 
$	
	\hat{P}_{\lambda = \pm}
	\equiv
	\frac{1}{2}
	(1 + \lambda \htau^3)
$
such that
\begin{equation}\label{eq:S_2nd_derivative2}
\begin{aligned}
	\frac{
	\delta^2  S_{\chi}^{(2)}[\hat{Q}_{\sfsp}]
	}{
	\delta A^{\mu}_{\cl,\Omega}
	\,
	\delta A^{\nu}_{\q,-\Omega}
	}
	\Bigg\vert_{A=0}
	=&\,
	v_F^2
	\int_{\textbf{p},\omega'}
	\sum_{\lambda,\lambda' = \pm}
	T_{\lambda \lambda'}
	\,
	\hat{v}_F^{\mu}(\textbf{p})
	\hat{v}_F^{\nu}(\textbf{p}),
\\
	T_{\lambda \lambda'}
	\equiv&\,
	\Tr
	\left[
	\hat{P}_{\lambda}\Gdress(\textbf{p},\Omega+\omega') 
	\hat{\gamma}_{\cl}(\Omega+ \omega',\omega')
	\hat{P}_{\lambda'}
	\Gdress(\textbf{p},\omega') 
	\hat{\gamma}_{\q}(\omega',\Omega+\omega')
	\right].
\end{aligned}
\end{equation}
The trace that appears in the integrand can be evaluated to give
\begin{align}
	T_{++}
	=
	F(\omega_2) \, t_{++},
\qquad
	T_{+-}
	=
	\left[F(\omega_1) - F(\omega_2)\right] 
	\, 
	t_{+-},
\qquad
	T_{-+}
	=
	0,
\qquad
	T_{--}
	=
	-
	F(\omega_1)
	\, 
	t_{--},
\end{align}
where
\begin{align}
	t_{\lambda \lambda'}
	\equiv
	\frac{
	2
	(\varE^2 + \Delta_0^2 \sin^2(\theta) + \varpi_{1,\lambda}\varpi_{2,\lambda'})
	}{
	\left[
		\varpi_{1,\lambda}^2 - \varE^2 - \Delta_0^2\sin^2(\theta)
	\right] 
	\left[
		\varpi_{2,\lambda'}^2 - \varE^2 - \Delta_0^2\sin^2(\theta) 	
	\right]
	}.
\end{align}
Here $\theta$ is the polar angle in $\textbf{p}$ space and
\begin{eqnarray}
	\varpi_{1,\pm}
	=
	\omega_1 
	\pm 
	\frac{i}{2\tel}
	\, 
	\Gamma(\omega_1),
\qquad
	\varpi_{2,\pm} 
	=
	\omega_2 
	\pm 
	\frac{i}{2\tel}
	\, 
	\Gamma(\omega_2).
\end{eqnarray}
We now convert $\int_{\textbf{p}} \rightarrow \nu_0 \int d\varE \int
\frac{d\Omega_{\hat{\textbf{p}}}}{4\pi} $. The $\varE$ integral can be done by the contour method. The results are
\bsub\label{eq:Is}
\begin{eqnarray}
\label{eq:Ipp}
	I_{++}
	&=&
	\int_{\varE} T_{++} 
	=
	\frac{
		2\pi i  
	}{
		\varpi_{1,+} - \varpi_{2,+}
	}
	\left[
	\frac{\varpi_{1,+} \, \sign(\omega_1)}{
		\sqrt{\varpi_{1,+}^2 - \Delta_0^2 \sin^2(\theta)}
	}
	-
	\frac{\varpi_{2,+} \, \sign(\omega_2)}{
		\sqrt{\varpi_{2,+}^2 - \Delta_0^2 \sin^2(\theta)}
	}
	\right]
	F(\omega_2),
\end{eqnarray}
\begin{eqnarray}
	\label{eq:Ipm}
	I_{+-}
	&=&
	\int_{\varE} T_{+-}
	=
	\frac{
		2\pi i 
	}{
		\varpi_{1,+} - \varpi_{2,-}
	}
	\left[
	\frac{\varpi_{1,+} \, \sign(\omega_1)}{
		\sqrt{\varpi_{1,+}^2 - \Delta_0^2 \sin^2(\theta)}
	}
	+
	\frac{\varpi_{2,-} \, \sign(\omega_2)}{
		\sqrt{\varpi_{2,-}^2 - \Delta_0^2 \sin^2(\theta)}
	}
	\right]
		\left[
			F(\omega_1) - F(\omega_2) 
		\right],
\\
\label{eq:Imm}
	I_{--}
	&=&
	\int_{\varE} T_{--}
	=
	\frac{
		2\pi i
	}{
		\varpi_{1,-} - \varpi_{2,-}
	}
	\left[
	\frac{\varpi_{1,-} \, \sign(\omega_1)}{
		\sqrt{\varpi_{1,-}^2 - \Delta_0^2 \sin^2(\theta)}
	}
	-
	\frac{\varpi_{2,-} \, \sign(\omega_2)}{
		\sqrt{\varpi_{2,-}^2 - \Delta_0^2 \sin^2(\theta)}
	}
	\right]
	F(\omega_1). 
\end{eqnarray} 
\esub
Notice that $\frac{
	\delta^2  (S_{\chi}^{(1)}[\hat{Q}_{\sfsp}])^2
}{
	\delta A^{\mu}_{\cl,\Omega}
	\,
	\delta A^{\nu}_{\q,-\Omega}
}
\Big\vert_{A=0} = 0$ and therefore does not contribute to $\sigma_{\sfbb,R}^{\mu \nu}(\Omega)$. 
By combining the results in Eq.~(\ref{eq:Kubo_def}) and Eqs.~(\ref{eq:S_2nd_derivative2})--(\ref{eq:Is}), 
the real part of the optical conductivity can be readily obtained as
\begin{equation}\label{eq:bulk_sigma_exact}
\begin{aligned}
\re \,\sigma_{\sfbb,R}^{\mu \nu}(\Omega)
&=
- \delta^{\mu \nu}\times
\frac{e^2}{i\Omega}
\frac{1}{2i}
\left\lbrace 
v_F^2
\,
\nu_0 
\int_{\omega'}
\int \frac{d\Omega_{\textbf{p}}}{4\pi}
\hat{v}_F^{\mu}(\textbf{p})
\hat{v}_F^{\nu}(\textbf{p})
\,
{\cal S}(\Omega + \omega',\omega',\theta)
\times
2\pi
\left[
F(\Omega + \omega') - F(\omega')
\right]
\right\rbrace,
\end{aligned}
\end{equation}
where function the ${\cal S}$ is defined via
\begin{equation}
	{\cal S}(\omega_1,\omega_2,\theta)
	=
	\im
	\left\{
	\begin{aligned}
	&\,
	\dfrac{1}{\varpi_{1,+} - \varpi_{2,+}
	}
	\left[
	\dfrac{\varpi_{1,+} \, \sign(\omega_1)}{
		\sqrt{\varpi_{1,+}^2 - \Delta_0^2 \sin^2(\theta)}
	}
	-
	\dfrac{\varpi_{2,+} \, \sign(\omega_2)}{
		\sqrt{\varpi_{2,+}^2 - \Delta_0^2 \sin^2(\theta)}
	}
	\right]
	\\
	-
	&\,
	\dfrac{
		1
	}{
		\varpi_{1,+} - \varpi_{2,-}
	}
	\left[
	\dfrac{\varpi_{1,+} \, \sign(\omega_1)}{
		\sqrt{\varpi_{1,+}^2 - \Delta_0^2 \sin^2(\theta)}
	}
	+
	\dfrac{\varpi_{2,-} \, \sign(\omega_2)}{
		\sqrt{\varpi_{2,-}^2 - \Delta_0^2 \sin^2(\theta)}
	}
	\right]
	\end{aligned}
	\right\}.
\end{equation}
\end{widetext}

We can perform the angular integrations exactly to 
finally obtain
\begin{align}\label{eq:bulk_sigma_ang}
\!\!\!\!\!
	\frac{\re \,\sigma_{\sfbb,R}^{\mu \nu}(\Omega)}{\sigma_{\mathsf{dc}}}
	=
	\frac{3}{2 \Omega}
	\int_{-\infty}^{\infty} d\omega	
	&\,
		{\cal J}^{\mu \nu}
		(\Omega+\omega,\omega)
\nonumber\\
	&\,
	\times
	\left[
		F(\Omega + \omega) - F(\omega)
	\right],
\!\!\!
\end{align}
where $\sigma_{\mathsf{dc}} = e^2 (2 \nu_0) D$ is the normal state dc conductivity
and  
$D = v_F^2 \tau_{\mathsf{el}}/3 $ is the diffusion constant.
The kernel in Eq.~(\ref{eq:bulk_sigma_ang}) is given by
\begin{multline}
\label{eq:bulk_sigma_ang_calJ}
	\!\!\!\!
	{\cal J}^{\mu \nu} (\omega_1,\omega_2)
\\
	\,
	=
	\Gel
	\im
	\left\{
	\begin{aligned}
	&\,
		\left[	
		\frac{
			\Phi^{\mu \nu}\left(\frac{\varpi_{1,+}}{\Delta_0}\right) - \Phi^{\mu \nu}\left(\frac{\varpi_{2,+}}{\Delta_0}\right)
		}{
			\varpi_{1,+} - \varpi_{2,+}
		}
		\right]
	\\
		-	
	&\,
		\left[	
		\frac{
			\Phi^{\mu \nu}\left(\frac{\varpi_{1,+}}{\Delta_0}\right) - \Phi^{\mu \nu}\left(\frac{\varpi_{2,-}}{\Delta_0}\right)
		}{
			\varpi_{1,+} - \varpi_{2,-}
		}
		\right]
	\end{aligned}
	\right\},
\end{multline}
where
\bsub\label{eq:disO_sigma_Phi}
\begin{align}
	\Phi^{\mu\mu}
	(\upsilon)
	\equiv&\,
	\left\lbrace 
	\frac{\upsilon}{4}
	\left[
	-\upsilon
	+
	\left(
	1 + \upsilon^2
	\right)
	\coth^{-1}
	\upsilon
	\right]
	\right\rbrace,
\end{align}
for $\mu \in \{x,y\}$,
\begin{align}
	\Phi^{zz}
	(\upsilon)
	\equiv&\,
	\left\lbrace 
	\frac{\upsilon}{2}
	\left[
	\upsilon
	+
	\left(
	1 - \upsilon^2
	\right)
	\coth^{-1}
	\upsilon
	\right]
	\right\rbrace,
\end{align}
\begin{align}
	\Phi^{\mu\nu}(\omega)
	=
	0,
	\quad
	\mu \neq \nu.
\end{align}
\esub
Notice that Eq.~(\ref{eq:disO_sigma_Phi}) implies that the bulk optical conductivity is purely diagonal. 
Eqs.~(\ref{eq:bulk_sigma_ang})--(\ref{eq:disO_sigma_Phi}) are
the main results for the bulk optical conductivity of a weakly disordered $p+ip$ WSC.
Our results are similar to those reported in Ref.~\cite{p_wave_opt_resp_Hirschfeld}, 
despite some differences in the details.
In the metallic limit, $\Delta_0 \rightarrow 0$, $\Gamma(\omega) \rightarrow 1$ and $\Phi^{yy}, \Phi^{zz} \rightarrow 1/3$. As a result, 
$
{\cal J}^{\mu \nu}(\omega_1,\omega_2) \rightarrow \frac{1}{3}\delta^{\mu \nu}
\frac{1}{1 + (\omega_1 - \omega_2)^2 \tau_{\mathsf{el}}^2}
$ and we recover the Drude conductivity
\begin{equation}\label{eq:Drude}
	\re \, \sigma^{\mu \nu}_{\mathsf{Drude}}(\Omega)
	=
	\delta^{\mu \nu}
	\,
	\sigma_{\mathsf{dc}}
	/
	\left[
	1 + (\Omega \tau_{\mathsf{el}})^2\right].
\end{equation}	

In Fig.~\ref{fig:p_wave_sigma_bulk}, we plot the 
zero-temperature
optical conductivity based on 
Eqs.~(\ref{eq:bulk_sigma_ang})--(\ref{eq:disO_sigma_Phi})
with different disorder strengths $\Gel$. 
For weak disorder [Fig.~\ref{fig:p_wave_sigma_bulk}(a)], the behavior of 
$\re \, \sigma_{\sfbb,R}^{\mu \nu}(\Omega)$ 
is similar to the Mattis-Bardeen result for $s$-wave superconductors \cite{s_wave_MB_original}.
This is despite the fact that for the WSC studied here, the response is nonzero even when $\Omega < 2\Delta_0$ due to the 
point-node nature of the bulk pairing. 
For large frequencies ($\Omega \gg \Delta_0$), the pairing gap does not play a role anymore and 
$\re \, \sigma_{\sfbb,R}^{\mu \nu}(\Omega)$ approaches $\re \, \sigma_{\mathsf{Drude}}^{\mu \nu}(\Omega)$.
As the disorder strength becomes stronger, impurities gradually fill the gap and the scattering rate 
$\sim \Gamma(\omega)$ is no longer gapless, as shown in Fig.~\ref{fig:SCBA_p_wave_Gamma}. 
Consequently, the zero-frequency response is no longer vanishing and 
$\re \, \sigma_R^{\mu \nu}(\omega)$ becomes more Drude-like [Fig.~\ref{fig:p_wave_sigma_bulk}(b)--(d)].
For strong disorder, the optical conductivity approaches the Drude result and the $p$-wave features are completely suppressed. 
This makes physical sense because the impurity strength $\Gel$ dominates over the pairing gap $\Delta_0$ in this regime, 
and the system behaves as a dirty ``thermal metal.'' 

The Mattis-Bardeen optical conductivity for dirty $s$-wave superconductors can be efficiently derived within
the same Keldysh framework articulated here, as we show in Appendix \ref{sec:app_Mattis_bardeen}.


\section{Discussion and Conclusion}\label{sec:discussion_conclusion}

To conclude, we studied the optical absorbance in a $p + ip$ Weyl superconductor by examining the contribution from the bulk and surface. 
Fig.~\ref{fig:combine_A} summarizes all of the main results presented in this paper, which were discussed in 
Sec.~\ref{sec:ResultsSum}. 
In the clean limit, we showed that the absorbance is solely contributed by the surface-bulk response. 
The frequency dependence of the absorbance depends on both the coherence length and penetration length.
On the other hand, in the presence of disorder, we demonstrated that the bulk response is nonzero using Keldysh response theory. 
The optical conductivity from the bulk is Mattis-Bardeen-like in the weak disorder limit, 
and becomes more Drude-like as the system becomes dirtier. 
The overall optical absorbance of the system depends on the interplay between the penetration depth, coherence length and disorder strength. 
In the weak disorder limit, we found that the surface-bulk absorbance is more pronounced in the type I superconductor regime.
For the type II regime, the disordered bulk plays a more important role and one would need a relatively clean system to observe the surface-bulk effect.
In both regimes, as long as the disorder strength is weak enough, the characteristic frequency-dependence of the surface-bulk absorbance can provide 
an additional indicator for the surface states.

In WSC candidate materials, there can in general be multiple pairs of Weyl nodes and thus additional chiral Majorana fluid states.
These can wrap around different facets of the sample, depending upon the orientation of the Fermi arcs.  
Additional arcs provide more channels for the surface-bulk transition and amplify the topological anomalous skin effect. 
For radiation with a nonzero incident angle, the surface-bulk absorption is still effective as long as the radiation is
shone upon a facet with surface states. If polarization of the radiation is parallel to the plane of incidence, 
we expect to have maximum absorption at the Brewster angle at which the reflectance is minimized \cite{Jackson}.

When evaluating the total absorbance, we have restricted ourselves to the weak-disorder regime for simplicity.
In general, disorder can influence both the bulk and surface states \cite{rare_WSM_Jed2,Slager2017}. 
On one hand, the chiral Majorana surface states are expected to remain Anderson delocalized along the dispersive
direction, perpendicular to surface Fermi arc. On the other hand, bulk states away from the Weyl points
are converted by arbitrarily weak disorder into weakly multifractal scattering states, giving 
rise to thermal quasiparticle diffusion. 
A simple approach to incorporating disorder into the surface states is to introduce a finite scattering rate $\kappa_{\sfb}$ 
into Eq.~(\ref{eq:broad_delta_fn_bulk}).
However, this does not capture the modification of the spatial profile of the surface wavefunctions (e.g., in along 
the direction of the Fermi arc), due to the disorder.  
A more precise theoretical description for analyzing the consequences of disorder on position-dependent surface-bulk responses 
warrants future investigation.

Unlike $s$-wave superconductors, $p$-wave superconductors are not protected by Anderson's theorem \cite{s_wave_Andersons_thm}. 
As a result, the pairing gap is vulnerable to disorder and superconductivity can be destroyed.
The $p$-wave features of the system are therefore expected to survive only in the weak-disorder limit.

Our work opens a new door for studying responses that originate from the interplay between bulk and surface states in topological materials. 
An important extension of this work would be performing material-specific and first-principle calculations 
in order to make quantitative predictions for the surface-bulk absorption peak in candidate WSC materials,
such as UTe$_2$ 
\cite{WSC_UTe2_Aoki_NMR,
	WSC_UTe2_JF_high_field,
	WSC_UTe2_JF_specific_heat,
	WSC_UTe2_JP1_surface_resistivity,
	WSC_UTe2_JP2_Science,
	WSC_UTe2_JP3_thermal_transport,
	WSC_UTe2_JP4_Kerr_rotation,
	WSC_UTe2_angular_specific_heat_Machida,
	WSC_UTe2_STM_Jiao_Nature,
	WSC_UTe2_JP5_muSR,
	WSC_UTe2_JP6_high_field,
	WSC_UTe2_DFT1_Andriy,
	WSC_UTe2_DFT2_Ishizuka,
	WSC_UTe2_DFT3_Xu,
	WSC_UTe2_Wray_ARPES_DFT,
	WSC_UTe2_Agterberg,
	WSC_UTe2_Yanase}.

\begin{acknowledgments}
	We thank Andriy Nevidomskyy and Sergey Syzranov for helpful discussions.
	T.\ C.\ W.\ and M.\ S.\ F.\ acknowledge support by NSF CAREER Grant No.~DMR-1552327, 
	and by the Welch Foundation Grant No.~C-1809.
	H.\ K.\ P.\ acknowledges support from IRCC, IIT Bombay (RD/0518-IRCCSH0-029).
\end{acknowledgments}


\appendix


\section{Bulk scattering states}\label{sec:app_eigenstates_bulk}

The bulk scattering states are obtained by solving the Schr\"{o}dinger's equation with the Hamiltonian given by 
Eq.~(\ref{eq:model_ham}) in the main text
\begin{equation}
	\hat{h}(-i\partial_x,\textbf{k})
	\,
	\psi^{\sfb}_{\lambda}(q,\textbf{k};x)
	=
	\lambda 
	\,
	E^{\sfb}_{q,\textbf{k}}
	\,
	\psi^{\sfb}_{\lambda}(q,\textbf{k};x),
\end{equation}
where $\lambda = \pm 1$, $\textbf{k} = (k_y,k_z)$,
$q \geq 0$ is a standing-wave momentum in the $x$-direction, 
and the bulk eigenenergy is
\begin{equation}
	E^{\sfb}_{q,\textbf{k}}
	=
	\sqrt{
	\varE_{q,\textbf{k}}
	+
	\Delta^2 (q^2 + k_y^2)
	},
	\quad
	\varE_{q,\textbf{k}}
	=
	\frac{q^2 + \textbf{k}^2 - k_F^2}{2m}.
\end{equation}
By imposing the boundary condition that
\begin{equation}
\psi^{\sfb}_{\lambda}(q,\textbf{k};x = 0)
=
0
\end{equation}
and 
requiring $\psi^{\sfb}_{\lambda}(q,\textbf{k};x \rightarrow \infty)$ to be finite, we obtain the bulk states in the following two cases:
\vspace{1mm}
\\
(i) $k_F^2 - k_y^2 - k_z^2 - 2m^2 \Delta^2 > 0$
\vspace{1mm}
\\
As discussed in Sec.~\ref{subsec:model}, we have to consider degeneracy in this case. For $\qmin < q \le q_0$, we choose the following
positive-energy orthonormal states

\begin{eqnarray}
\Psi^{\sfb (1)}_{\lambda = +1} (q,\textbf{k};x)
&=&
\frac{
\psi^{\sfb<}(q,\textbf{k};x)
+
i
\phi^{\sfb<}(q,\textbf{k};x)
}{
	\sqrt{{\cal N}^{\sfb (1)}_{q,\textbf{k}}}
}
,
\qquad
\\
\Psi^{\sfb(2)}_{\lambda = +1} (q,\textbf{k};x)
&=&
\frac{
	\psi^{\sfb<}(q,\textbf{k};x)
	-
	i
	\phi^{\sfb<}(q,\textbf{k};x)
}{
	\sqrt{{\cal N}^{\sfb (2)}_{q,\textbf{k}}}
},
\end{eqnarray}
where the normalization constants are 
\begin{eqnarray}
\frac{1}{
	{\cal N}^{\sfb (1)}_{q,\textbf{k}}
}
&=&
\frac{1}{2 -  2\frak{a}(q,\textbf{k})},
\\
\frac{1}{
{\cal N}^{\sfb (2)}_{q}
}
&=&
\frac{1}{2 +  2\frak{a}(q,\textbf{k})},
\end{eqnarray}
and the function $\frak{a}(q,\textbf{k})$ will be defined below. 
$\psi^{\sfb<}(q,\textbf{k};x)$ and $\phi^{\sfb<}(q,\textbf{k};x)$ are two non-orthogonal but independent positive-energy 
bulk states given by
\begin{equation}
\begin{aligned}
&\psi^{\sfb<}(q;x)
=
\\
&\frac{1}{\sqrt{N^{\sfb}_{q,\textbf{k}}}}
\begin{Bmatrix}
\begin{bmatrix}
f_1(q,\textbf{k})
\\
f_2(q,\textbf{k})
\end{bmatrix}
e^{iqx}
+
\begin{bmatrix}
f_1^*(q,\textbf{k})
\\
-f_2(q,\textbf{k})
\end{bmatrix}
e^{-iqx}
\\
+
\begin{bmatrix}
-f_1(q_{-},\textbf{k})
\\
-f_2(q_{-},\textbf{k})
\end{bmatrix}
e^{i q_{-} x}
+
\begin{bmatrix}
-f_1^*(q_{-},\textbf{k})
\\
f_2(q_{-},\textbf{k})
\end{bmatrix}
e^{-i q_{-} x}
\end{Bmatrix}
\end{aligned}
\end{equation}
and
\begin{equation}
\begin{aligned}
&\phi^{\sfb<}(q;x)
=
\\
&
\frac{1}{\sqrt{M^{\sfb}_{q,\textbf{k}}}}
\begin{Bmatrix}
\begin{bmatrix}
-g_2(q,\textbf{k})
\\
f_1^*(q,\textbf{k})
\end{bmatrix}
e^{iqx}
+
\begin{bmatrix}
g_2(q,\textbf{k})
\\
f_1(q,\textbf{k})
\end{bmatrix}
e^{-iqx}
\\
+
\begin{bmatrix}
g_2(q_-,\textbf{k})
\\
-f_1^*(q_-,\textbf{k})
\end{bmatrix}
e^{i q_- x}
+
\begin{bmatrix}
-g_2(q_-,\textbf{k})
\\
-f_1(q_-,\textbf{k})
\end{bmatrix}
e^{-i q_- x}
\end{Bmatrix}\!.
\end{aligned}
\end{equation}
Here, we have defined the following functions
\begin{align}
	f_1(q,\textbf{k})
	=&\,
	\frac{(ik_y - q) m\Delta}{q} ,
\\
	f_2 (q,\textbf{k})
	=&\,
	\frac{m(\varE_{q,\textbf{k}} - E^{\sfb}_{q,\textbf{k}})}{q},
\\
	g_2 (q,\textbf{k})
	=&\,
	\frac{m(\varE_{q,\textbf{k}} + E^{\sfb}_{q,\textbf{k}})}{q},
\\
	q_{-}(q,\textbf{k}) 
	=&\,
	\sqrt{2k_F^2 - 2k_y^2 - 2k_z^2 - q^2 - 4m^2\Delta^2},
\end{align}
and the normalization constants are
\begin{equation}
	\frac{1}{N^{\sfb}_{q,\textbf{k}}}
	=
	\left[
	\begin{aligned}
	&\,
		|f_1 (q,\textbf{k})|^2 
		+ 
		f_2(q,\textbf{k})^2 
	\\	
		+
	&\,
		\dfrac{q_{-}}{q}
		(
		|f_1(q_{-},\textbf{k})|^2 
		+ 
		f_2 (q_{-},\textbf{k})^2
		)
	\end{aligned}
	\right]^{-1},
\end{equation}
\begin{equation}
	\frac{1}{M^{\sfb}_{q,\textbf{k}}}
	=
	\left[
	\begin{aligned}
	&\,
		|f_1 (q,\textbf{k})|^2 
		+ 
		g_2(q,\textbf{k})^2 
	\\
		+ 
	&\,
		\dfrac{q_{-}}{q}
		(
		|f_1(q_{-},\textbf{k})|^2 
		+ 
		g_2 (q_{-},\textbf{k})^2
		) 
	\end{aligned}
	\right]^{-1}.
\end{equation}

$\psi^{\sfb<}(q,\textbf{k};x)$ and $\phi^{\sfb<}(q,\textbf{k};x)$ satisfy the following properties
\begin{align}
	\int_0^{\infty}
	dx 
	\,
	\psi^{\sfb< \dagger}(q',\textbf{k};x)
	\,
	\psi^{\sfb<}(q,\textbf{k};x)
	=&\,
	2\pi \delta(q - q'),
\\
	\int_0^{\infty}
	dx 
	\,
	\phi^{\sfb< \dagger}(q',\textbf{k};x)
	\,
	\phi^{\sfb<}(q,\textbf{k};x)
	=&\,
	2\pi \delta(q - q'),
\end{align}
and have nonzero overlap
\begin{equation}
	\int_0^{\infty}
	dx 
	\,
	\psi^{\sfb< \dagger}(q',\textbf{k};x)
	\,
	\phi^{\sfb<}(q,\textbf{k};x)
	=
	i \frak{a}(q) \, 2\pi \delta(q - q'),
\end{equation}
where
\begin{equation}
\frak{a}(q,\textbf{k})
=
\frac{
	\im f_1(q,\textbf{k})
}
{
	\sqrt{
		N^{\sfb}_{q,\textbf{k}} M^{\sfb}_{q,\textbf{k}}
	}
}
\left[
g_2(q,\textbf{k}) 
-
f_2(q,\textbf{k}) 
\right]
\left[
1
+
\frac{q}{q_{-}}
\right].
\end{equation}
The states $\Psi^{\sfb(1) \dagger}_{\lambda}$ and $\Psi^{\sfb(2) \dagger}_{\lambda }$ are constructed in such a way that they are orthonormal according to
\begin{equation}
\int_0^{\infty} dx \,
\Psi^{\sfb(i) \dagger}_{\lambda} (q',\textbf{k};x)
\Psi^{\sfb (j)}_{\lambda} (q,\textbf{k};x)
=
2\pi \delta(q - q') \delta_{ij}.
\end{equation}

For $q > q_0$, there is no degeneracy. The bulk state in this case is
\begin{equation}
\label{eq:app_bulk_state_greater}
\begin{aligned}
\psi_{ +1}^{\sfb>}(q,\textbf{k};x)
&=
\frac{1}{
	\sqrt{N^{\sfb>}_{q,\textbf{k}}}
}
\begin{Bmatrix}
\frakc_{+,q,\textbf{k}}
\begin{bmatrix}
u_{q,\textbf{k}}
\\
v_{q,\textbf{k}}  e^{i\phi_{q,\textbf{k}}}
\end{bmatrix}
e^{i qx}
\\\\
+
\frakc_{-,q,\textbf{k}}
\begin{bmatrix}
u_{q,\textbf{k}}
\\
-v_{q,\textbf{k}} e^{-i\phi_{q,\textbf{k}}}
\end{bmatrix}
e^{-i qx}
\\\\
-
\begin{bmatrix}
\Theta_{q,\textbf{k}}
\\
\Phi_{q,\textbf{k}}
\end{bmatrix}
e^{-\lambda_{q,\textbf{k}} x}
\end{Bmatrix}\!,
\end{aligned}
\end{equation}
where
\begin{align}
	u_{q,\textbf{k}}
	=&\,
\sqrt{
	\frac{1}{2}
	\left(
	1 + \frac{\varE_{q,\textbf{k}}}{
		E_{q,\textbf{k}}^{\sfb}
	}
	\right)
},
\\
	v_{q,\textbf{k}}
	=&\,
\sqrt{
	\frac{1}{2}
	\left(
	1 - \frac{\varE_{q,\textbf{k}}}{
		E_{q,\textbf{k}}^{\sfb}
	}
	\right)
},
\\
	\lambda_{q,\textbf{k}}
	=&\,
\sqrt{
	q^2 + 2k_y^2 + 2k_z^2  - 2k_F^2 + 4m^2 \Delta^2 
},
\\
	\phi_{q,\textbf{k}}
	=&\,
	\arg(q + ik_y),
\\
	\frakc_{\pm,q,\textbf{k}}
	=&\,
	\frac{m}{2}
	\left[
	\begin{aligned}
	&\,
		\dfrac{
			\Delta(\lambda_{q,\textbf{k}} - k_y) e^{\mp i\phi_{q,\textbf{k}}}
		}{
			u_{q,\textbf{k}}
		}
	\\
		\mp
	&\,
		i
		\dfrac{
			\varE_{q,\textbf{k}} + E_{q,\textbf{k}}^{\sfb} + 2 m\Delta^2
		}{
			v_{q,\textbf{k}}
		}
	\end{aligned}
	\right],
\\
	\Theta_{q,\textbf{k}}
	=&\,
	\left(\frakc_{+,q,\textbf{k}} + \frakc_{-,q,\textbf{k}}\right) u_{q,\textbf{k}},
\\
	\Phi_{q,\textbf{k}}
	=&\,
	\left(
		-
		\frakc_{-,q,\textbf{k}}e^{-i\phi_{q,\textbf{k}}}
		+
		\frakc_{+,q,\textbf{k}}e^{i\phi_{q,\textbf{k}}}
	\right) 	
	v_{q,\textbf{k}},
\end{align}
and
\begin{equation}
\frac{1}{
	{N^{\sfb>}_{q,\textbf{k}}}
}
=
\frac{2}{
	|\frakc_{+,q,\textbf{k}}|^2
	+
	|\frakc_{-,q,\textbf{k}}|^2
}
\end{equation}
is the normalization constant. 

In all the cases, the negative energy solution is related to the positive one via particle-hole symmetry
\begin{equation}\label{eq:app_eigenstate_bulk_neg}
\psi_{-1}^{\sfb}(q,\textbf{k};x)
=
\hat{\sigma}^1 
\left[
\psi_{+1}^{\sfb}(q,-\textbf{k};x)
\right]^*.
\end{equation}
(ii) $k_F^2 - k_y^2 - k_z^2 - 2m^2 \Delta^2 \le 0$
\vspace{1mm}
\\
In this regime, there is no degeneracy for all $q >0$. 
The bulk states are simply given by Eq.~(\ref{eq:app_bulk_state_greater}).


\section{Meissner effect and temperature dependence of the penetration depth}\label{sec:app_Meissner}

In this Appendix, we corroborate the discussion of the Meissner effect in WSCs in Sec.~\ref{sec:intro_Mesiser} 
by considering the 
spinless WSC model defined by Eq.~(\ref{eq:hamiltonian}).
We compute the temperature $T$-dependence of the correction to the London penetration depth due to the presence of surface states.
Since the $T$-dependence of the penetration depth originates from the paramagnetic current-current correlation function, 
it is sufficient to study the $T$-dependence of the latter \cite{He3B_Meissner}.


\subsection{Temperature dependence due to the bulk response}

The $yy$ component of the bulk paramagnetic current-current correlation function is given by
\begin{equation}
\begin{aligned}
&\Pi_{1,\sfbb}^{yy}(i\Omega_n = 0, \textbf{q}=0)
\\
&=
-
\frac{1}{2}
\left(
\frac{e}{m}
\right)^2
T\sum_{\omega_n}
\int_\textbf{k}
\mathsf{Tr}
\bigg[
k_y 
\hG_{\sfb}(i\omega_n,\textbf{k})
k_y
\hG_{\sfb}(i\omega_n,\textbf{k})
\bigg]
\\
&=
\frac{1}{2}
\left(
\frac{e}{m}
\right)^2
\frac{\beta}{2}
\int_\textbf{k}
k_y^2
\,
\text{sech}^2
\left(
\frac{
\beta E^{\sfb}_{\textbf{k}}
}
{2}
\right),
\end{aligned}
\end{equation}
where $\int_{\textbf{k}} = \int\frac{d^3 k}{(2\pi)^3}$, $\beta = 1/T$ 
and 
$ E^{\sfb}_{\textbf{k}} = \sqrt{\varE_{\textbf{k}}^2 + \Delta^2 (k_x^2 + k_y^2)}$.
At low temperature, the dominate contribution comes from the Weyl nodes. 
We can thus linearize $E^{\sfb}_{\textbf{k}}$ 
[cf.\ Eq.~(\ref{eq:hamiltonian_Weylpoint})] 
and rescale the coordinates to remove the velocity anisotropy such that
\begin{equation}
\begin{aligned}
&
\Pi_{1,\sfbb}^{yy}(i\Omega_n = 0, \textbf{q}=0)
\\
&=
\frac{e^2}{2m^2} 
\frac{\beta}{2}
\frac{1}{\Delta^4  (k_F/m)}
\int_{\delta  \tilde{\textbf{k}}}
\delta \tilde{k}_y^2
\,
\text{sech}^2
\left(
\frac{
	\beta \delta \tilde{k}
}
{2}
\right)
\times 2,
\end{aligned}
\end{equation}
where the factor of 2 accounts for the two Weyl nodes and 
$\delta \tilde{\textbf{k}}$ is rescaled by the velocity. 
Performing the integral and analytical continuation, we obtain the $yy$ component retarded current-current correlation function
\begin{equation}
\label{eq:app_Pi_yy_bulk00}
	\Pi_{\mathsf{bb},1,R}^{yy}( 0, \textbf{0})
	=
	-\frac{e^2}{2m} 
	\frac{7\pi^2}{45}
	\left(\frac{1}{\Delta^4 k_F \beta^4}\right)
	\sim 
	T^4.
\end{equation}
Similarly, we can show that the $zz$ component is
\begin{equation}
\label{eq:app_Pi_zz_bulk00}
	\Pi_{1,\sfbb,R}^{zz}(0, \textbf{0})
	=
	-
	\frac{e^2}{2m} 
	\left(\frac{4k_F}{\pi^2 \beta^2 \Delta^2}\right)
	\sim 
	T^2.
\end{equation}


\subsection{Temperature dependence due to the surface response}

We now turn to the paramagnetic current-current correlation function due to the surface states. 
The $yy$ component is given by
\begin{align}
	\!\!\!\!
	&\,
	\Pi_{1,\sfss}^{yy}(i\Omega_n = 0, \textbf{q}=0;x,x')
\nonumber\\
	&\,\,
	=
	-\frac{T}{2}
	\left(
	\frac{e}{m}
	\right)^2
\nonumber\\
	&\,\,
	\,\,\phantom{=}
	\times
	\sum_{\omega_n}
	\int_\textbf{k}
	\mathsf{Tr}
	\left[
	k_y
	\hG_{\sfs}(i\omega_n,\textbf{k};x,x')
	k_y
	\hG_{\sfs}(i\omega_n,\textbf{k};x',x)
	\right]
\nonumber\\
	&\,\,
	\simeq
	\frac{1}{2}
	\left(
	\frac{e}{m}
	\right)^2
	\frac{\beta}{4}
	\int_\textbf{k}
	k_y^2
	\,
	\text{sech}^2
	\left(
	\frac{
		\beta \Delta k_y
	}
	{2}
	\right)
	\Sigma^{\sfss}(\textbf{k};x,x'),
	\!\!\!
\end{align}
where $\textbf{k} = (k_y,k_z)$ and
\begin{equation}
	\Sigma^{\sfss}(\textbf{k};x,x')
	=
	\left|\psi^{\sfs}(\textbf{k};x)\right|^2
	\left|\psi^{\sfs}(\textbf{k};x')\right|^2.
\end{equation}
The surface state $\psi^{\sfs}(\textbf{k};x)$ is given by Eq.~(\ref{eq:eigenstate_surf}).
In the $T \rightarrow 0$ limit, we can approximate the retarded correlation function as
\begin{align}
	&\,
	\Pi_{1,\sfss,R}^{yy}( 0, \textbf{0};x,x')
\nonumber\\
	&\,\,
	\simeq
	-
	\frac{1}{2}
	\left(
		\frac{e}{m}
	\right)^2
	\frac{\beta}{4}
	\int_{-k_F}^{k_F} 
	\frac{dk_z}{2\pi}
	\int_{-\sqrt{k_F^2 - k_z^2}}^{\sqrt{k_F^2 - k_z^2}}
	\frac{dk_y}{2\pi}
	k_y^2
\nonumber\\
	&\,\,
	\qquad
	\times
	\sech^2
	\left(
		\frac{\beta \Delta k_y}{2}
	\right)
	\Sigma^{\sfss}(k_y = 0,k_z;x,x')
\nonumber\\
	&\,\,
	\simeq
	-
	\left(
	\frac{e}{m}
	\right)^2
	\frac{\beta}{2(2\pi)^2}
	\int_0^{\infty} dk_y
	\,
	k_y^2
	\sech^2
	\left(
	\frac{\beta \Delta k_y}{2}
	\right)
\nonumber\\
	&\,\,
	\qquad
	\times
	\int_0^{k_F}
	dk_z
	\,
	\Sigma^{\sfss}(k_y = 0,k_z;x,x')
\nonumber\\
	&\,\,
	=
	-
	\left(
	\frac{e}{m}
	\right)^2
	\frac{1}{12 \beta^2 \Delta^3}
	\int_0^{k_F}
	dk_z
	\,
	\Sigma^{\sfss}(k_y = 0,k_z;x,x')
\nonumber\\
	&\,\,
	\sim
	T^2.
\end{align}
The $zz$ component can be evaluated in a similar manner, but one has to be more careful.
We write 
\begin{align}\label{eq:app_Pizz_ss_00}
	\Pi_{1,\sfss,R}^{zz}( 0, \textbf{0};x,x')
	\simeq
	S_1 + S_2, 
\end{align}
where
\begin{equation}
\begin{aligned}[b]
	S_1
	\equiv&\,
	-
	\frac{1}{2}
	\left(
	\frac{e}{m}
	\right)^2
	\frac{\beta}{4}
	\int_{-k_F}^{k_F} 
	\frac{dk_z}{2\pi}
	\int_{-\sqrt{k_F^2 - k_z^2}}^{\sqrt{k_F^2 - k_z^2}}
	\frac{dk_y}{2\pi}
\\
	&\,
	\times
	k_z^2
	\,
	\sech^2
	\left(
	\frac{\beta \Delta k_y}{2}
	\right)
	\Sigma^{\sfss}(k_y = 0,k_z;x,x')
\end{aligned}
\end{equation}
and
\begin{equation}
\begin{aligned}[b]
	S_2
	\equiv&\,
	-
	\frac{1}{2}
	\left(
	\frac{e}{m}
	\right)^2
	\frac{\beta}{4}
	\int_{-k_F}^{k_F} 
	\frac{dk_z}{2\pi}
	\int_{-\sqrt{k_F^2 - k_z^2}}^{\sqrt{k_F^2 - k_z^2}}
	\frac{dk_y}{2\pi}
\\
	&\,
	\times
	k_z^2
	\,
	\sech^2
	\left(
	\frac{\beta \Delta k_y}{2}
	\right)
	\left(
	\frac{1}{2}
	\frac{\partial^2  \Sigma^{\sfss}}{\partial k_y^2}
	\bigg\rvert_{k_y = 0}
	k_y^2
	\right).
\end{aligned}
\end{equation}
First, consider
\begin{equation}
\begin{aligned}
	S_1
	&=
	-
	\left(
	\frac{e}{m}
	\right)^2
	\frac{\beta}{(2)(4)(2\pi)^2}
	\frac{2(4)}{\beta \Delta}
	\int_0^{k_F} dk_z
	\,
	k_z^2
\\
&\qquad
	\times
	\tanh
	\left(
	\frac{\beta \Delta}{2}
	\sqrt{k_F^2 - k_z^2}
	\right)
	\Sigma^{\sfss}(k_y = 0,k_z;x,x')
\\
&
\simeq
	-
	\left(
	\frac{e}{m}
	\right)^2
	\frac{1}{4\pi^2 \Delta}
	\int_0^{k_F} dk_z
	\,
	k_z^2
	\,
	\left(
	1 - 
	2e^{
		-\beta \Delta
		\sqrt{k_F^2 - k_z^2}
	}
	\right)
\\
&\qquad
	\times
	\Sigma^{\sfss}(k_y = 0,k_z;x,x').
\end{aligned}
\end{equation}
The first term just gives us a $T$-independent constant. 
One can show that the second term gives a strongly subleading 
$T^6$ dependence.
Meanwhile, for $T \rightarrow 0$,
\begin{align}
	\!\!\!\!
	S_2
	\simeq&\,
	-
	\frac{1}{2}
	\left(
	\frac{e}{m}
	\right)^2
	\frac{\beta}{2(2\pi)^2}
	\int_{0}^{k_F} 
	dk_z
	\,
	k_z^2
	\,
	\frac{\partial^2  \Sigma^{\sfss}}{\partial k_y^2}
	\bigg\rvert_{k_y = 0}
\nonumber\\
	&\,
	\times
	\int_{0}^{\infty}
	dk_y
	\,
	\sech^2
	\left(
	\frac{\beta \Delta k_y}{2}
	\right)
	k_y^2
\nonumber\\
	=&\,
	-
	\frac{1}{24}
	\left(
	\frac{e}{m}
	\right)^2
	\frac{1}{\beta^2 \Delta^3}
	\int_{0}^{k_F} 
	dk_z
	\,
	k_z^2
	\,
	\frac{\partial^2  \Sigma^{\sfss}}{\partial k_y^2}
	\bigg\rvert_{k_y = 0}
\nonumber\\
	\sim&\,
	T^2.
\end{align}
Hence, for $T \rightarrow 0$,
\begin{equation}
\Pi_{1,\sfss,R}^{zz}( 0, \textbf{0};x,x')
=
S_1 + S_2
\sim 
T^2,
\end{equation}
meaning that the $zz$ component of the surface current-current correlation function only 
renormalizes the coefficient of $T^2$ power law from the bulk [cf.\ Eq.~(\ref{eq:app_Pi_zz_bulk00})].


\subsection{Temperature dependence due to the surface-bulk cross terms}

We can also study the effects of the surface-bulk term by computing $\Pi_{1,\sfsb,R}^{\mu \nu}( 0, \textbf{0};x,x')$ 
using the exact bulk and surface states (as we did for the topological anomalous skin effect
computed in Sec.~\ref{sec:TASE}),
but the exact expression is unwieldy so we omit it here.
By focusing on the $T \rightarrow 0$ limit, we find that 
\begin{eqnarray}
	\Pi_{1,\sfsb,R}^{yy}( 0, \textbf{0};x,x')
	&\sim&	
	T^4,
\\
	\Pi_{1,\sfsb,R}^{zz}( 0, \textbf{0};x,x')
	&\sim&
	T^2,
\end{eqnarray}
i.e.\ the surface-bulk cross-terms take the same $T$-dependence as the bulk terms 
[Eq.~(\ref{Meissner:BB})]
and thus do not give new power-laws in $T$.


\section{The kernel $\hat{J}_{q,\textbf{k}}$ in the surface-bulk optical conductivity}\label{sec:app_double_overlap}

The integral appeared in Eq.~(\ref{eq:kernel_Jhat}) of the main text 
\begin{equation}
	\hat{J}^{\sfsb}_{\hat{q},\hat{\textbf{k}}}
	=
	k_F^2 
	\lcoh \int_{x,x' > 0}
	e^{-{\cal K} x}
	\,
	{\Sigma}^{\sfsb}_{+1}(q,\textbf{k};x',x)
	\,
	e^{-{\cal K} x'}
\end{equation}
can be evaluated exactly. 
Due to the possible degeneracy in the bulk band discussed in Sec.~\ref{subsec:model}, we consider the following two cases:
\vspace{1mm}
\\
\begin{widetext}
(i) $k_F^2 - k_y^2 - k_z^2 - 2m^2 \Delta^2 > 0$
\begin{equation}
\begin{aligned}
	\hat{J}^{\sfsb}_{\hat{q},\hat{\textbf{k}}}
	=&\,
	2\left(1 - \hat{\textbf{k}}^2\right)
	\,
	\theta\left(1 - \hat{\textbf{k}}^2\right)
	\,
	\times
	\,
	\begin{cases}
	0
	,
	\qquad
	&
	0 < \hat{q} < \hat{q}_{\mathsf{min}},
\\\\
	\nsum_{i = 1}^2
	\left|
	\nsum_{j = 1}^4
	\frak{j}_{<}^{ (i,j)}(\hat{q},\hat{\textbf{k}})
	\right|^2
	,
	\qquad
	&
	\hat{q}_{\mathsf{min}} < \hat{q} \le \hat{q}_0,
\\\\
	\left|
	\nsum_{j = 1}^3
	\frak{j}_{>}^{ (j)}(\hat{q},\hat{\textbf{k}})
	\right|^2
	,
	\qquad
	&
	\hat{q}_0 < \hat{q},
	\end{cases}
\end{aligned}
\end{equation}
where
\begin{equation}
\hat{k}^{\mu} 
=
k^{\mu} /k_F
,
\qquad
\hat{q}
=
q / k_F,
\end{equation}
\begin{equation}
	\frak{j}_{<}^{ (1,1)}(\hat{q},\hat{\textbf{k}})
	=
\frac{
R(\hat{q},\hat{\textbf{k}})
}{
	\sqrt{
		{\cal N}^{\sfb (1)}_{q,\textbf{k}}
	}
}
\left\{
\dfrac{1}{
	\sqrt{
		N^{\sfb}_{q,\textbf{k}}
	}
}
\left[
	i f_1^*(q,\textbf{k}) + f_2(q,\textbf{k})
\right]
+
\dfrac{
	i
}{
\sqrt{
	M^{\sfb}_{q,\textbf{k}}
}
}
\left[
	-i g_2(q,\textbf{k}) + f_1^*(q,\textbf{k})
\right]
\right\},
\end{equation}
\begin{equation}
	\frak{j}_{<}^{ (1,3)}(\hat{q},\hat{\textbf{k}})
	=
\frac{
	R(\hat{q}_{-},\hat{\textbf{k}})
}{
	\sqrt{
		{\cal N}^{\sfb (1)}_{q,\textbf{k}}
	}
}
\left\{
\dfrac{1}{
	\sqrt{
		N^{\sfb}_{q,\textbf{k}}
	}
}
\left[
	-i f_1^*(q_{-},\textbf{k}) - f_2(q_{-},\textbf{k})
\right]
+
\dfrac{i}{
	\sqrt{
		M^{\sfb}_{q,\textbf{k}}
	}
}
\left[
	i g_2(q_{-},\textbf{k}) - f_1(q_{-},\textbf{k})
\right]
\right\},
\end{equation}
\begin{equation}
	\frak{j}_{<}^{ (2,1)}(\hat{q},\hat{\textbf{k}})
	=
\frac{
	R(\hat{q},\hat{\textbf{k}})
}{
	\sqrt{
		{\cal N}^{\sfb (2)}_{q,\textbf{k}}
	}
}
\left\{
\dfrac{1}{
	\sqrt{
		N^{\sfb}_{q,\textbf{k}}
	}
}
\left[
i f_1^*(q,\textbf{k}) + f_2(q,\textbf{k})
\right]
-
\dfrac{i}{
	\sqrt{
		M^{\sfb}_{q,\textbf{k}}
	}
}
\left[	
	-i g_2(q,\textbf{k}) + f_1^*(q,\textbf{k})
\right]
\right\},
\end{equation}
\begin{equation}
	\frak{j}_{<}^{ (2,3)}(\hat{q},\hat{\textbf{k}})
	=
\frac{
	R(\hat{q}_{-},\hat{\textbf{k}})
}{
	\sqrt{
		{\cal N}^{\sfb (2)}_{q,\textbf{k}}
	}
}
\left\{
\dfrac{1}{
	\sqrt{
		N^{\sfb}_{q,\textbf{k}}
	}
}
\left[
	-i f_1^*(q_{-},\textbf{k}) - f_2(q_{-},\textbf{k})
\right]
-
\dfrac{i}{
	\sqrt{
		M^{\sfb}_{q,\textbf{k}}
	}
}
\left[
	i g_2(q_{-},\textbf{k}) - f_1(q_{-},\textbf{k})
\right]
\right\},
\end{equation}
\end{widetext}
and where 
\begin{align}
	\frak{j}_{<}^{ (1,2)}(\hat{q},\hat{\textbf{k}})
	=&\,
	-\frak{j}_{<}^{ (1,1)*}(\hat{q},\hat{\textbf{k}}),
\\
	\frak{j}_{<}^{ (1,4)}(\hat{q},\hat{\textbf{k}})
	=&\,
	-\frak{j}_{<}^{ (1,3)*}(\hat{q},\hat{\textbf{k}}),
\\
	\frak{j}_{<}^{ (2,2)}(\hat{q},\hat{\textbf{k}})
	=&\,
	-\frak{j}_{<}^{ (2,1)*}(\hat{q},\hat{\textbf{k}}),
\\		
	\frak{j}_{<}^{ (2,4)}(\hat{q},\hat{\textbf{k}})
	=&\,
	-\frak{j}_{<}^{ (2,3)*}(\hat{q},\hat{\textbf{k}}).
\end{align}
We also define
\begin{equation}
R(\hat{q},\hat{\textbf{k}})
=
\dfrac{
	1
}{
	\left[ 
		i\hat{q}  +  (k_F\lcoh)^{-1} + \hat{\cal K}
	\right]^2 
	-
	\hat{\kappa}_{\hat{\textbf{k}}}^2
}
\end{equation}
where
\begin{equation}
	\hat{\kappa}_{\hat{\textbf{k}}}	
	=
	\kappa_{\textbf{k}}/k_F,
	\quad
	\hat{\cal K}	
	=
	{\cal K}/k_F,
\end{equation}
(ii) $k_F^2 - k_y^2 - k_z^2 - 2m^2 \Delta^2 \le 0$
\begin{equation}
	\hat{J}^{\sfsb}_{\hat{q},\hat{\textbf{k}}}
	=
	2\left(1 - \hat{\textbf{k}}^2\right)
	\,
	\theta\left(1 - \hat{\textbf{k}}^2\right)
	\,
	\left|
	\nsum_{j = 1}^3
	\frak{j}_{>}^{ (j)}(\hat{q},\hat{\textbf{k}})
	\right|^2,
\end{equation}
where
\begin{align}
	\frak{j}^{(1)}_>(\hat{q},\hat{\textbf{k}})
	=&\,
	\frac{
	R(-\hat{q},\hat{\textbf{k}})
	}{
	\sqrt{
		N_{q,\textbf{k}}^{\sfb>}
	}
	}
	\frakc_{+,q,\textbf{k}} 
	\left(
	u_{q,\textbf{k}}
	+
	i v_{q,\textbf{k}}e^{i\phi_{q,\textbf{k}}}
	\right), 
\\
	\frak{j}^{(2)}_>(\hat{q},\hat{\textbf{k}})
	=&\,
	\frac{
	R(\hat{q},\hat{\textbf{k}})
	}{
	\sqrt{
		N_{q,\textbf{k}}^{\sfb>}
	}
	}
	\frakc_{-,q,\textbf{k}} 
	\left(
	u_{q,\textbf{k}} 
	-
	i  v_{q,\textbf{k}}e^{-i\phi_{q,\textbf{k}}}
	\right),
\\
	\frak{j}^{(3)}_>(\hat{q},\hat{\textbf{k}})
	=&\,
	\frac{
	R(-i\hat{\lambda}_{\hat{q},\hat{\textbf{k}}},\hat{\textbf{k}})
	}{
	\sqrt{
		N_{q,\textbf{k}}^{\sfb>}
	}
	}
	\left(
	\Theta_{q,\textbf{k}}
	+
	i \Phi_{q,\textbf{k}}
	\right),
\end{align}
where
\begin{equation}
\hat{\lambda}_{\hat{q},\hat{\textbf{k}}}
=
\lambda_{q,\textbf{k}}/k_F.
\end{equation}


\begin{widetext}

\section{Mattis-Bardeen formula in dirty $s$-wave superconductors}
\label{sec:app_Mattis_bardeen}

In this Appendix, we compute the $s$-wave superconductor saddle point $\QspSwave$
and rederive the Mattis-Bardeen optical conductivity in the dirty limit \cite{s_wave_MB_original},
using the Keldysh formalism.

One can repeat the derivation presented in Sec.~\ref{subsec:Keldysh_formalism} by introducing BCS singlet pairing interactions \cite{s_wave_NLSM5_Yunxiang}. 
At the static mean-field level, the saddle point equation 
is still given by Eq.~(\ref{eq:Q_saddle_pt_eqn}), 
but now with  
$\hat{h}_{\bdg} = \varE_{\textbf{k}} \, \hsig^3 + \Delta_0 \, \hsig^1$, 
where 
$\Delta_0$ is the $s$-wave BCS gap. 
An important observation is that in this case, 
the denominator of Eq.~(\ref{eq:Q_saddle_pt_eqn}) can be diagonalized via
\cite{s_wave_NLSM1_Larkin,s_wave_NLSM2_Lerner}
\begin{equation}\label{eq:DiagU}
	\hat{U}_{\omega}^{-1}
	\left[
		\hsig^3 \hat{\omega}
		+
		i\eta \htau^3 \hsig^3
		+
		\hsig^3 
		\left(
		\Delta_0 \hsig^1
		\right)
	\right]
	\hat{U}_{\omega}
	=
	\hat{\Xi}_{\omega},
\end{equation}
with $\hat{\Xi}_{\omega}$ a momentum-independent diagonal matrix.
As a result, the Green's function can be diagonalized by $\hat{U}_{\omega}$ as well,
\begin{equation}\label{eq:app_G_swave_trans}
\begin{aligned}
	\Gdress^{\swave}
	=
	\frac{1}{
	\hsig^3 \hat{\omega}
	+
	i\eta \htau^3 \hsig^3
	-
	\hsig^3 \hat{h}_{\bdg} + i \QspSwave
	}
=
	\hat{U}_{\omega}
	\frac{
		1
	}{
		\left[
		\hat{\Xi}_{\omega}
		-
		\varE_{\textbf{k}} 
		+ 
		i \hat{U}_{\omega}^{-1} \QspSwave \hat{U}_{\omega}
		\right]
	}
	\hat{U}_{\omega}^{-1}.
\end{aligned}
\end{equation}
Using the knowledge of the normal metal saddle point $\Qsp^{\mathsf{metal}} =\frac{1}{2\tel}  \htau^3 \hsig^3$, we have
\begin{equation}
\begin{aligned}
	\QspSwave(\omega)
	= 
	\frac{1}{2\tel}
	\hat{U}_{\omega} \htau^3 \hsig^3 \hat{U}_{\omega}^{-1}
	=
	\frac{1}{2\tel}
	\frac{\sign(\omega) \, \htau^3}{
		\sqrt{(\omega + i \eta \htau^3)^2 - \Delta_0^2}
	}
	\begin{bmatrix}
	\omega & -\Delta_0
	\\ 
	\Delta_0 & -\omega
	\end{bmatrix}_{\sigma}.
\end{aligned}
\end{equation}
In the $\Delta_0 = 0$ limit, we recover the normal metal saddle point $\Qsp^{\mathsf{metal}}$. 
Note that $\QspSwave$ can alternatively be obtained by solving the Usadel equation \cite{s_wave_NLSM4_Kamenev}.

The linear response optical conductivity is again given by Eq.~(\ref{eq:Kubo_def}).
At the saddle point level, we again just expand the action up to second of $\BigA$ and evaluate the 
expression in Eq.~(\ref{eq:S_2nd_derivative1}), but with $\Gdress \rightarrow \Gdress^{\swave}$ given 
by Eq.~(\ref{eq:app_G_swave_trans}).
By converting $\int_{\textbf{p}} \rightarrow \nu_0 \int d\varE \int
\frac{d\Omega_{\hat{\textbf{p}}}}{4\pi} $ and using the integral
\begin{equation}\label{eq:app_loop_integral}
	\int_{\textbf{p}}
	\frac{
	v_F^{\mu}(\textbf{p})
	v_F^{\nu}(\textbf{p})
	}{
	\left[
	\Xi_{\omega + \omega'}\sigma^3
	+
	i\Gel \tau^3\sigma^3 
	-
	\varE_\textbf{p}
	\right]
	\left[
	\Xi_{\omega'} \sigma^{3'}
	+
	i\Gel \tau^{3'} \sigma^{3'} 
	-
	\varE_\textbf{p}
	\right]
	}
\simeq
	2\pi \nu_0 D 
	\left( 
	1 
	- 
	\delta_{\tau^3 \sigma^3,\tau^{3'}\sigma^{3'}}
	\right)
	\delta^{\mu \nu},
\end{equation}
valid in the diffusive regime, we simplify Eq.~(\ref{eq:S_2nd_derivative1}) in the $s$-wave case, 
and obtain the optical conductivity in the dirty limit
\begin{equation}\label{eq:app_sigma_s_wave}
	\sigma_{\sfbb,R}^{\sfs,\mu \nu}(\Omega)
=
-
\frac{e^2}{i\Omega}
\frac{1}{2i}
\left\lbrace 
\frac{\pi \nu_0 D}{4}
\int_{\omega'}
\Tr
\left[
\hat{U}_{\Omega + \omega'}
\htau^3 \hsig^3
\hat{U}^{-1}_{\Omega + \omega'}
\hat{\gamma}_{\cl}(\Omega+ \omega',\omega')
\hat{U}_{\omega'}
\htau^3 \hsig^3
\hat{U}^{-1}_{ \omega'}
\hat{\gamma}_{\q}(\omega',\Omega+\omega')
\right]
\right\rbrace 
\delta^{\mu \nu},
\end{equation}
where $\int_{\omega'} = \int \frac{d\omega'}{2\pi}$.
By recognizing that 
$\frac{1}{2\tel}\hat{U}_{\omega}
\htau^3 \hsig^3
\hat{U}^{-1}_{\omega}$ 
is nothing but the $s$-wave saddle point $\QspSwave$, we arrive the Mattis-Bardeen result
\begin{equation}
	\sigma_{\sfbb,R}^{\sfs,\mu \nu}(\Omega)
	=
	\frac{
		\pi \sigma_{\mathsf{dc}}
	}{
	\Omega
	}
	\left\lbrace 
	\int_{\omega'}
	\left\{
	\frac{
		\left[
			\Delta_0^2 
			+ 
			(\Omega + \omega')
			\omega'
		\right]
		\sign(\omega')
		\,
		\sign(\Omega + \omega') 
	}{
		\sqrt{(\Omega+\omega')^2 - \Delta_0^2}
		\sqrt{\omega'^2 - \Delta_0^2}
	}
	\right\}
	\left[
	F(\Omega + \omega')
	-
	F(\omega')
	\right]
	\right\rbrace 
	\delta^{\mu \nu},
\end{equation}
where it is understood that the integral over $\omega'$ excludes regions with 
$\omega'^2 - \Delta_0^2 < 0$ 
and 
$(\Omega+\omega')^2 - \Delta_0^2 < 0$. 
Here, 
$\sigma_{\mathsf{dc}} = e^2 (2 \nu_0) D$ is the Drude conductivity in the normal state. 
In the $T = 0$ limit, the above integral can be performed analytically, resulting in
\begin{equation}
\sigma_{\sfbb,R}^{\sfs,\mu \nu}(\Omega)
=
\sigma_{\mathsf{dc}}
\left[
\left(
1
+
\frac{2\Delta_0}{\Omega}
\right)
E
\left(
\frac{\Omega - 2\Delta_0}{\Omega + 2\Delta_0}
\right)
-
\frac{4\Delta_0}{\Omega}
K
\left(
\frac{\Omega - 2\Delta_0}{\Omega + 2\Delta_0}
\right)
\right]
\theta(\Omega - 2\Delta_0)
\,
\delta^{\mu \nu},
\end{equation}
where $E$ and $K$ are complete elliptical integrals. 
We note that a similar derivation based on the $\hat{Q}$-matrix non-linear $\sigma$ model 
in the Matsubara formalism also appeared in a recent study \cite{s_wave_NLSM3_Mirlin}. 
\end{widetext}


\end{document}